\documentclass[fleqn,usenatbib]{mnras}

\usepackage[T1]{fontenc}

\DeclareRobustCommand{\VAN}[3]{#2}
\let\VANthebibliography\thebibliography
\def\thebibliography{\DeclareRobustCommand{\VAN}[3]{##3}\VANthebibliography}

\usepackage{graphicx}	
\usepackage{amsmath}	
\usepackage{booktabs}
\usepackage{txfonts}
\usepackage{natbib}
\usepackage{pdflscape}
\usepackage[normalem]{ulem}
\usepackage{microtype}
\usepackage{float}
\PassOptionsToPackage{breaklinks=true}{hyperref}
\usepackage{capt-of}
\usepackage{caption}

\usepackage{xcolor}

\definecolor{patricia}{RGB}{80, 200, 120}

\definecolor{peter}{RGB}{150,75,0}

\newcommand{\kms} {\,km\,s$^{-1}$}

\def\dn1{$\delta\nu_{01}$\xspace}
\def\dn2{$\delta\nu_{02}$\xspace}

\title[Tracing the evolution of Z\,Vul]{Unravelling Mass Transfer in Algols from Surface Abundances.\\
I. Z Vulpeculae\thanks{Based on observations made with the Mercator Telescope, operated on the island of La Palma by the Flemish Community, at the Spanish Observatorio del Roque de los Muchachos of the Instituto de Astrofísica de Canarias. Based on observations obtained with the {\sc hermes} spectrograph, which is supported by the Research Foundation - Flanders (FWO), Belgium, the Research Council of KU Leuven, Belgium, the Fonds National de la Recherche Scientifique (F.R.S.-FNRS), Belgium, the Royal Observatory of Belgium, the Observatoire de Genève, Switzerland and the Thüringer Landessternwarte Tautenburg, Germany.}
}

\author[Dervi{\c{s}}o{\v{g}}lu\,et\,al.]
{A.\,Dervi{\c{s}}o{\v{g}}lu,$^{1,2}$\thanks{E-mail: ahmet.dervisoglu@atauni.edu.tr}
S.\,Adalalı,$^{2}$
F.\,G\"{u}ney,$^{3,4}$
B.\,Hoyman,$^{5}$
T.\,{\c{S}}ahin,$^{3,4}$
Ö.\,{\c{C}}akırlı,$^{5}$
K.\,Pavlovski,$^{6}$
\newauthor D.\,Mkrtichian,$^{7}$
P.\,De Cat,$^{8}$
and P.\,Lampens$^{8}$
\\
$^{1}$Department of Astronomy and Space Sciences, Atat\"{u}rk\,University, \,Faculty of Science, \,Erzurum, \,25050, \,T\"{u}rkiye \label{inst:ATU}\\
$^{2}$T\"{u}rkiye National Observatories, \,DAG, \,25050, \,Erzurum, \,T\"{u}rkiye \label{inst:TNO}\\
$^{3}$Department of Space Sciences and Technologies, Faculty of Science, Blok-B, Akdeniz University, Antalya 07058, \,T\"{u}rkiye\\
$^{4}$Graduate Institute of Natural and Applied Sciences, Akdeniz University, 07058, Antalya, \,T\"{u}rkiye\\
$^{5}$Department of Astronomy and Space Sciences, Faculty of Science, Ege University, 35100 Bornova, Izmir, \,T\"{u}rkiye\\
$^{6}$Department of Physics, Faculty of Science, University of Zagreb, 10\,000 Zagreb, Croatia\\
$^{7}$National Astronomical Research Institute of Thailand, 191 Siriphanich Bldg., Huay Kaew Rd., Chiang Mai 50200, Thailand\\
$^{8}$Royal Observatory of Belgium, Ringlaan 3, B-1180 Brussel, Belgium}

\date{Accepted XXX. Received YYY; in original form ZZZ}

\pubyear{\the\year{}}

\begin{document}
\label{firstpage}
\pagerange{\pageref{firstpage}--\pageref{lastpage}}
\maketitle

\raggedbottom
\begin{abstract}
The photospheric abundance pattern of the components is a sensitive probe of mass and angular momentum transfer in Algols.  With the aim to trace the evolutionary history and disclosed efficiency of mass transfer in Algols  we collected new high-resolution spectra for about a dozen Algols with the {\sc hermes} spectrograph at the Mercator Telescope on La Palma, Canary Islands. In the present paper, which is the first in the series, we present the results of a comprehensive analysis of Z Vul, a hot Algol-type system. We derived the absolute stellar quantities and elemental abundances for the components. The carbon-to-nitrogen (C/N) ratio is the most sensitive tracer of thermonuclear processing and mass transfer. It is found to be within expected value for gainer but donor lacks the severe C/N inversion traditionally expected for a deeply stripped star. An extensive grid of over three million evolutionary models was calculated  with the Cambridge \texttt{STARS} code which allows determination of the best-fitting and all surviving models. Constraints from the He and CNO abundances indicate that Z Vul underwent an episode of nearly conservative mass transfer during the Main Sequence life-time of the originally more massive component (Case A). Our detailed chemical tagging confirms that the surface abundances of the mass-losing component represent the delicately stripped outer layers still not reaching the CNO-rich core. This underscores the need of coupling high-resolution spectroscopic abundances with interior stellar profiles to accurately reconstruct the evolutionary history of interacting binaries.
\end{abstract}
 



\section{Introduction}
Algol-type binary systems present intriguing evolutionary challenges. A less massive (sub)giant star orbits a more massive main-sequence star, a configuration that is seemingly inconsistent with stellar evolution. This discrepancy in mass \citep{Struve_1948}, known as the `Algol paradox', is resolved by invoking the large-scale mass transfer (MT) process between the components \citep{Crawford_1955}. Initially, the more massive star is evolving faster than its less massive companion, but has only limited space for expansion once it reaches the (sub)giant evolutionary phase. 
After having filled its Roche lobe (RL = critical equipotential), defined by the mass ratio between the components, it begins rapidly losing mass to its less massive companion. The result of this almost cataclysmic process is a reversal in the mass ratio as observed in the classical Algols.
\par
The impact of MT extends beyond the reversal of the mass ratio. It significantly alters the photospheric chemical compositions of both stars because deeper layers of the mass-losing component become exposed while its former surface layers are deposited on its companion. 
Thus, the abundances of helium (He) and carbon-nitrogen-oxygen (CNO) can be used to trace the evolutionary history of Algols. In the course of the CNO cycle during the main sequence evolution, the relative abundances of both these species and helium, which is a product of this nucleosynthesis cycle, are significantly altered. Indeed, He and N are enhanced while C and O are depleted \citep{Sarna_1996}. Pioneering observational studies of the abundances of the He and CNO species in the spectra of binary stars confirmed the theoretical expectations and serve as an important proof of large-scale MT in Algols \citep{Plavec_1982, Plavec_1983, Parthasarathy_1983, Dobias_1985, Balachandran_1986, Tomkin_1989, Cugier_1989, Tomkin_1993}. 
The mass transferred from the mass-losing star (donor) also carries a large amount of specific angular momentum (AM), which is imparted to the mass-gaining star (gainer) upon accretion \citep{Dervisoglu2010}. 
Modeling AM evolution requires calculating the efficiency of MT and the subsequent loss of mass and AM from the system. 
The total AM of the system ($J_t$) is the sum of the orbital ($J_{orb}$) and stellar AM ($J_g$, $J_d$) contributions \citep{Deschamps2014}. The evolution of angular momentum is crucial because its conservation or loss dictates several key aspects of the dynamics of close binary systems. Angular momentum evolution is intrinsically linked to the efficiency and nature of mass transfer, distinguishing between conservative and nonconservative evolution. Determining the efficiency of mass transfer, quantified by the mass transfer efficiency parameter ($\beta$), is a primary source of uncertainty in the modeling binary star evolution. Nonconservative mass transfer results in a loss of the total system mass and angular momentum. Studies have inferred that prototype Algols like $\beta$ Per must have lost at least 15$\%$ of their initial total mass and 30$\%$ of their total angular momentum during evolution \citep{Deschamps2015}. One of the main uncertainties in modeling binary star evolution is calculating the efficiency of mass transfer and the subsequent mass and angular momentum loss from the system.
\par
The evolution of AM is crucial because its conservation or loss dictates several key aspects of the dynamics of close binary systems. AM evolution is intrinsically linked to the nature and the efficiency of MT, distinguishing between conservative and non-conservative evolution. Determining the efficiency of MT, quantified by the MT efficiency parameter ($\beta$), is a primary source of uncertainty in the modeling of binary star evolution.Non-conservative MT results in a loss of the total system mass and AM. Studies have inferred that prototype Algols like $\beta$ Per must have lost at least 15$\%$ of their initial total mass and 30$\%$ of their total AM during their evolution \citep{Deschamps2015}. One of the main uncertainties in the modeling of the evolution of binary stars is calculating the efficiency of MT (i.e. determining $\beta$) and the subsequent mass and AM loss from the system.
\par
Determination of the chemical abundances in the spectra of binary systems is hampered by ever-changing line blending of the components' spectra in the course of the orbital cycle and, in particular for Algols, by the large light ratio between the components. The mass-losing component in the inner pair of the $\beta$ Per triple system  is contributing barely 1\% to the total light of the Algol system, and its spectral lines were detected fairly late \citep{Tomkin_1978}. The technique of spectral disentangling (\texttt{SPD}) \citep{Simon_1994, Hadrava_1995} overcomes the difficulties of line blending between the spectra of the components and makes it possible to reconstruct the individual component spectra \citep{Hensberge_2000}, which then can be analysed with all spectroscopic tools as for single-star spectra \citep{Pavlovski_2005, Pavlovski_Southworth_2009, Pavlovski_2009}. 
The power of \texttt{SPD} is illustrated with the separation of the complete optical spectrum of the giant mass-losing component in the Algol system \cite{Kolbas_2015} and the secondary's spectrum in the binary system $\alpha$~Dra in which the spectral lines are almost washed out due to high rotational velocity \citep{Pavlovski_2022}. 
\par
Elemental abundances have been determined from the disentangled spectra of the components in several Algol-type systems: RZ~Cas \citep{Tkachenko_2009}, TW~Dra \citep{Tkachenko_2010}, u~Her \citep{Kolbas_2014}, Algol \citep{Kolbas_2015}, XZ~Cep \citep{Martins_2017} and $\delta$~Lib \citep{Dervisoglu_2018}, whilst synthetic composite spectra were used in the analysis of TX~UMa \citep{Glazunova_2011}. The detailed abundance analyses listed above confirm the general trend of carbon underabundance in mass-gaining components of some classical Algols found by \citet{Ibanoglu2012} in a comprehensive observational study of the C\,{\sc II} line at $\lambda$\,4267\,{\AA}. Extensive evolutionary analyses based on the carbon-to-nitrogen (C/N) abundance ratio for the hot Algol-type system u~Her \citep{Kolbas_2014} and the classical Algol $\delta$~Lib \citep{Dervisoglu_2018} favour a conservative to very mildly non-conservative mass-transfer regime, accompanied by little mass loss from the system.
\par
Recent work has extended the use of surface abundances to a much broader domain than just Algols. Homogeneous surveys of large samples of Galactic O-type stars have uncovered both apparently single stars and single-lined spectroscopic binaries (SB1) whose helium and nitrogen surface patterns simply cannot be reproduced by single-star evolutionary models. The most plausible explanation is that these objects are mass gainers from past interaction events, polluted by helium- and nitrogen-rich material transferred during Roche-lobe overflow in their earlier lives, while their stripped donor companions have now become too faint to detect \citep{Martinez_2025, Simon_2026, Martinez_2026}. Thus, photospheric He and N abundances serve as powerful flags for binary interaction products even among seemingly isolated massive stars, and they offer valuable constraints on the mass and angular changes that occur in interacting binaries.
\par
The efficiency of the MT process has also broader relevance because the same process, at higher masses, governs the formation of compact-object binaries and their gravitational-wave (GW) signals. Detailed binary models identify stable Case~A transfer in short-period massive systems as a channel to the merging of binary black holes \citep{Xu_2025}; matching the observed orbital periods and mass ratios of massive Algols points to near-conservative MT in some systems and highly inefficient transfer in others \citep{Sen_2022, Sen_2026}, while the post-interaction separation is set by the specific AM carried off by the ejected matter \citep{Klencki_2026, Olejak_2025}. Because the donor's surface nitrogen enhances toward its CNO-equilibrium value through the semi-detached phase as processed layers are uncovered \citep{Sen_2022}, the photospheric C/N ratio records the mass-transfer history.
\par
In the present paper, we extend our previous analysis \citep{Kolbas_2014, Kolbas_2015, Pilecki2017, Dervisoglu_2018} to the hot Algol-type binary system Z~Vul. This work aims to confront high-quality observational data with theoretical evolutionary models of binary systems, in order to unravel the complex processes that shape the evolution of these systems and, in particular, to characterise the properties of the MT between the components. The analysis of Z~Vul presented here represents the first in a series of planned publications dedicated to this topic. The paper is organised as follows. A short overview of the Algol-type binary system Z~Vul is given in Section~\ref{sec:Z_Vul}, and satellite photometric and the new spectroscopic observations are described in Section~\ref{sec:Obs}. In Section~\ref{sec:ANALYSIS}, we describe the methodology, with emphasis on the separation of the individual component spectra and the feedback between the different steps of the analysis; the light-curve solution, the radial velocities (RVs) measured from our spectra and the detailed abundance analysis from the disentangled component spectra are also reported. The calculation of the theoretical evolutionary models and their confrontation with the measured He and CNO abundances are presented and discussed in Section~\ref{sec:Evolutionary Analyses}. We summarise our conclusions in Section~\ref{sec:Concluding}.

\section{Overview of Z Vul}
\label{sec:Z_Vul}
Z\,Vul (HD\,181987, TIC\,90171962) is an eclipsing double-line spectroscopic binary system with the characteristics of Algol-type systems. It consists of a main-sequence star of spectral type B3V-B5V and a Roche-lobe filling star of spectral type around A5IV in a circular orbit with period $P = 2.45$\,d \citep{Lazaro2009}.
\par 
The first photoelectric observations of Z\,Vul were secured by \citet{Popper_1957b} in 1951 and 1956 at the Lick and Palomar observatories, respectively. He used an instrumental photometric system but provided relations to transform it to the standard Johnson $UBV$ system. Also, \citet{Popper_1957a} obtained 18 spectra in 1946 at the McDonald Observatory, and in 1955 and 1956 with the Coud\'e and grating spectrographs of the 100-inch and 60-inch at the Mt Wilson Observatory, respectively. He measured the RVs and eventually adopted the RVs semi-amplitudes $K_1 = 92\pm2$\kms\ and $K_2 = 219\pm4$\kms\ as a compromise with an early result by \citet{Plaskett1920}, which gives the mass ratio, $q = 0.42\pm0.02$. \citet{Popper_1957a} solved the light curves with the Russell-Merrill method and obtained absolute parameters for the components of Z\,Vul, confirming the Algol characteristics with a hotter star of almost twice the mass of its cooler giant companion. 
\par
Further photoelectric photometry was gathered by \citet{Broglia1964}. He did photometric observations with a 0.60-m telescope at the Observatoire du Pic du Midi and secured altogether about 3000 measurements in the standard Johnson $UBV$ system. The light curves published by \citet{Broglia1964} were analysed for the first time by \citet{Cester1977} using a novel, straightforward synthesis approach that approximates the stars as triaxial ellipsoids \citep{Wood_1971, Wood_1973}. \citet{Cester1977} confirmed a semi-detached configuration for Z\,Vul but noticed that the primary (hotter) component does not follow the mass-luminosity relation. Indeed, it is undermassive for its luminosity, spectral type and its luminosity class. In general, their solution is in accordance with the solution of \citet{Popper_1957a}, except for the primary's effective temperature.
\par
New photometric and spectroscopic observations were reported by \citet{Lazaro2009}. They secured almost 2\,000 near-infrared (NIR) photometric measurements in the $JHK$ passbands. In addition, 11 spectra were collected with the IDS spectrograph on the 2.5-m Isaac Newton Telescope (INT) at the Roque de los Muchachos Observatory with a spectral dispersion of about 0.5 \AA/pixel. The spectra cover a spectral range from about 4\,200 to 9\,100\,{\AA}. 
\par
In a comprehensive analysis, \citet{Lazaro2009} combined all available photometry (instrumental $UBV$ by \citet{Popper_1957b}, standard $UBV$ by \citet{Broglia1964} and their $JHK$ photometry) as well as the RVs measurements (28 spectra from \citet{Popper_1957a} and their own measurements from the INT spectra). Their own code {\sc BinaRoche} \citep{Lazaro_2002} for the simultaneous multipassband light curves solution in the Roche geometry was employed in the calculations. The authors slightly revised the mass ratio to $q = M_2/M_1 = 0.44\pm0.02$, probably because they took into account the Rossiter-McLaughlin effect \citep[i.e. the distortion of the spectral line profiles in the course of the eclipses][]{Cegla_2016, Kunovac_2020}. A simultaneous fit of the combined $UBVJHK$ light curves gives the absolute parameters for the components of Z\,Vul: $M_1 = 5.29\pm0.50$\,M$_\odot$, $R_1 = 4.93\pm0.18$\,R$_\odot$, and $T_{\rm eff,1} = 18\,000\pm300$\,K for the primary; $M_2 = 2.33\pm0.25$\,M$_\odot$, $R_1 = 4.67\pm0.17$\,R$_\odot$, and $T_{\rm eff,2} = 9300\pm150$\,K for the secondary.
\par 
It should be noted that the solution of \citet{Popper_1957a} is within the 1-$\sigma$ uncertainties of the solution found by \citet{Lazaro2009}. The solution given by \citet{Cester1977} deviates considerably in the estimated effective temperature for both components. The absolute dimensions, in particular for the primary component, found by \citet{Ghoreyshi2008}, are considerably different from the values published by others, probably because they used substantially different RV semi-amplitudes.
\par
In several papers, the authors presented evidence for still ongoing MT between the components of the Z\,Vul binary system. We thus cite, for example:

- An increase of the orbital period was reported by \citet{Ibanoglu2012} with the rate $\dot{P} = (3.95\pm0.46) \times 10^{-8}$\,yr$^{-1}$. This period change translates into the rate $\dot{M} = (5.49\pm0.68) \times 10^{-8}$\,M$_\odot\text{yr}^{-1}$, indicating a modest but slow ongoing MT from the Roche-lobe filling component to the more massive star. An early result of the period change was reported by \citet{Simon1999}.

- The measurements of the equivalent width of the spectral line C\,{\sc ii} $\lambda4267$\,{\AA} of the mass-gaining component in Z\,Vul showed a significantly depleted carbon abundance relative to the solar value, as found by \citet{Ibanoglu2012}. The authors interpreted this finding in terms of contamination of matter due to the MT.

- \citet{Lazaro2009} examined the infrared $JHK$ light curves and the RVs and found the presence of asymmetries or deviations from a purely geometric binary model, suggesting ongoing MT processes in the system. A simultaneous analysis of the visual (\textit{UBV}) and infrared (\textit{JHK}) light curves revealed a notable distortion: the ingress of primary eclipse appeared slightly depressed compared to the light-curve model, which otherwise provided a good fit to the visual light curves. This flux depression reached its maximum at orbital phase 0.86, approximately. The amplitude of the effect appeared to diminish toward the end of the primary eclipse, suggesting a localised, phase-dependent distortion likely associated with circumstellar material or spot activity.

- Z\,Vul is included in the compilation of linear polarisation measurements in close binaries \citep{Pfeiffer1977} with the mean value $\overline{P}(\%) = 0.9\pm0.1$ per cent as measured by \citet{Shakhovskoi_1965}. Detection of polarisation is attributed to scattering in circumbinary or circumstellar material, likely resulting from mass flows and redistribution of matter associated with MT within the system.

- \citet{Peters_1998} analysed the far-ultraviolet (FUV) spectrum of Z\,Vul  acquired in 1996 during the ORFEUS-SPAS II mission on the Space Shuttle STS-80. The spectra span the region 900–1220\,{\AA} and have a resolution of approximatelly 0.33\,{\AA}. They found absorption and emission lines of highly ionised species of N\,{\sc v}, C\,{\sc iv} and Si\,{v}, indicating the presence of a high-temperature circumstellar plasma within the system. Z\,Vul is a close binary system with a relatively large mass-accreting component, and there is no room for the formation of an accretion disk. Instead, a direct impact of the mass stream from the Roche-lobe-filling component is expected, and the FUV spectrum is direct evidence. Further evidence for an absence of an accretion disk is the absence of emission in the H$\alpha$ line.

- Z\,Vul has been detected as an IRAS source, IRAS\,19196+2529 in the 60\,$\mu$m band and with upper limits in 12, 25 and 100\,$\mu$m \citep{Friedemann_1996}. Hence, no clear long-IR excess is present. The binary was also not detected in a radio survey at 10.6\,GHz by \citet{Woodsworth_1977}

\section{THE PHOTOMETRIC AND SPECTROSCOPIC OBSERVATIONS}
\label{sec:Obs}
\subsection{TESS Photometry}
\label{sec:Photometry_Obs}
Space photometry on-board the NASA Transiting Exoplanet Survey Satellite\footnote{\url{https://tess.mit.edu}} (TESS; \citet{Ricker2014}) is achieving high quality precision. Z\,Vul was observed in five sectors: 14 (July 2019), 40 (June 2021), 41 (July 2021), 54 (July 2022), and 81 (July 2024). Sectors 40, 41, 54, and 81 provide short cadence sampling at 120\,s and are in perfect agreement whereas Sector~14 is available only at the 1800\,s cadence of the full frame images. In homogeneity, we therefore base the light curve solution on the four short cadence sectors and exclude Sector~14. All selected data were retrieved from the NASA Mikulski Archive for Space Telescopes (MAST)\footnote{\url{https://mast.stsci.edu/portal/Mashup/Clients/Mast/Portal.html}}. To extract the TESS flux, we downloaded the \textit{target pixel files} (TPFs) for all observed sectors using the \texttt{Lightkurve} Python package \citep{Lightkurve2018}. The TPFs contain the raw pixel-level data within the predefined masks acquired by the spacecraft, together with the Science Processing Operations Center (SPOC) pipeline aperture definitions \citep[e.g.][]{Jenkins_2016}. We tested a range of aperture masks for each sector, beginning with the pipeline defined mask and proceeding to custom apertures of varying sizes and evaluated the resulting Simple Aperture Photometry (SAP) light curves for photometric quality and potential contamination from nearby sources, selecting the optimal aperture on a per sector basis. The SAP light curves of the four short cadence sectors (Sectors 40, 41, 54, and 81) were downloaded from MAST and prepared by converting the timestamps to MJD and extracting the flux. Each sector was normalised independently by dividing its flux by the out of eclipse maximum. Before combination, the geometric configurations of the individual sectors were found to be mutually consistent within their uncertainties. The four normalised light curves were then phase folded with a single linear ephemeris of Z\,Vul and combined into one phase distribution then averaged within uniform phase bins. Each populated on average by approximately fifteen measurements, yielding a master phased light curve of $5\,000$ points at a resolution of $\Delta\phi = 2\times10^{-4}$ for the subsequent synthesis.
\par
The TESS pixels subtend a large angle (21\,arcsec), so the extracted light curves could be contaminated with `third' light. We queried the Gaia DR3 database\footnote{\url{https://vizier.cds.unistra.fr/viz-bin/VizieR?-source=I/355/gaiadr3}} for all sources within 1\,arcmin of Z\,Vul. Many sources were returned due to the proximity of the Galactic plane. All are much fainter than our target ($G = 7.372$\,mag) except one with $G = 11.484$\,mag, which is at a distance of about 13\,arcsec from Z\,Vul. It is an optical double listed in the Washington Catalogue \citep{Mason_2001}. This source lies near the edge of the adopted apertures for all sectors, and its faintness relative to our target ensures that its contribution to the measured flux is negligible. Any residual signal is further suppressed because the pixel selection for the optimal apertures is based on the Pixel Response Function (PRF). We therefore find no evidence for a significant third light ($l_3$) component in our light curve solution, and the contribution of this neighbour is expected to be below the detection threshold of the binary modelling.

\subsection{New {\sc hermes} high-resolution spectroscopy}
\label{sec:Spectroscopy_Obs}
A set of high signal-to-noise (S/N), high-resolution e\'chelle spectra of Z\,Vul was obtained between July 2020 and March 2023 with the High-Efficiency and high-Resolution Mercator Echelle Spectrograph ({\sc hermes}) attached to the 1.2-m Mercator Telescope at the Roque de los Muchachos Observatory on La Palma (Canary Islands, Spain)\footnote{\url{https://www.mercator.iac.es/instruments}}. {\sc hermes} is a fiber-fed high-resolution ($R = 85\,000$) spectrograph that is very efficient \citep{Raskin_2011}. These spectra cover the entire optical and NIR spectral range from 3\,770 to 9\,000\,{\AA} in 55 spectral orders. A total of 29 spectra were obtained with S/N between 130 and 240 around $\lambda\,5\,500$\,{\AA}. The corresponding observing log is given in Table \ref{tab:speclog}. 
\par
The raw \'echelle frames were processed with the dedicated automated {\sc hermes} reduction pipeline \citep{Raskin_2011}, which performs bias and inter-order background subtraction, flat-fielding with order-localised blaze and pixel-to-pixel corrections, optimal order extraction, wavelength calibration, cosmic-ray removal, and barycentric velocity correction. Accurate determination of the continuum level in the observed spectra is crucial in the \texttt{SPD} method, which is an important part of our analysis. Any error in the continuum determination of the observed spectra would manifest itself as a wavy pseudo-continuum in the disentangled component spectra, which in turn affects the line depths. For the continuum normalisation we used the Python-based interface {\sc pegasus}\footnote{\url{https://github.com/dervisoglu/pegasus}}, which was employed for the order-extracted spectra and for the continuum normalisation of the merged \'echelle spectra. This order-by-order approach enables a reliable normalisation of the orders that contain the broad Balmer lines, which span approximately 100 to 150\,{\AA} and and facilitated a robust merging of the \'echelle orders in the spectrum merging stage. The software determines the continuum level automatically while preserving the profiles of the Balmer and other strong lines. For each \'echelle order, the blaze function was determined by interpolating the blaze profiles of the adjacent orders, as was used in \citet{Kolbas_2015}. It also yields a robust combination of the orders at the merging stage. Telluric features were removed simultaneously during normalisation through the integrated algorithm of {\sc pegasus}.
\par
In addition, a high-resolution spectrum acquired in May 2010 using the Coud\'{e} Echelle Spectrometer (CES) at the 1.5-m RTT150 telescope of the Türkiye National Observatories (TRG\"{o}z), has also been used \citep{Musaev_2000}. This spectrograph provides a broad wavelength coverage from 3\,700 to 10\,000\,{\AA} in 85 \'echelle orders achieving a spectral resolution of $R = 125\,000$. This spectrum was already used in the analysis by \citet{Ibanoglu2012}.
\par
Furthermore, we found another 10 spectropolarimetric spectra of Z\,Vul in the PolarBase Database \citep{Petit2014}. These spectra were obtained with the Echelle SpectroPolarimetric Device for the Observation of Stars (ESPaDOnS) at the Canada-France-Hawaii Telescope (CFHT). ESPaDOns is a bench-mounted high-resolution stellar spectropolarimeter which operates across the optical wavelength domain (from 3\,700 to 10\,500\,{\AA}) and provides high-resolution spectroscopic and polarimetric data ($R = 81\,000$) \citep{Donati1997}. The archival spectra were secured on the same night in Sept. 2014. Hence, we stacked them to enhance the S/N.

\section{ANALYSIS}
\label{sec:ANALYSIS}
\subsection{Light curves analysis}
\label{sec:Photometry}
First, we collected the available photometric measurements with an emphasis on well-standardised photometric systems. The light curves in a broad wavelength range are needed for the determination of the wavelength-dependent light ratio between the components. After examination of all available photometry of Z\,Vul, the following data sets were selected for the analysis:
\begin{enumerate}
\item Photoelectric measurements in the standard Johnson $UBV$ system by \citet{Broglia1964}. He used the 0.60-m telescope at the Observatoire du Pic du Midi and in the seasons 1955 (29 nights) and 1957 (2 nights) secured 1001 measurements in the $U$, 1059 in the $B$, and 995 in the $V$ passbands.
\item Near-infrared photometric measurements in the Johnson $JHK$ passbands by \citet{Lazaro2009} obtained in June 1997 with the 1.5-m CST telescope at the Observatorio del Teide, Canary Islands. A cooled broadband filter CVF photometer with an InSb detector was used. The collected number of measurements is 224 in the $J$ passband, 231 in $H$ and 231 in $K$.
\item The TESS detector employs a single broad bandpass spanning approximately 6\,000 to 10\,000 {\AA}, which we modelled directly using the TESS response function implemented in the 2015 version of the Wilson-Devinney (\texttt{WD}) code, band 95\footnote{\url{https://faculty.fiu.edu/~vanhamme/lcdc2015}} \citep{Wilson1971, Wilson2014}. At a cadence of 120\,s, the four sectors comprise 20\,309 (Sector 40), 19\,149 (Sector 41), 18\,890 (Sector 54), and 19\,174 (Sector 81) measurements. Following the procedure described in Sect. \ref{sec:Photometry_Obs}, these were combined into a single phase distribution and averaged within uniform phase bins, yielding the 5\,000 point master phased light curve adopted for the synthesis.
\end{enumerate}
\par
For the analysis, we used the 2015 version of the \texttt{WD} code \citep{Wilson1971, Wilson2014} and the \texttt{PyWD2015} program, which is an interface developed for this code \citep{Güzel2020}. The analysis was conducted under the assumption of a semi-detached system (Mode 5), in which the secondary component fills its Roche lobe. Based on this assumption, the surface potential ($\Omega_{\rm 2}$) of the secondary component is assumed to be equal to the critical Roche potential. In addition, the relative luminosity (L$_{\rm 2}$) of the secondary component was calculated automatically, based on other system parameters. The logarithmic limb-darkening coefficients ($x_{\rm 1, 2}$ and $y_{\rm 1, 2}$) of the components were calculated using the tables of \cite{van1993} with the internal routines of the \texttt{WD} code. Initially, the bolometric albedos (A$_{\rm 1, 2}$) and gravitational darkening coefficients (g$_{\rm 1, 2}$) of the components were assumed to be 1.0 \citep{ Rucinski_1969, Lucy_1967}. The calculations were started, assuming synchronous rotation of the components ($F_{\rm 1, 2}$ = 1.0). Before combining the four sectors, we solved each TESS sector light curve independently with the \texttt{WD} code, obtaining four distinct light fractions $l_{fi}$ at the TESS passband. The four solutions returned geometric configurations that agree within their $1\sigma$ uncertainties, and in every sector the third light was negligible. We retained all four epoch values $l_{fi}$ as independent points in the light fraction function $l_f(\lambda)$ (Sect. \ref{sec:Atmos}).
\par
A preliminary solution was obtained with the primary's effective temperature fixed at the value obtained by \citet{Ibanoglu2012}, and the mass ratio determined by \citet{Popper_1957a}. The $l_{fi}$ ratios obtained from this step were subsequently used in renormalisation of the disentangled spectra and the determination of the atmospheric parameters for the components, as will be described in Sect.\,\ref{sec:Disent} and Sect.\,\ref{sec:Atmos}. Two parameters come from the solution of the RV curve: the mass ratio, $q=M_2/M_1$, and the semi-major axis, $a$. The mass ratio defines the size of the Roche-lobe filling component and the system configuration. These parameters were determined in another iteration with \texttt{WD} code using the RVs listed in Table\,\ref{tab:speclog}. 
\par
One point to consider during the computations is that, because the \texttt{WD} code uses a semi-major axis as one of the input parameters, when the RV data are not used, a change in orbital inclination ($i$) will cause the semi-major axis not to adapt to this change. This not only changes the calculated radii of the components but also affects the radius of the critical Roche potential surface; thus, in the case of a semi-detached solution, it will also affect the constrained radius of the component filling its Roche lobe. This adds an error to the solution. Therefore, a change was made to the \texttt{PyWD2015} code, and instead of the semi-major axis, the value $a \sin(i)$ was entered into the program. The semi-major axis was calculated according to the instantaneous $i$ value and written to the input files of the \texttt{WD} code, thereby allowing the program to dynamically change the semi-major axis based on orbital inclination. In subsequent runs, these parameters were fixed, and the solutions were sought only for the light curves. All light-curve solutions were recomputed with the improved atmospheric parameters (cf. \ref{sec:Atmos}). 

\begin{table*}
\begin{center}
\captionof{table}{Comparison of the orbital and geometric parameters of Z\,Vul derived from different modeling configurations across diverse light curve datasets. The tabulated values contrast the solutions obtained via (i) attenuation region model, (ii) the starspot optimization approach, and (iii) the definitive radial velocity curve analysis.}
\label{tab:LC_RV}
\small
\resizebox{\textwidth}{!}{%
\begin{tabular}{lllllll}
\hline
\hline
Parameter/Solution                                    & Attenuation Region (TESS) & Spot (TESS)          & \cite{Broglia1964}   & \cite{Lazaro2009}      &   RV \\ 
\hline
\textit{i} ($^\circ$)                                 & 86.97 $\pm$ 0.04          & 87.27  $\pm$ 0.39    & 88.57 $\pm$ 0.90     & 88.87 $\pm$ 0.04       &  87.27 (fix) \\ 
$q~(=M{_2}/M{_1})$                                    & 0.3911 $\pm$ 0.0073       & 0.3911 $\pm$ 0.0073  & 0.3911 $\pm$ 0.0073  & 0.3911 $\pm$ 0.0073    &  0.3911 $\pm$ 0.0073 \\ 
$a$~({\mbox{$R_{\odot}$}})                            & 15.746 $\pm$ 0.115        & 15.746 $\pm$ 0.115   & 15.746 $\pm$ 0.115   & 15.746 $\pm$ 0.115     & 15.746 $\pm$ 0.115 \\ 
$V_{\gamma}$  (km s$^{-1}$)                           & -18.92 $\pm$ 0.91         & -18.92 $\pm$ 0.91    & -18.92 $\pm$ 0.91    & -18.92 $\pm$ 0.91      & -18.92 $\pm$ 0.91 \\
$T_{1}$ (K)                                           & 16\,760       (fix)       & 16\,760 (fix)        & 16\,760 (fix)        & 16\,760 (fix)          & 16\,760  (fix) \\ 
$T_{2}$ (K)                                           & 9\,020 $\pm$ 150          & 9\,198 $\pm$ 390     & 8\,500 $\pm$ 250     & 8\,900 $\pm$ 500       & 9\,198   (fix) \\ 
$\Omega$                                              & 3.7010 $\pm$ 0.0409       & 3.7450 $\pm$ 0.0640  & 3.8574 $\pm$ 0.1274  & 3.8050 $\pm$ 0.1150    & 3.7450  $\pm$ 0.0640 \\ 
$l_1 /\left(l_1+l_2\right)_{\mathrm{\textit{U}}}$     & -                         & -                    & 0.930 $\pm$ 0.018    & -                      &  - \\
$l_1 /\left(l_1+l_2\right)_{\mathrm{\textit{B}}}$     & -                         & -                    & 0.868 $\pm$ 0.018    & -                      &  - \\
$l_1 /\left(l_1+l_2\right)_{\mathrm{\textit{V}}}$     & -                         & -                    & 0.847 $\pm$ 0.016    & -                      &  - \\
$l_1 /\left(l_1+l_2\right)_{\text{\textit{TESS}}}$    & 0.762 $\pm$ 0.006         & 0.756 $\pm$ 0.008    & -                    & -                      &  - \\
$l_1 /\left(l_1+l_2\right)_{\mathrm{\textit{J}}}$     & -                         & -                    & -                    & 0.654 $\pm$ 0.015      &  - \\
$l_1 /\left(l_1+l_2\right)_{\mathrm{\textit{H}}}$     & -                         & -                    & -                    & 0.627 $\pm$ 0.016      &  - \\
$l_1 /\left(l_1+l_2\right)_{\mathrm{\textit{K}}}$     & -                         & -                    & -                    & 0.623 $\pm$ 0.016      &  - \\
$l_3$                                                 & -                         & -                    & -                    & -                      &  - \\
\hline
$\phi_p^{\circ}$                                      & -                         & 105 $\pm$ 23         & 105 (fix)            & 105 (fix)              & 105 (fix)\\ 
$\lambda_p^{\circ}$                                   & -                         & 360 $\pm$ 2          & 360 (fix)            & 360 (fix)              & 360 (fix)\\ 
$R_p^{\circ}$                                         & -                         & 30 $\pm$ 11          & 30  (fix)            & 30 (fix)               & 30 (fix)\\
$f_{T,p}$                                             & -                         & 0.85 $\pm$ 0.05      & 0.92 $\pm 0.04$      & 0.82 $\pm$ 0.02        & 0.85 (fix)\\
\hline
$\phi_s^{\circ}$                                      & -                         & 143 $\pm 24$         & 143 (fix)            & 143 (fix)              & 143 (fix)\\ 
$\lambda_s^{\circ}$                                   & -                         & 358 $\pm 3$          & 358 (fix)            & 358 (fix)              & 358 (fix)\\ 
$R_s^{\circ}$                                         & -                         & 65  $\pm 23$         & 65  (fix)            & 65  (fix)              & 65  (fix)\\
$f_{T,s}$                                             & -                         & 0.86 $\pm$ 0.05      & 0.87 $\pm$ 0.12      & 0.96 $\pm$ 0.02        & 0.86 (fix)\\ 
\hline
\end{tabular}}
\end{center}
\end{table*}

\par
The observed light curve of Z Vul often presents asymmetries or features, such as flux depressions during eclipse ingress/egress, that deviate significantly from predictions based on a purely geometric binary model. These deviations provide direct evidence of circumbinary or circumstellar material, such as an accretion disc or dense gas stream, resulting from MT. While the \texttt{WD} code \citep{Wilson2014} includes provisions for "Circumstellar Light-attenuating Regions" to model such features, the physical parameters governing these regions, specifically the wavelength-dependent attenuation law exponent ($\alpha$), mean molecular weight ($\mu_e$), and electron density ($n_e$), are non-optimizable within the standard differential corrections (DC) algorithm. Consequently, deriving a solution that accurately determines stellar geometric and thermal properties while accounting for physical obscuration through fixed regions is beyond the scope of this study. To overcome this limitation and ensure a physically grounded yet computationally flexible solution, we adopted a validated workaround that is frequently utilized in the literature. Rather than assuming a fixed, unoptimized geometry for the circumstellar material, we approximated the morphological impact of localized obscuration by introducing stellar spot parameters into our model. Although spot formation due to magnetic activity was not expected, given the masses and evolutionary states of the components, this pragmatic strategy simulates the photometric effect of accumulated matter from MT using localized thermal anomalies. This approach has been successfully applied to the analysis of other Algol-type binaries \citep{Dervisoglu_2018, Lehmann2020}. The O’Connell effect in Algol systems, characterized by brightness asymmetries between maxima, is often attributed to localized temperature anomalies caused by circumstellar material \citep{Liu2003, Deng2025}. Theoretical models suggest that as stars orbit within an extended envelope, their forward hemispheres capture material, converting kinetic energy into thermal energy, resulting in localized heating. In the Algol-like system V455 Car \citep{Deng2025}, this effect was attributed to the presence of spots on the components caused by accretion from the donor star. 
\par
\begin{figure}
\centering
    \includegraphics[width=\columnwidth]{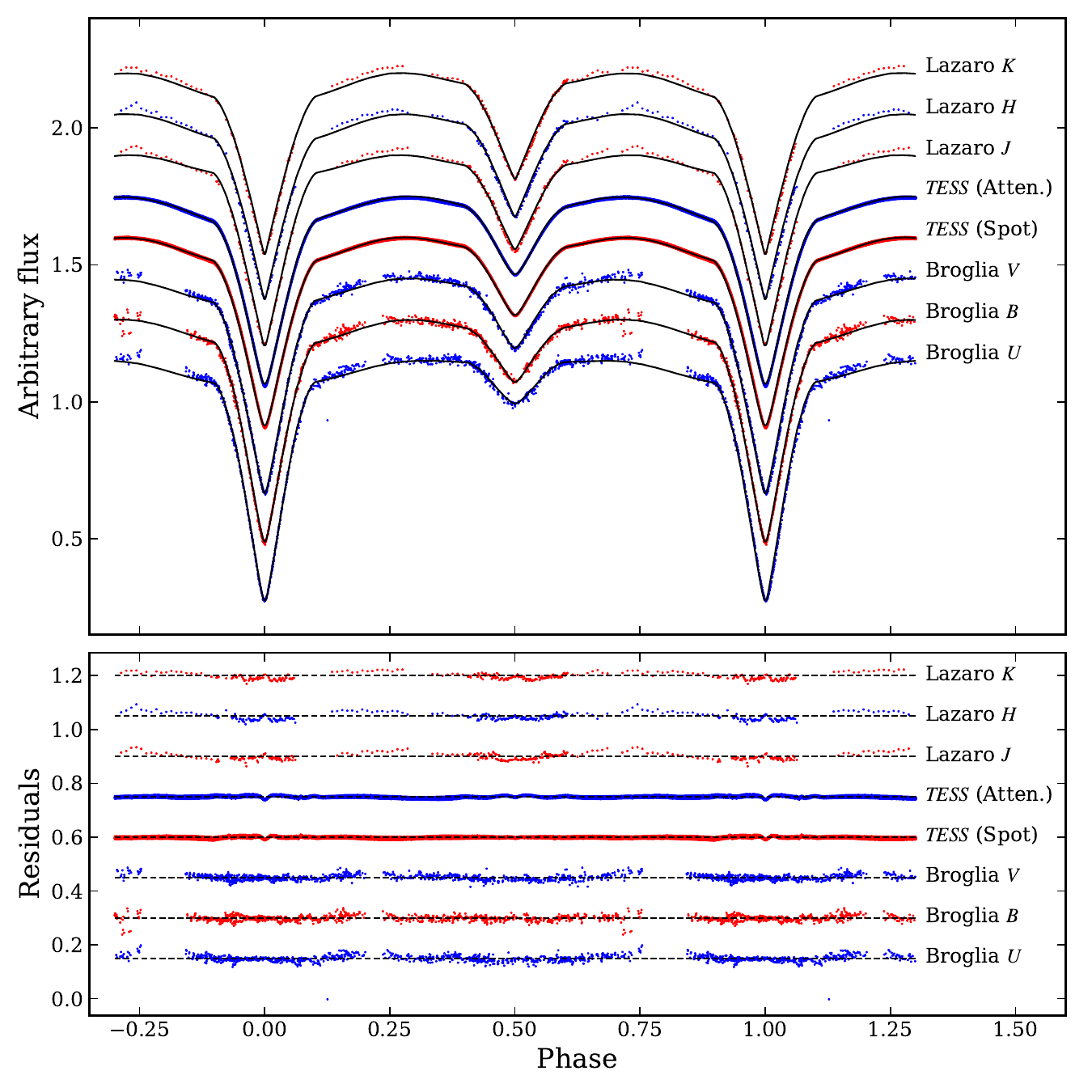}
    \caption{Observed multi-wavelength light curves of Z\,Vul represented by blue and red symbols for the respective datasets, overlaid with the best-fit synthetic models (solid lines), along with the corresponding solution residuals ($O-C$) displayed in the bottom panels. The data suite used in the modeling consists of ground-based optical Johnson $UBV$ photoelectric measurements from \citet{Broglia1964}, ground-based near-infrared Johnson $JHK$ photometry from \citet{Lazaro2009}, and space-based TESS photometry with attenuation and spot profile.}
    \label{fig:model_curve}
\end{figure}

To test this hypothesis for Z\,Vul, we first modelled the light curve using the \texttt{WD} code's ``attenuation region'' model, placing spherical areas in a Cartesian coordinate system centered on the primary component. This distribution, illustrated in Figure \ref{fig:attenuation}, shows that the observed structure can be explained by matter that accumulated on the orbital plane. We then compared these results with a solution using spot parameters to determine whether a similar effect could be achieved. During our simultaneous light-curve and radial velocity analysis, we implemented a \texttt{MCMC} approach, treating the spot parameters, latitude ($\phi_i^{\circ}$), longitude ($\lambda_i^{\circ}$), radius ($R_i^{\circ}$), and temperature factor ($f_{T_i}$), as free parameters for each component. This allowed the algorithm to explore the full parameter space for both hot-spot (accretion impact) and cool-spot (localized dimming/obscuration) configurations. Our optimal \texttt{MCMC} results converged consistently toward a cool-spot configuration for both the primary and secondary components, effectively simulating the localized flux deficit caused by the intervening accretion structure rather than intrinsic magnetic activity. The convergence and posterior distributions for these spot parameters are presented in Figure \ref{fig:LC_mcmc} (upper right panel for the primary and lower left panel for the secondary component). As shown in Table \ref{tab:LC_RV}, the system parameters derived in this spot-optimization approach remain consistent with the unoptimized attenuation region model within $1\sigma$ error limits, confirming the robustness of this methodology for subsequent evolutionary modelling. Figure \ref{fig:LC_mcmc} (lower right panel) shows the modeled component shapes and the cold spot regions at 0.25 orbital phase. Figure \ref{fig:model_curve} overplots the synthetic light curves on the \textit{UBV}, TESS, and \textit{JHK} photometry, together with the corresponding $(O-C)$ residuals. For the TESS band we present two solutions, a spot model and an attenuation model, which differ in their treatment of the out-of-eclipse asymmetry. The radial velocity solution was repeated for the spot case and remained compatible within the uncertainties (Table \ref{tab:LC_RV}, Figure \ref{fig:RV_Curve}). We therefore adopted the spot optimised solution in all subsequent steps.
\par
Thus, the final solution of the light curves incorporates the final iteration for the spectroscopic parameters, including fixed spot parameters. Then, the final light curve solutions were calculated keeping the spectroscopic parameters ($q$, $a$, and $\gamma$) fixed. These final orbital and geometrical solutions are given in Table\,\ref{tab:LC_RV}. Furthermore, the definitive absolute physical parameters derived from the TESS spot solution synthesis of these fully optimized configurations are compiled in Table\,\ref{tab:absolute}.

Figure \ref{fig:model_curve} illustrates the model curves applied to these diverse light curve datasets and the resulting residuals, while Table \ref{tab:LC_RV} provides a detailed comparison of the parameters derived from these data. The photometric solutions were found to be consistent with the atmospheric parameters within a $2\sigma$ uncertainty range of the atmospheric parameters. To ensure the derivation of realistic error values and robust system parameters, a comprehensive analysis was performed using the \texttt{LC$_{\texttt{MCMC}}$} code \citep{Pavlovski_2022}. While the traditional Differential Corrections (DC) algorithm within the \texttt{WD} code cannot optimize the parameters governing localized surface features, our \texttt{MCMC} approach successfully optimized two spots simultaneously. In total, the sampler explored a 12-dimensional parameter space, which included the mass ratio ($q$) and semi-major axis ($a$), along with their respective error values obtained from the radial velocity solutions. For the execution of the \texttt{MCMC} algorithm, we utilized $N_{walkers}=160$. While the standard rule of thumb suggests employing at least twice as many walkers as the number of dimensions ($N_{dim}=12$), we specifically chose 160 walkers to leverage higher CPU performance and ensure that the sampler thoroughly explored the parameter space, well beyond the minimum requirements. The ensemble was initialized from wide distributions centered on physically motivated values and typically converged after approximately 160\,million iterations for each passband. Owing to the high dimensionality of the solution, the resulting corner plot was divided into three separate sections to maintain visual clarity. The detailed \texttt{MCMC} results and specific spot locations derived from the best-fit model are presented in Figure \ref{fig:LC_mcmc}.
\par

\begin{table}
\caption{Absolute parameters of Z Vul. 
\label{tab:absolute}}
\centering
\begin{tabular}{lccc}
\hline\hline
Parameter      & Unit         & Primary           & Secondary  \\
\hline
$M$            & M$_{\odot}$  & 6.26 $\pm$ 0.08   & 2.44 $\pm$ 0.03  \\
$R$            & R$_{\odot}$  & 4.84 $\pm$ 0.11   & 4.73 $\pm$ 0.02 \\
$\log g$       & dex          & 3.88 $\pm$ 0.07   & 3.45 $\pm$ 0.36 \\
$T_{\rm eff}$  & K            & 16\,760 $\pm$ 270 &  9\,690 $\pm$ 250 \\
$\log L$       & L$_{\odot}$  & 3.21 $\pm$ 0.02   & 2.24 $\pm$0.01 \\
$v \sin i$     & km\,s$^{-1}$ & 111 $\pm$ 7       & 95 $\pm$ 15     \\
\hline
\end{tabular}
\end{table}

\subsection{RVs measurements and spectroscopic orbit}
\label{sec:RVs}
Radial velocities were measured using the Cross-Correlation Function (CCF) technique \citep{Simkin_1974, Tonry_1979}  as implemented in the {\sc RaveSpan} code \citep{Pilecki2017}. All 31 high-resolution spectra were used for the computation CCF (Table \ref{tab:speclog}). 
\par
\begin{figure}
	\includegraphics[width=\columnwidth]{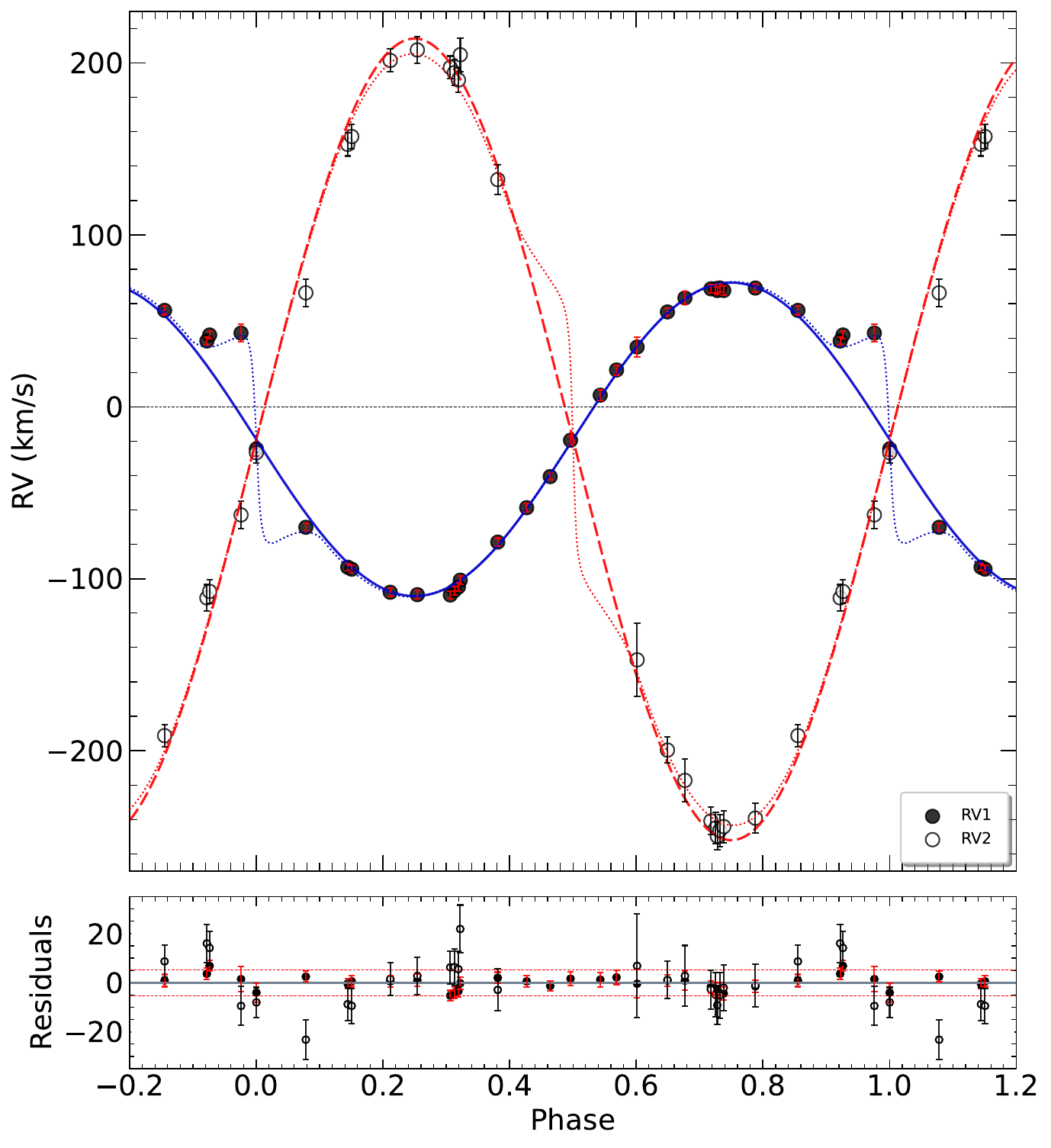}
    \caption{Radial velocity measurements of the Z Vul. Solid points with error bars represent the primary component ($RV_1$), and the open points with error bars correspond to the secondary component ($RV_2$). The dashed curve indicates the best-fit Keplerian orbital solution, and the dotted curve shows the synthetic model, which includes tidal effects. The lower panel displays the residuals between the observed radial velocities and the respective model predictions.}
    \label{fig:RV_Curve}
\end{figure}

For the initial measurements, synthetic spectra were generated based on the approximate values of effective temperature (\textit{T$_{\rm eff}$}) and surface gravity (log \textit{g}) obtained for each component from the literature. With improved values for $T_{\rm eff}$ and $\log g$ emerging in the analysis, the template spectra were recomputed and the RVs measurements repeated. Numerous measurements were performed at spectral segments of~50\,{\AA} in the spectral range from 4\,000 to 8\,000\,{\AA}. Figure \ref{fig:RV_Curve} presents the final radial velocity curves of Z Vul, combining all measurements for both the primary and secondary components, with their corresponding uncertainties. The Rossiter-McLaughlin effect due to the spectra distorted during the eclipses is detected. The radial velocity curve solutions were calculated for three cases: (i) with spots, (ii) with attenuation regions, and (iii) without spots or attenuation regions. Although we solved the RVs with 2015 version of the \texttt{WD} code (which optimizes separation and mass ratio), the equivalent RVs semi-amplitudes are $K_1 = 91.15 \pm 1.53$\kms\ and $K_2 = 233.05 \pm 2.46$\kms\, which yields the mass ratio $q = 0.3911 \pm 0.0073$ with the adopted spot solution displayed in Figure \ref{fig:RV_Curve}. Our new determination of semi-amplitudes are within 1-$\sigma$ uncertainty with $K_1$ derived by \citet{Popper_1957a} but is incompatible (about 14\,\kms) with his $K_2$.

\subsection{Spectral Disentangling}
\label{sec:Disent}
In the spectral disentangling method, a time series of the observed binary star spectra enables the separation of the component spectra \citep{Simon_1994, Hadrava_1995}. A further advantage of {\sc spd} is that the S/N of the disentangled component spectra is increased, making the spectroscopic analysis of such spectra more reliable and confident \citep{Pavlovski_2005}.
\par
The \texttt{FDBINARY} code\footnote{\url{http://sail.zpf.fer.hr/fdbinary/}} \citep{Ilijic2004} is used to disentangle the observed {\sc hermes} spectra of Z\,Vul. It is a basic assumption of {\sc spd} that the spectral line profiles do not change except for the dilution factor in the course of the orbital cycle. Thus, including the spectra obtained in the course of ingress or egress of the eclipses in which the line profiles are distorted due to the Rossiter-McLaughlin effect would be a violation of that assumption. Thus, only the spectra obtained out of the eclipses were used. In our sample, there are 23 out-of-eclipse spectra well distributed in orbital phase (Table \ref{tab:speclog}, Figure\,\ref{fig:RV_Curve}). To avoid undulations in the disentangled spectra, the observed spectra were carefully normalised with special care for the broad Balmer lines (Sect.\,\ref{sec:Spectroscopy_Obs}). 
\par
The orbital parameters for disentangling were fixed to the values determined from the spectroscopic orbit in Sect. \ref{sec:RVs}. In particular, the radial velocity semi-amplitudes, $K_1$ and $K_2$, were held at their best-fitting values, ensuring a consistent set of orbital elements throughout the analysis. We opted to perform {\sc spd} in short spectral segments of approximately 100 \AA. The spectral intervals were selected so that their boundaries coincided as closely as possible with the continuum regions; otherwise, when applying the Fourier-transform-based {\sc spd}, additional sinusoidal-like oscillations could appear in the disentangled spectra. To define these intervals, all spectra were first merged into a single integrated spectrum containing all spectral lines that shifted in position across orbital phases. Then a Median+Max analysis was applied to this integrated spectrum to identify the continuum regions. The resulting intervals were approximately 100\,$\AA$ wide, with the exception of broad features such as the H and He lines, for which this rule was relaxed.
\par
To propagate the uncertainties from the spectroscopic orbital solution into the disentangled spectra, we employed a Bayesian approach using the \texttt{MCMC} algorithm implemented by \citet{Barbaros2023}\footnote{\url{https://doi.org/10.55064/tjaa.1203660}}, hereafter referred to as \texttt{FDBeMC}. Custom scripts were developed, and the Python \texttt{emcee} module \citep{Foreman-Mackey2013}, which implements the Affine Invariant ensemble sampler instead of the traditional Metropolis algorithm, was integrated into our code. Rather than re-determining $K_1$ and $K_2$, the \texttt{FDBeMC} analysis was restricted to sampling within the error margins of the semi-amplitudes obtained from the RVs solution (Sect. \ref{sec:RVs}). This allowed us to quantify how the uncertainties in $K_1$ and $K_2$ propagate into the separated component spectra. The resulting variation in the disentangled spectra provides the uncertainties required for the subsequent atmospheric parameter determination, for which reliable error estimates are essential.
\par
\begin{figure}
    \centering
    \includegraphics[width=\columnwidth]{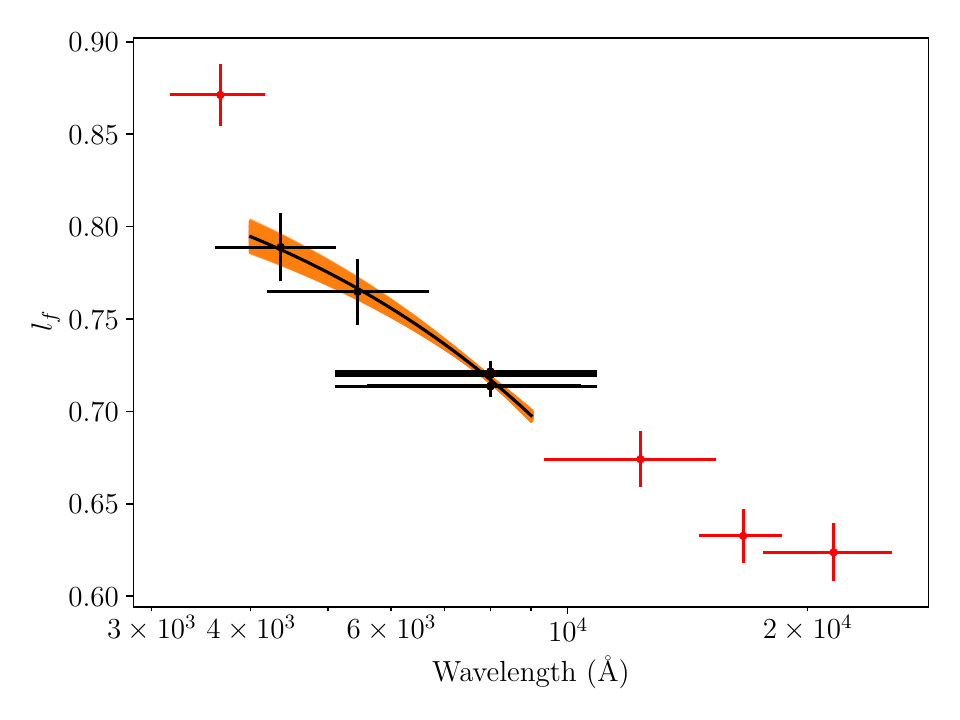}
    \caption{
    Fractional light contribution of the primary component in Z\,Vul derived from the multi-wavelength light curve analysis. The entire ensemble of data points (combined red and black symbols) represents the full suite of photometric light curves modelled in this study, which includes Johnson $UBV$ \citep{Broglia1964}, TESS (four sectors), and Johnson NIR $JHK$ \citep{Lazaro2009}. Black symbols specifically denote the 4\,000 to 8\,000 {\AA} wavelength regime covered by the optical spectra utilized in the spectroscopic analysis.
    }
    \label{fig:lffunction_pri}
\end{figure}

With the orbital parameters known {\sc spd} is performed in separation mode with generic light factors \citep{Pavlovski_2010}. Separated spectra of the components are released in common continuum of binary system. For the atmospheric diagnostics of the components (determination of the effective temperature $T_{\rm eff}$, surface gravity, metallicity [M/H] and projected rotational velocity) separated spectra of individual components still should be renormalised to their own continua. This means the wavelength dependent light ratio $l_f(\lambda)$ between the components must be known (Figure \ref{fig:lffunction_pri}, Table \ref{tab:LC_RV}).

These renormalised components' spectra are ready for the spectroscopic analysis; determination of the atmospheric parameters and subsequently detail abundance analysis.


\subsection{Atmospheric Parameters for the Components}
\label{sec:Atmos}
Since {\sc spd} was performed in pure separation mode the disentangled spectra are still in the common continuum of the binary system. The light ratio between the components is needed for a proper renormalisation to their own continua \citep{Pavlovski_2005, Tkachenko_2010}. Usually, the light ratio can be determined with high precision from the light curve analysis \citep{Hensberge_2000, Pavlovski_Southworth_2009, Pavlovski_2018, Dervisoglu_2018}. The fractional light contribution of the component to the total light of the binary system is preserved in disentangled or separated spectra and could be determined from the optimal fitting of these spectra to synthetic spectra \citep{Tamajo_2011, Tkachenko_2015}. In simultaneous fitting of both disentangled spectra furter constrain could be applied: the sum of the fractional light contributions of the components must be equal to unity. In some partially eclipsing systems, due to the ambiguity in the determination of the radii, the light ratio cannot be determined with sufficient accuracy. In such cases, the spectroscopic light ratio is helpful in breaking the degeneracy in the ratio of the radii \citep{Pavlovski_2009, Pavlovski_2018, Pavlovski_2023} or could be used for a direct determination of this ratio  \citep{Tkachenko_2015}. In non-eclipsing systems, a spectroscopic determination of the light ratio is the only way to determine absolute stellar parameters \citep{Torres_2015, Torres_2025}.

\begin{table*}
\caption{Atmospheric parameters for the primary and secondary components of Z Vul derived across four sequential \texttt{MCMC} optimization stages (Runs 1–4). The final "Best" parameters reported correspond to the minimum $\chi^2$ solution from the complete parameter space explored in Run 4. 
\label{tab:Atmospheric_Par}}
\centering
\begin{tabular}{lccccccc}
\hline
\hline
Run            & Star & $T_{\rm eff}$     & $\log g$        & [M/H]            & $l_{fa}$               & $l_{fb}$          & $v\sin i$  \\
               &      & K                 & [cgs]           & dex              & frac                   &  frac             & km\,s$^{-1}$ \\
\hline
Run 1          & Pri. & $16\,940 \pm 350$ & $3.88 \pm 0.07$ & $-0.18 \pm 0.12$ & $-0.00026 \pm 0.00012$ & $0.900 \pm 0.056$ & $108 \pm 8$ \\
               & Sec. & $8\,895  \pm 650$ & $3.05 \pm 0.36$ & $-0.15 \pm 0.34$ & $0.00011  \pm 0.00007$ & $0.123 \pm 0.040$ & $94 \pm 15$ \\
\hline
Run 2          & Pri. & $16\,950 \pm 290$ & $3.89 \pm 0.02$ & $-0.18 \pm 0.12$ & $-0.00027 \pm 0.00012$ & $0.901 \pm 0.055$ & $108 \pm 7$ \\
               & Sec. & $9\,530  \pm 300$ & $3.45 \pm 0.04$ & $0.15  \pm 0.18$ & $0.00011  \pm 0.00007$ & $0.123 \pm 0.040$ & $93 \pm 15$ \\
\hline
Run 3          & Pri. & $17\,000 \pm 280$ & $3.89 \pm 0.02$ & $-0.31 \pm 0.27$ & $-0.00020 \pm 0.00002$ & $0.869 \pm 0.017$ & $104 \pm 10$ \\
               & Sec. & $9\,870  \pm 250$ & $3.45 \pm 0.04$ & $0.16  \pm 0.18$ & $0.00019  \pm 0.00003$ & $0.128 \pm 0.025$ & $95 \pm 15$ \\
\hline
Run 4          & Pri. & $16\,790 \pm 270$ & $3.89 \pm 0.02$ & $0.01  \pm 0.04$ & $-0.00020 \pm 0.00002$ & $0.867 \pm 0.017$ & $106 \pm 10$ \\
               & Sec. & $9\,860  \pm 250$ & $3.45 \pm 0.04$ & $0.01  \pm 0.03$ & $0.00019  \pm 0.00003$ & $0.128 \pm 0.025$ & $93 \pm 15$ \\
\hline
Best             & Pri. & $16\,760$         & $3.88$          & $0.00$           & $-0.00021$             & $0.870$           & $111$ \\
               & Sec. & $9\,690$          & $3.45$          & $0.00$           & $0.00016$              & $0.105$           & $95$ \\
\hline
\end{tabular}
\end{table*}

The light curve analysis and spectroscopic analysis of disentangled spectra are interrelated and an mutual iterative process. The light curves provide the light ratio between the components and complemented with the RVs allow determination of the fundamental stellar quantities for both components (mass $M$, radius $R$ and luminosity $L$). The effective temperature, at the least for one of the components, should be known from other sources and here feedback from the atmospheric diagnostics of disentangled spectra are the most reliable way for setting the temperature scale of the components. Or vice versa, disentangled spectra could be renormalised with the wavelength dependent light ratio emerged in the light curves solutions. Moreover, fixing the surface gravities for the hot components would lift degeneracy between the effective temperature and surface gravities and lead to confident solution. Usually, in couple of iterations stable solutions in both, the light curves solutions and atmospheric analysis, are achieved \citep{Pavlovski_2018, Pavlovski_2023}. 

The atmospheric parameters are determined using \texttt{iSpec$_{\texttt{MCMC}}$} \citep{Dervisoglu_2018}. 
The core strength of \texttt{iSpec$_{\texttt{MCMC}}$} lies in its Bayesian approach using the Markov Chain Monte Carlo (\texttt{MCMC}) optimisation method, which is implemented via the Python-based \texttt{emcee} module. The software incorporates a broad range of extensive synthetic spectral grids, such as ATLAS, BStar, MARCS, and TLUSTY \citep{Blanco-Cuaresma2019} yielding a comprehensive search over wide parameter space of $T_{\rm eff}$, $\log g$ and [M/H] to obtain the best-fitting values. Employing the \texttt{MCMC} method, the code determines the best-fitting parameters and calculates uncertainties while simultaneously revealing parameter correlations. 
The results of the \texttt{MCMC} task, including the revealed parameter correlations, are presented as corner diagrams for both components in Figure \ref{fig:Atmospheric_corner}.
\par
\texttt{iSpec$_{\texttt{MCMC}}$} is primarily used in the context of binary star system analysis, offering functionality optimized for both single-star and binary-star modes. In the single-star mode, the total $l_{f}$ is assumed to be unity, making it suitable for isolated stars or when analyzing a single component without need to consider the companion. In binary mode, the software simultaneously optimizes both stellar components. This mode operates in two sub-modes: constrained and unconstrained modes as introduced in \citet{Tamajo_2011}. In the constrained mode, the $l_{fi}$ of the two components is normalized such that their sum equals unity (i.e., $l_{f_p} + l_{f_s} = 1$), reflecting a physically consistent total system light. In the unconstrained mode, the $l_{fi}$ values were treated as free parameters, allowing them to vary independently. This can be useful for exploratory analyses or when normalization is handled externally.
\par
\cite{Dervisoglu_2018} have continually developed and updated \texttt{iSpec$_{\texttt{MCMC}}$} to address challenges specific to analyzing Algol-type binaries, where the components often have highly unequal luminosities. \texttt{iSpec$_{\texttt{MCMC}}$} was enhanced to simultaneously fit the two components of a binary system. This is critical because the normalization (and thus the line depth accuracy) of the separated spectrum of the fainter component is highly sensitive to the component's $l_{f}$ ratio. The software can integrate a wavelength-dependent function for the $l_{f}$ ratio. Because the $l_{f}$ ratio is dependent on the wavelength, especially for hot stars whose Spectral Energy Distribution (SED) peaks in the ultraviolet region, \texttt{iSpec$_{\texttt{MCMC}}$} allows the $l_{fi}$ ratio to be treated as a linear function of wavelength, $l_{f_p} = l_{f_pa} \times \lambda + l_{f_pb}$. The $l_{fi}$ ratios of the component stars are crucial for correctly normalizing the separated spectra obtained via the SPD. In interacting binary systems, particularly those containing hot stars, such as the Algols being studied, the relative luminosity of each component varies significantly with wavelength because the stars have different effective temperatures and, thus, different SEDs. When analyzing high-resolution \'echelle spectra, which typically cover the visible range (4\,000 to 6\,800 $\AA$), the SED characteristics of the components imply that the $l_{fi}$ ratio of the primary component ($l_{f_p}$) changes almost linearly across this interval. Conversely, the secondary component's $l_{fi}$ ratio ($l_{f_s}$) increases linearly in this range because the total normalized light is 1 ($l_{f_s} = 1 - l_{f_p}$). This change is mathematically modeled as a linear function of wavelength ($\lambda$) within the \texttt{iSpec$_{\texttt{MCMC}}$} code: $l_{f_p} = l_{f_pa} \times \lambda + l_{f_pb}$ (i.e., $y = ax + b$). The linear trend is used within an enhanced version of \texttt{iSpec$_{\texttt{MCMC}}$} that simultaneously optimizes both the component spectra. By treating the $l_{fi}$ ratio as a wavelength-dependent linear function, the software only needs to determine the coefficients $l_{f_pa}$ and $l_{f_pb}$. This approach ensures that the renormalization step is integrated smoothly and dynamically during the atmospheric parameter fitting (\texttt{MCMC} process), leading to more accurate and reliable atmospheric parameters for both stars.

\begin{figure*}
\centering
	\includegraphics[width=0.85\textwidth]{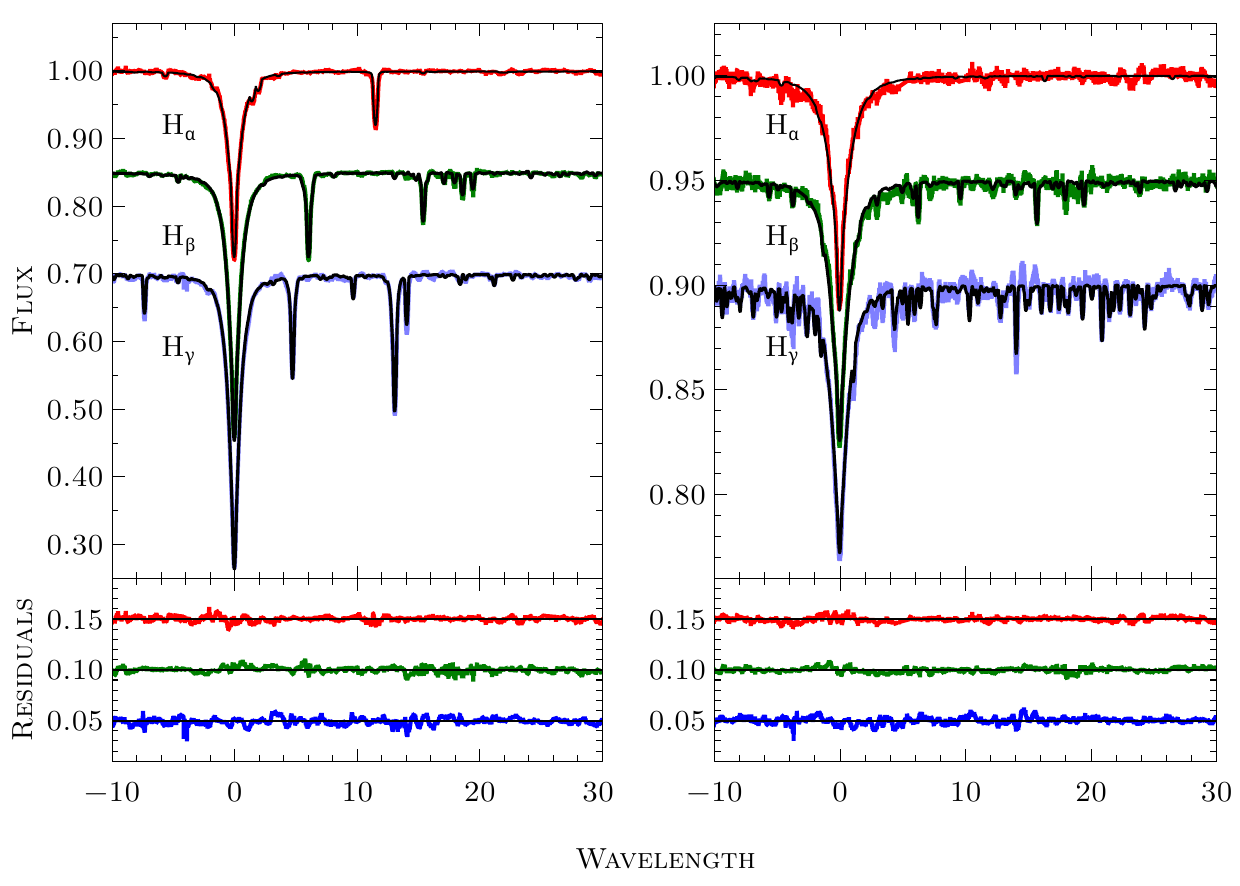}
    \caption{Comparison of the best-fitting model and the renormalised observed disentangled spectra of the primary (left) and secondary (right) components, centred around Balmer lines H$\alpha$, H$\beta$ and H$\gamma$. The residuals are shown in the bottom panel.
    }
    \label{fig:Balmer_lines_fits}
\end{figure*}

In the analysis of disentangled spectra for Z\,Vul we adopted profile fitting of the spectra within the wavelength range of 4\,000-6\,800 $\AA$. As part of the application of the \texttt{iSpec$_{\texttt{MCMC}}$} software, a series of execution steps were defined to determine the atmospheric parameters of both components of the Z Vul binary system. To achieve this, each step applies progressive constraints on key parameters, such as the effective temperature ($T_{\rm eff}$), surface gravity ($\log g$), metallicity ([M/H]), and fractional light contribution coefficients ($l_{f_{p,s}a},\,l_{f_{p,s}b}$). To refine this process, we implemented a four-stage iterative optimization strategy. 
\par 
\textit{Run 1}: all parameters were treated as free, with no constraints applied. This step explored the entire parameter space and served as a baseline. The results derived from this extensive exploration of the parameter space are presented in Figure \ref{fig:Atmospheric_corner} and Table \ref{tab:Atmospheric_Par}. To ensure visual clarity and facilitate a detailed inspection of the posterior distributions, the resulting corner plots were partitioned to represent each stellar component individually. The $l_{fi}$ ratios derived from the Run 1 solution were found to be consistent within $2\sigma$ with those obtained from the photometric results. 
\par
\textit{Run 2}: the $\log g$ values were constrained using the results obtained from the combined RVs and light curve analysis, thereby narrowing the parameter space during the \texttt{MCMC} exploration. This constraint is documented in Table \ref{tab:Atmospheric_Par} and is visually represented in Figure \ref{fig:Atmospheric_corner} by the yellow circles. Fixing the surface gravity of the components lifts the degeneracy with $T_{\rm eff}$ present in hot stars. 
\par
\textit{Run 3}: in addition to the constraints on $\log g$, the fractional $l_{fi}$ was constrained based on the $l_f$ function derived from the light fractions reported in Table \ref{tab:LC_RV} and Figure \ref{fig:lffunction_pri}. This fitting process was restricted to the 4\,000 to 8\,000 $\AA$ wavelength range. We also derived uncertainties for the coefficients using \texttt{MCMC} optimization, which are reported in Table \ref{tab:Atmospheric_Par} and indicated in Figure \ref{fig:Atmospheric_corner} by blue triangles. 
\par
\textit{Run 4}: in the final execution, $\log g$, $l_f$, and the metallicity of each component were constrained simultaneously, restricting the solution to a physically consistent and observationally supported region of the parameter space. The atmospheric parameters of each component were taken from the global $\chi^2$ minimum, the maximum-likelihood solution of the Run 4 \texttt{MCMC} chains, marked by the red squares in Figure \ref{fig:Atmospheric_corner}, and their uncertainties were derived from the corresponding marginalised posterior distributions. These values, with their $1\sigma$ uncertainties, are listed in Table \ref{tab:Atmospheric_Par} for both components, where the light contribution of each component is now given as its relative light fraction $l_f$. These optimised atmospheric parameters were subsequently held fixed and adopted for the abundance analysis. A comparison between the synthetic spectra generated from these parameters and the observed stellar spectra is presented in Figure~\ref{fig:Balmer_lines_fits}.
\par
When reviewing the literature on Z Vul, we examined the effective temperatures calculated in previous studies to contextualise our findings. \cite{Ibanoglu2012} used the reddening-free colour index $[u-b]$ from Str\"{o}mgren photometry published by \citet{Hilditch_1975}. They obtained an effective temperature of 16\,300\,K for the system. \cite{Lazaro2009} calculated the effective temperature of the primary component to be 18\,000 $\pm$ 300\,K. Both previous determinations of the effective temperature of the primary component are within $2-\sigma$ of the effective temperatures calculated in this study. It is important to note that while previous determinations of the $T_{\rm eff}$ are based on photometric indices and various calibrations, we derived the temperature directly from the spectroscopic analysis of high-resolution individual disentangled spectra of the components in Z\,Vul binary system, providing a more robust determination of the stellar atmospheric quantities.

\begin{figure}
	\includegraphics[width=9cm]{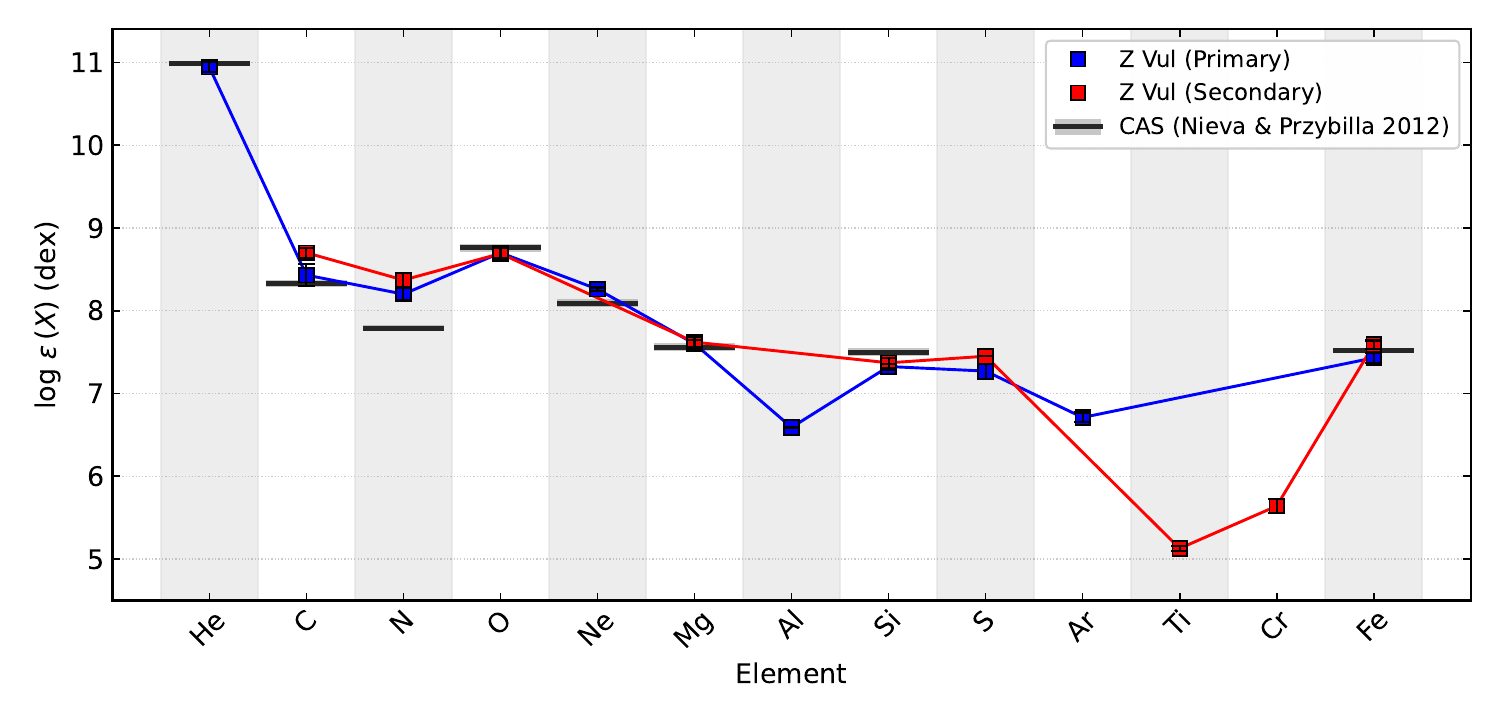}
    \caption{Chemical abundance patterns of the Z\,Vul system. The primary and secondary components are shown with blue and red squares, respectively. Error bars indicate the derived uncertainties. Grey bands mark the present-day cosmic abundance standard of \citet{nieva2012present}.}
    \label{fig:all_abundances}
\end{figure}

\subsection{Abundance Analysis}
For the abundance analysis, the renormalized spectra of the Z Vul components were employed. Elemental abundance determinations involved the \textsc{TLUSTY} model atmosphere code \citep{hubeny1995non}, the \textsc{SYNSPEC} spectrum synthesis program \citep{hubeny1995non}, and the BSTAR2006 NLTE model grids \citep{lanz2007grid}. The TLUSTY version 205 \citep{Hubeny1988, Lanz2003, Lanz2007}, widely used in the literature for modelling the stellar atmospheres of hot stars, is a one-dimensional stellar atmosphere code that allows for both LTE and non-LTE model atmosphere calculations.  
\par
We employed a recently compiled comprehensive atomic line list for hot B-type stars \citep{Guney2026, sahin2025vizier} in our spectroscopic analysis. This list incorporates transitions from 13 key photospheric elements (He, C, N, O, Ne, Mg, Al, Si, P, S, Cl, Ar, and Fe). Its reliability has been established through previous benchmarking using the standard star HR\,1765 \citep{sahin2019} and the Algol-type system u\,Her \citep{Kolbas_2014}. Similar methodological rigor has been applied to investigations of chemically peculiar hot stars \citep{csahin2018a} and hot post-AGB candidates \citep{csahin2018b}, where careful line selection and NLTE treatment are essential for reliable abundance determinations. In a recent study, we further validated this line list by analysing the rapidly rotating Algol binary ET\,Cru \citep{Yucel_2026}.

\begin{table}
\caption{Comparison of elemental abundances derived for the primary and secondary components of Z\,Vul.
         $N_{lines}$: Number of spectral lines. Solar reference: \citet{asplund2009chemical}. CAS: present-day cosmic abundance standard of \citet{nieva2012present}.} 
\label{tab:summary_abundances}
\centering
\resizebox{\columnwidth}{!}{%
\begin{tabular}{lcccc}
\hline\hline
              & Primary                                    & Secondary                            & Sun                     & CAS\\
 \cline{2-5}
Spec.         & $\log\epsilon(X)$ LTE | NLTE ($N_{lines}$) & $\log\epsilon(X)$ LTE ($N_{lines}$)  & $\log\epsilon(X)_\odot$ & $\log\epsilon(X)_\odot$ \\
\hline
\ion{He}{i}   & 11.03 $\pm$ 0.07 | 10.94 $\pm$ 0.06 (7)    & ---                                  & 10.93 $\pm$ 0.01        & 10.99 $\pm$ 0.01\\
\ion{C}{i}    & ---                                        & 8.70  $\pm$ 0.06 (5)                 & 8.43 $\pm$ 0.05         & 8.33 $\pm$ 0.04\\
\ion{C}{ii}   & 8.23 $\pm$ 0.15 | 8.43 $\pm$ 0.13 (3)      & ---                                  & 8.43 $\pm$ 0.05         & 8.33 $\pm$ 0.04\\
\ion{N}{i}    & ---                                        & 8.37  $\pm$ 0.08 (3)                 & 7.83 $\pm$ 0.05         & 7.79 $\pm$ 0.04\\
\ion{N}{ii}   & 8.08 $\pm$ 0.07 | 8.20 $\pm$ 0.07 (9)      & ---                                  & 7.83 $\pm$ 0.05         & 7.79 $\pm$ 0.04\\
\ion{O}{i}    & ---                                        & 8.69 $\pm$ 0.08 (3)                  & 8.69 $\pm$ 0.05         & 8.76 $\pm$ 0.05\\
\ion{O}{ii}   & 8.61 $\pm$ 0.06 | 8.70 $\pm$ 0.00 (6)      & ---                                  & 8.69 $\pm$ 0.05         & 8.76 $\pm$ 0.05\\
\ion{Ne}{i}   & 8.38 $\pm$ 0.06 | 8.26 $\pm$ 0.02 (5)      & ---                                  & 7.93 $\pm$ 0.10         & 8.09 $\pm$ 0.05\\
\ion{Mg}{ii}  & 7.48 $\pm$ 0.06 | 7.60 $\pm$ 0.05 (3)      & 7.62 $\pm$ 0.06 (3)                  & 7.60 $\pm$ 0.04         & 7.56 $\pm$ 0.05\\
\ion{Al}{iii} & 6.48 $\pm$ 0.04 | 6.59 $\pm$ 0.01 (2)      & ---                                  & 6.45 $\pm$ 0.03         & ---\\
\ion{Si}{ii}  & 7.22 $\pm$ 0.08 | 7.38  $\pm$ 0.04 (3)     & 7.37 $\pm$ 0.06 (3)                  & 7.51 $\pm$ 0.03         & 7.50 $\pm$ 0.05\\
\ion{Si}{iii} & 7.31 $\pm$ 0.09 | 7.24  $\pm$ 0.05 (4)     & ---                                  & 7.51 $\pm$ 0.03         & 7.50 $\pm$ 0.05\\
\ion{S}{ii}   & 7.28 $\pm$ 0.09 | 7.27  $\pm$ 0.09 (12)    & 7.45  (1)                            & 7.12 $\pm$ 0.03         & ---\\
\ion{Ar}{ii}  & 6.60 $\pm$ 0.00 | 6.71  $\pm$ 0.06 (2)     & ---                                  & 6.40 $\pm$ 0.13         & ---\\
\ion{Ti}{i}   & ---                                        & 5.12 $\pm$ 0.04 (2)                  & 4.95 $\pm$ 0.05         & ---\\
\ion{Ti}{ii}  & ---                                        & 5.14 $\pm$ 0.05 (4)                  & 4.95 $\pm$ 0.05         & ---\\
\ion{Cr}{i}   & ---                                        & 5.65  (1)                            & 5.64 $\pm$ 0.04         & ---\\ 
\ion{Cr}{ii}  & ---                                        &  5.64 $\pm$ 0.08 (4)                 & 5.64 $\pm$ 0.04         & ---\\
\ion{Fe}{i}   & ---                                        &  7.61 $\pm$ 0.07 (9)                 & 7.50 $\pm$ 0.04         & 7.52 $\pm$ 0.03\\
\ion{Fe}{ii}  & ---                                        &  7.57 $\pm$ 0.07 (10)                & 7.50 $\pm$ 0.04         & 7.52 $\pm$ 0.03\\
\ion{Fe}{iii} & 7.47 $\pm$ 0.05 | 7.43  $\pm$ 0.06 (4)     & ---                                  & 7.50 $\pm$ 0.04         & 7.52 $\pm$ 0.03\\
\hline
\end{tabular}%
}
\end{table}

A consistent approach to spectral line broadening was applied in our analysis. The Stark broadening of the hydrogen Balmer lines was calculated using the unified theory profiles available in \textsc{SYNSPEC} \citep{vidal1973hydrogen}, with the treatment for higher series members adopted from \citet{hubeny1994nlte}. For helium, broadening tables were employed as follows: \citet{barnard1974broadening} for the 4471\,\AA\ line, \citet{shamey1969stark} for the 4026, 4387, and 4922\,\AA\ lines, and \citet{dimitrijevic1984stark} for all remaining He\,{\sc i} transitions. All metal lines were synthesized as Voigt profiles, incorporating radiative, van der Waals, and Stark broadening using established constants.
\par
The \textsc{TLUSTY} model atmosphere for the Z Vul primary was calculated in full non-local thermodynamic equilibrium (NLTE), incorporating opacity contributions from both bound-free and bound-bound transitions of H, He, C, N, O, Ne, Mg, Si, S, and Fe. The corresponding model atoms were extensive, including, for instance, \ion{H}{i} (9 levels), \ion{He}{i} (24), \ion{C}{ii} (22), \ion{C}{iii} (46), \ion{N}{ii} (42), \ion{O}{ii} (48), \ion{Mg}{ii} (25), \ion{Si}{ii/iii/iv} (40, 30, 23), \ion{S}{ii/iii/iv} (33, 41, 38), and 
\ion{Fe}{ii/iii/iv}. 
\par
For the abundance determination, we restricted our analysis to weak and medium-strength metal lines to minimise uncertainties from damping and saturation effects.
\par
In B-star atmospheres, the NLTE effects are significant for many species. In particular, \ion{He}{i}, \ion{Ne}{i}, and \ion{Si}{ii/iii} require corrections owing to photoionization and radiative pumping \citep[e.g.,][]{auer1973analyses}. Sulphur (\ion{S}{ii/iii}) is prone to overpopulation through collisions and radiative transitions \citep[for example,][]{vrancken1996non}, while iron (\ion{Fe}{iii}) is subject to overionization, which may lead to substantial abundance underestimation if neglected \citep[for example,][]{nieva2012present}. To ensure reliable results, detailed NLTE line formation calculations were performed for all species reported here.
\par
For \ion{He}{i} and \ion{Ne}{i}, these NLTE conditions cause a significant strengthening of their optical lines relative to the LTE predictions. In contrast, the departures for \ion{Si}{ii/iii} originate from the complex interplay of metastable level populations and ultraviolet line-blocking effects \citep[e.g.,][]{auer1973analyses, alexeeva2020neon, mashonkina2020non}.
\par
The derived photospheric elemental abundances for the primary and secondary components of Z\,Vul are presented in Table~\ref{tab:summary_abundances} and are compared in Figure~\ref{fig:all_abundances}. For reference, Table \ref{tab:summary_abundances} lists the standard solar abundances of \citet{asplund2009chemical} alongside the present-day cosmic abundance standard (CAS) of \citet{nieva2012present}; the latter is also shown in Figure \ref{fig:all_abundances}. The CAS was derived in NLTE from 29 sharp-lined early B-type stars in the solar neighbourhood, with a star to star scatter of only 0.05 dex or less. It therefore represents the current composition of the local star-forming material and provides a much more appropriate baseline for the young Z\,Vul system than the solar values. By contrast, the solar abundances reflect the interstellar medium as it was 4.6 Gyr ago, since modified by Galactic chemical evolution, the radial migration of the Sun, and photospheric diffusion. Note that aluminium, sulphur, argon, titanium, and chromium are included in the CAS, so their entries are left blank.
\par
Spectroscopic analysis of the Z Vul system reveals marked chemical differentiation between the hot primary and cooler secondary components,  bearing clear signatures of its evolutionary history dominated by MT. For the primary, the NLTE effects were significant. After applying NLTE corrections, the helium abundance ($\log \epsilon$(He) = 10.94 $\pm$ 0.06) was virtually solar, indicating no strong helium enhancement. Carbon shows an increase of $\approx$0.2 dex in NLTE, reaching a near-solar value, while nitrogen is slightly enriched ([N/H] $\approx$ +0.37). Oxygen aligns precisely with solar abundance. Silicon exhibits contrasting NLTE behavior between ionization stages, underscoring the importance of detailed line formation calculations. The NLTE iron abundance was close to that of the Sun. Overall, the primary is characterized by mild nitrogen enrichment and near-solar carbon, consistent with a gainer star that has been superficially polluted by material processed via the CNO cycle.
\par
When we compare the primary's composition with the CAS, its nitrogen enhancement stands out at $[\mathrm{N/H}]_{\rm CAS} \approx +0.41$, a departure that exceeds the combined $1\sigma$ uncertainties by a factor of five. In contrast, carbon ($[\mathrm{C/H}]_{\rm CAS} \approx +0.10$), along with oxygen, magnesium, and iron, matches the cosmic reference within $1.3\sigma$. Silicon, meanwhile, continues to show the ionization-stage dispersion noted earlier. The neon abundance pattern appears enhanced by $+0.33$\,dex when judged against the solar scale, but this excess shrinks to $+0.17 \pm 0.05$\,dex when compared with the directly measured cosmic neon abundance. In other words, roughly half of the apparent excess is simply an artefact of the chosen reference frame, not an intrinsic feature of the star. Finally, the helium abundance we derive, $\log\epsilon(\mathrm{He}) = 10.94 \pm 0.06$, is in excellent agreement with the CAS value of $10.99 \pm 0.01$. This confirms that the primary shows no helium enhancement relative to either reference scale.
\par
In contrast, the chemical signature of the secondary component is distinct and noteworthy when compared directly to the primary. The primary (gainer) is analysed in NLTE and shows near-solar carbon ($\log\epsilon(\mathrm{C}) = 8.43\pm0.13$, [C/H]$\approx+0.00$) and mild nitrogen enrichment ($\log\epsilon(\mathrm{N}) = 8.20\pm0.07$, [N/H]$\approx+0.37$), while oxygen, Mg, Si, and Fe are approximately solar. The secondary (donor) is analysed in LTE (NLTE effects being negligible at its effective temperature) and exhibits significantly higher abundances of both carbon ($\log\epsilon(\mathrm{C}) = 8.70\pm0.06$, [C/H]$\approx+0.27$) and nitrogen ($\log\epsilon(\mathrm{N}) = 8.37\pm0.08$, [N/H]$\approx+0.54$), with the difference in carbon exceeding the combined $1\sigma$ uncertainties. Oxygen, iron-group elements (Fe, Cr), and $\alpha$-elements (Mg, Si) remain near-solar in the secondary, similar to the primary. This pattern, i.e. enrichment in both C and N relative to the primary, and a donor C/N ratio ($\sim2.14$) that is actually higher than that of the gainer ($\sim1.70$), does not match the classic expectation of a CNO-processed donor (carbon depletion, C/N $\ll 1$). 

Against the CAS, the secondary's absolute carbon and nitrogen abundances  are elevated by $+0.37$ dex and $+0.58$\,dex, respectively. However, because the CNO cycle conserves the total number of catalyst nuclei and cannot raise carbon above its initial abundance, these high absolute values warrant caution. We therefore regard the absolute abundances of the faint secondary, whose light contribution is only $l_{f,s} \approx 0.11$ to $0.13$, as the quantities most susceptible to renormalization and LTE systematics. These effects enter multiplicatively through the continuum and largely cancel out when abundance ratios are formed within the same spectrum. Consequently, we base our evolutionary interpretation on the C/N ratio, which is independent of the adopted reference scale.

The CAS itself provides an initial ratio of $(\mathrm{C/N})_{\rm CAS} \approx 3.5$, consistent with values of 3.6 to 4.0 derived from non-interacting high-mass binaries \citep{Pavlovski_2018} and from the solar scale. The donor's C/N ratio of $2.14^{+0.55}_{-0.44}$ thus falls 0.21\,dex below the initial value, yet remains far above the deeply processed regime exemplified by the u Her gainer, which exhibits at $\mathrm{C/N} = 0.9$ \citep{Kolbas_2014}. The absence of such a deep C/N inversion, typical of completely stripped Algol donors, indicates that the transferred layers in Z Vul are less processed. As our evolutionary models confirm (Section \ref{sec:Concluding}, Figure \ref{fig:donorprofile}), the donor has been stripped only to an intermediate region where the CNO cycle is incomplete, leaving the C/N ratio moderately altered but not inverted. The alternative possibility of an initially supersolar metallicity for the whole system is ruled out by the near-solar abundances of iron and $\alpha$-elements in both components, which also agree with the CAS values within the uncertainties.
\par
In summary, the chemical profiles of both components confirmed that MT altered their surface compositions. The primary exhibits moderate pollution, while the secondary presents an abundance pattern that challenges the standard model of a CNO-processed donor. This implies that the MT dynamics and internal evolution of Z Vul may differ from those of other Algol systems, warranting further investigation with detailed evolutionary models to explain its distinct nucleosynthetic signature.
\par
The reliability of the spectral synthesis was confirmed by detailed line profile fitting. The close match between the observed and synthetic spectra for both components, illustrated in Figure~\ref{fig:abundanceI} and \ref{fig:abundanceII}, demonstrates the accuracy of the adopted atmospheric parameters and abundance determinations.

\section{Evolutionary Analyses}
\label{sec:Evolutionary Analyses}
Binary star evolution involves the transfer of mass between component stars, profoundly altering their properties and evolutionary paths. One star, termed the donor, expands during its evolution and fills its Roche lobe, initiating MT onto its companion (the gainer). The evolution of Algol-type systems resolves the 'Algol paradox' by positing that the star that was initially the more massive component became the mass donor when it evolved first, expanding and filling its Roche lobe, and is referred to as the current secondary component ($M_d^f$=$M_2$). This MT process often results in a mass ratio reversal, where the star that was initially less massive, the gainer (or mass-accreting star), becomes the more massive component ($M_g^f$=$M_1$). Modern computational tools, such as the Eggleton \texttt{STARS} code \citep{Eggleton1971, Stancliffe2009}, \texttt{MESA} \citep{Paxton2015}, \texttt{TWIN} \citep{Mink2007}, and \texttt{Brussels} codes \citep{Van2016}, allow for detailed modelling of these interactions by simultaneously solving the stellar structure equations for both stars while accounting for mass and AM changes. The theoretical modelling of binary star evolution has fundamentally shifted from the early assumptions of strictly conservative mass and AM transfer toward a "liberal" evolutionary framework, necessitated by the observed mass and AM deficits in evolved systems \citep{Eggleton2000}. The prerequisite loss of matter, carrying a distinct amount of AM, is required to properly model the formation of semi-detached Algol systems exhibiting accretion discs \citep{Dervisoglu2010}. Contemporary research efforts have focused on overcoming this AM challenge by implementing complex physical modifications, including finely tuned tidal interactions, localized hot spots, and magnetic stellar wind braking, within advanced binary evolution codes \citep{Van2016, Deschamps2013, Deschamps2015, Dervisoglu2010, Van2008}. Despite these technical advances, significant uncertainty persists regarding the exact mechanisms and efficiencies of mass and AM removal, particularly during the initial phase of interaction, known as Rapid Mass Transfer (RMT). Unlike the current stage of evolution, the Slow Mass Transfer (SMT) phase, which manifests through observable orbital period changes, the RMT phase is brief ($\approx 10^5 - 10^6$ years); thus, this phase has not yet been directly observed. Consequently, modelling the RMT within evolutionary codes relies on simplified a priori recipes that parameterize the mass and AM loss. \cite{Soberman1997} compiled and classified these dynamic prescriptions into four principal loss "modes": Fast (Jeans's mode), Isotropic Re-emission, Intermediate (Ring Formation), and Slow (Accretion). Their comprehensive analysis established a fundamental stability criterion for Roche Lobe Overflow (RLOF)-triggered MT: the specific AM ($h_e$) carried away by the lost matter must not exceed the AM per reduced mass of the system. This stability convention is universally adopted across major stellar evolution codes (\texttt{MESA}, \texttt{STARS}, \texttt{TWIN}, and \texttt{Brussels}). However, the specific choice of AM loss mode for RMT differs substantially. \texttt{MESA} typically allows user definition; Cambridge \texttt{STARS} implements the prescription of \cite{Hurley2002}, where lost matter is assumed to carry the specific AM of the mass donor (Fast Mode); conversely, \texttt{TWIN} and \texttt{Brussels} typically employ the Isotropic Re-emission assumption, where the matter carries the specific AM of the mass gainer. Although most current binary evolution codes adopt this convention to thread mass loss from systems, the choice of AM loss during RMT differs among them. Because we cannot observe the RMT phase in the system, we must rely on these a priori assumptions. Logically concluding that the loss mechanism is neither entirely conservative (slow mode) nor dominated by highly unstable intermediate mode structures (ring or disc loss), we proceeded with the two remaining modes: the Fast Case (where the donor's specific AM is lost) and Isotropic Re-emission (where the gainer's specific AM is lost). To achieve a comprehensive evolutionary solution for Z Vul, we constructed a binary evolution grid tailored to test both dominant AM loss hypotheses: the Fast Case and Isotropic Re-emission.
\par
The evolutionary analysis of Algol-type binary systems, such as Z Vul, begins by addressing the fundamental stellar paradox of an evolved, less massive star orbiting a younger, more massive companion. In the case of Z Vul, this translates to an A3III-type giant acting as the mass donor to a hotter B4V component, which is the mass gainer. This configuration is explained by a prior phase of MT, initiated when the originally more massive component filled its Roche Lobe (RLOF) and rapidly lost mass to its companion, leading to a reversal of the mass ratio. The primary objective of evolutionary modeling is the inverse problem: establishing the initial system properties; specifically, the initial masses ($M_d^i, M_g^i$), initial mass ratio ($q^i$), and initial orbital period ($P^i$), that evolved to match the observed final absolute parameters of the Z Vul system.
\par
The situation becomes more complex if the mass is lost from the system during the transfer. This non-conservative evolution is often parametrized using the MT efficiency parameter $\beta$, which represents the fraction of mass lost from the donor that is \textit{not} accreted by the gainer \citep{Dervisoglu_2018}. This parameter describes the ratio of mass lost from the system to that lost from the donor star ($\dot{M}_{t} = \beta \cdot |\dot{M}_{d}|$). Conservative MT assumes $\beta=0$, where all matter leaving the donor's Lagrange $L_1$ point is accreted by the gainer, conserving the total system mass and orbital AM ($J_{t}$). Conversely, nonconservative evolution ($\beta > 0$) implies mass loss from the system, resulting in a corresponding loss of $J_{t}$. Therefore, non-conservative models for Algol systems are typically constrained to an upper limit, defined here as $\beta \le 0.9$. When mass loss from the binary system is introduced ($\beta$), it becomes clear that the parameter space for our search to determine the initial absolute parameter of the progenitor system is two-dimensional. For a given value of $\beta$, the AM carried away by the escaping mass must be considered, and the initial orbital period of the system must be recalculated as follows: The necessity of exploring this non-conservative regime is dictated by prior studies indicating significant mass and AM loss, establishing a mandatory dual-parameter search involving both the initial mass ratio ($q^i = M_d^i / M_g^i$) and $\beta$.

\begin{figure*}
	\includegraphics[width=\columnwidth]{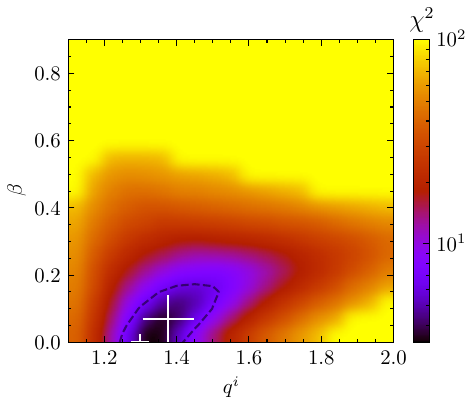}
    \includegraphics[width=\columnwidth]{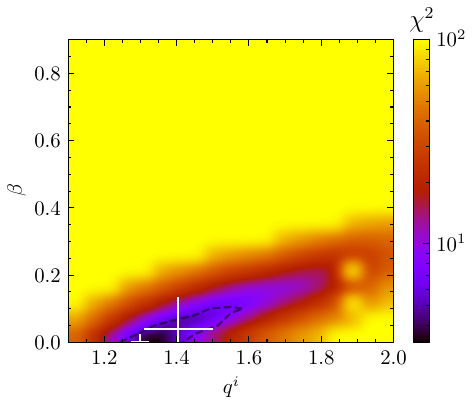}
    \caption{$\chi^2$ topology maps within the initial mass ratio versus mass-transfer efficiency ($q^{\rm i}$--$\beta$) parameter space for the evolutionary grid of Z\,Vul. The left panel displays the Fast Case (AM$_1$), and the right panel illustrates the Isotropic Re-emission Case (AM$_2$). In both panels, the global $\chi^2$ minimum (best-fit model) and the statistical confidence intervals (e.g., $1\sigma$) utilized to isolate the surviving progenitor models are overplotted.}
    \label{fig:evol_ahdII}
\end{figure*}

However, the complexity stems from the fact that the initial systemic parameters, including $M_d^i, M_g^i$ and $P^i$, are unknown. This methodology generally requires working inverse problems from currently observed parameters. Evolutionary models typically treat $q^i$ and $\beta$ as the free parameters. Even under the simplest assumption of conservative evolution, where both the total system mass and orbital AM are conserved, determining $q^i$ is necessary to set the initial parameter space for the merger. A principal source of difficulty arises when considering non-conservative MT, specifically in quantifying the AM carried away by the mass lost from the system. Therefore, to accurately calculate $P^i$, one must know the specific AM (angular momentum per unit mass) of the escaping material, $h_e$.This necessity is addressed through different angular momentum loss (AML) prescriptions as follows.  The description of non-conservative evolution introduces a critical dependency: the loss rate of $J_{t}$ depends directly on $h_e$, which is carried away by the escaped matter. \cite{Heuvel1994} derived an equation describing the change in the system's specific orbital AM $h_0$, based on the mass-loss rate and $h_e$. The change in $h_0$ is governed by the Equation \ref{equation1}.
\begin{equation}
\frac{dh_{0}}{h_{0}}=\frac{h_{e}-h_{0}}{h_{0}}\frac{d(M_{1}+M_{2})}{M_{1}+M_{2}}
\label{equation1} 
\end{equation}
The effect of this mass loss on the orbital period ($P^i$) is determined by comparing $h_e$ to the specific AM per unit mass of the entire binary system, $h_0$ ($J_{t}$ divided by total systemic mass $M_t$). The first limiting case is the Fast Mode (or Jeans' Mode, following van den Heuvel), which posits that the escaping matter carries the specific AM of the mass donor ($h_e = h_d$). It can be demonstrated that $h_d = q^i \cdot h_0$, and because $q^i > 1$ for Algol progenitors, $h_d > h_0$. This results in a higher AM loss relative to the system average, typically leading to the prediction of larger progenitor periods ($P^i$) than conservative values. This approach is consistent with the standard prescriptions adopted by codes such as \texttt{STARS} and has been previously applied to systems such as $\delta$ Librae \citep{Dervisoglu_2018}. The second case is Isotropic Re-emission, where the mass is presumed to be ejected from the vicinity of the accreting component, carrying the specific AM of the mass gainer ($h_e = h_g$). When $h_g$ is  related to $h_0$, the ratio depends solely on the mass ratio $q$. In this context, $q$ is defined as the mass of the donor divided by that of the gainer. For Algol progenitors, the donor star that evolves and begins MT is initially more massive. Therefore, the mass ratio $q^i > 1$ and a similar discussion for $h_d$ yields the relation $h_g = \frac{1}{q^i} \cdot h_0$. It is assumed that an element of mass ($dM_d^i$) lost by the donor star is captured by the gainer star and subsequently ejected isotropically from the gainer. Since $q^i > 1$, leading to $h_g < h_0$. The reduced AM loss results in smaller predicted $P^i$ values, as the loss is comparatively less severe. This mode is adopted in codes such as the TWIN and Brussels codes.

\begin{figure*}
	\includegraphics[width=\columnwidth]{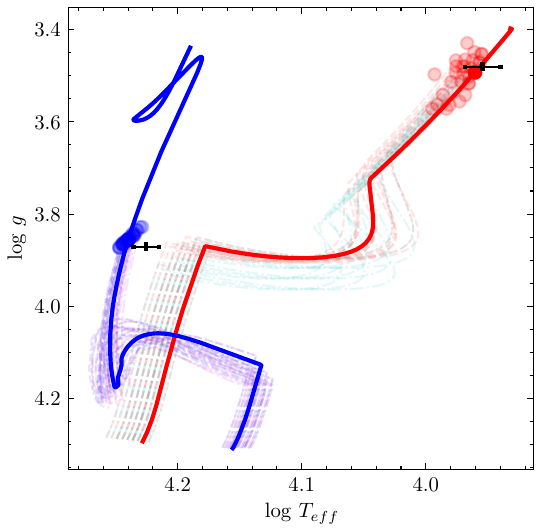}
    \includegraphics[width=\columnwidth]{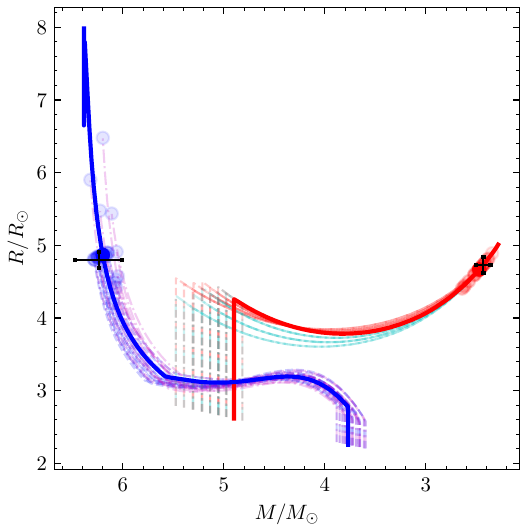}
    \caption{Evolutionary diagnostic diagrams for Z\,Vul: the Kiel diagram ($\log g$ vs. $\log T_{\rm eff}$; left panel) and the mass--radius diagram (right panel). In both panels, solid lines represent the best-fitting models, while dashed lines denote the surviving models within the confidence limits. The color-coding uniformly distinguishes the stellar components across the two AM loss regimes: blue and red lines represent the gainer ($M_{\rm g}$) and donor ($M_{\rm d}$) stars for the AM$_1$ model, whereas magenta and cyan lines correspond to the gainer ($M_{\rm g}$) and donor ($M_{\rm d}$) stars for the AM$_2$ model, respectively.}
    \label{evol_ahdIII}
\end{figure*}

The calculation of the total initial system mass ($M_t^i$) required to begin the evolutionary track is derived from mass conservation laws applied across the MT event and is independent of the AM loss formalism chosen i.e. same for both AM loss regimes. Following \cite{Dervisoglu_2018}, the initial total mass of the system ($M_t^i$) can be calculated for a given set of final parameters ($M_d^f$=$M_2$, $M_g^f$=$M_1$), initial mass ratio ($q^i$), and MT efficiency ($\beta$) using the relation derived by \cite{Giuricin_1981}:
\begin{equation} 
M_t^i = M_t^f \frac{(1+q^i)}{(1+q^f)} \frac{[1+q^f (1-\beta)]}{[1+q^i (1-\beta)]}
\label{equation2} 
\end{equation}
\par
However, the initial period ($P^i$) calculation depends explicitly on the adopted AM-loss regime. For the Fast Case ($h_e = h_d$), derived from the assumption that the lost matter carries the donor's specific AM, the initial period is given by \cite{Dervisoglu_2018}: 
\begin{equation} 
P^i = P^f \left(\frac{M_t^f}{M_t^i}\right)^2 \left(\frac{M_g^f}{M_g^i}\right)^3 \left(\frac{M_d^f}{M_d^i}\right)^{3(1-\beta)}
\label{equation3} 
\end{equation}
For the Isotropic Re-emission Case ($h_e = h_g$), on the other hand, based on the AM loss derived by \cite{Bhattacharya1991} and \cite{Soberman1997}:
\begin{equation} P^i = P^f \left(\frac{M_t^f}{M_t^i}\right)^2 \left(\frac{M_g^f}{M_g^i}\right)^{3/(1-\beta)}\left(\frac{M_d^f}{M_d^i}\right)^{3}
\label{equation4} 
\end{equation}
Notably, both equations converge to the same expression for the conservative case ($\beta=0$), yielding the familiar relation $P M_g^3 M_d^3 = \text{const}$ \citep{Paczynski1966}.
\par
The investigation of Z Vul was conducted by generating comprehensive binary evolution grids tailored to the two distinct AM loss regimes. The first methodology (AM$_1$, $h_e = h_d$) adopted for binary star evolution calculations within the scope of this study is described in \cite{Dervisoglu_2018}. The second methodology (AM$_2$, $h_e = h_g$) adopted for binary star evolution calculations within the scope of this study is described in \cite{Mink2007}. The \texttt{STARS} evolutionary code \citep{Eggleton1971, Stancliffe2009} was used for calculations. The \texttt{STARS} code allows for the simultaneous calculation of the evolutionary states of the components of any binary star system, given its initial mass and period. During the calculations, the internal structure equations of the components were solved together with the orbital and rotational AM equations under conditions that varied depending on the MT process that occurred when one of the components filled its Roche lobe. The in-house Python script {\sc BinGrid} was employed to automate the calculation of initial parameters across a predefined grid of $q^i=[1.1, 2.0]$ (in $\delta q^i=0.05$ increments) and $\beta=[0.0, 0.9]$ (in $\delta\beta=0.05$ increments), which is similar to the distribution obtained observationally for detached binaries \citep{Torres2010, Ibanoglu2012}. For the MT efficiency, we covered a very wide  range of values from conservative ($\beta=0.0$) to highly non-conservative ($\beta=0.9$), yielding $[19x19x2]=722$ distinct initial parameter combinations for each regime, as calculated from Equ. \ref{equation2}-\ref{equation3}-\ref{equation4}. This computational effort resulted in a massive database of over three million internal stellar structure models for comparison purposes. During the calculation, a critical physical filter was applied: if the calculated initial period ($P^i$) fell below a defined limit period ($P_{lim}$) for the given masses, the model was terminated. This constraint was applied based on the $P_{\rm lim}$ required for the initial components to fit within their Roche lobes at the zero-age main sequence, following equation 3 of \citet{Nelson2001}. Any calculated initial model with $P^i < P_{lim}$ was excluded from further calculation, as this indicates a configuration in which the smaller component would revolve inside the massive one, violating physics.

\begin{figure*}
	\includegraphics[width=\columnwidth]{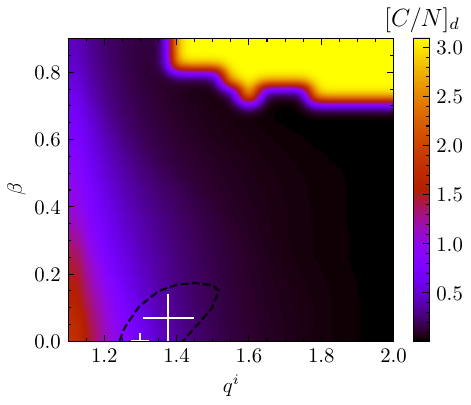}
    \includegraphics[width=\columnwidth]{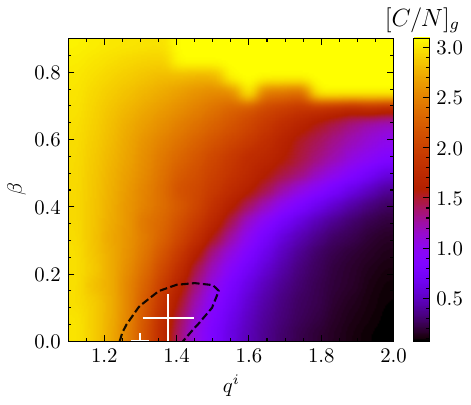}
    \caption{
    Expected surface carbon-to-nitrogen ($C/N$) ratio (by mass fraction) as a function of evolutionary path for the mass donor (left panel) and mass gainer (right panel) components of Z\,Vul, modeled under the Fast Case (AM$_1$) scenario without the effects of thermohaline mixing. The tracks illustrate the depletion of the donor's surface abundance as MT strips away the outer envelope, exposing deep, CNO-processed layers and driving the $C/N$ ratio down from the baseline cosmic value ($\sim 3.2$) to $\sim 0.5$, the surface of the mass-gainer component seemed mildly affected.}
    \label{evol_ahdIV}
\end{figure*}

\begin{figure*}
	\includegraphics[width=\columnwidth]{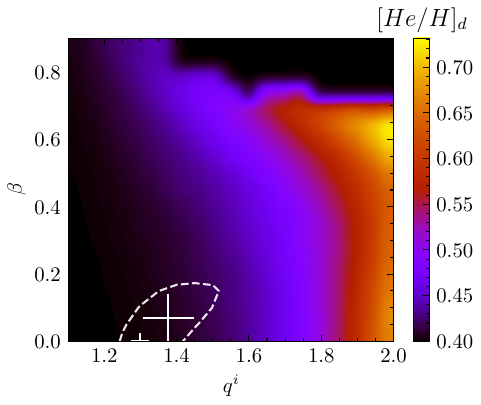}
    \includegraphics[width=\columnwidth]{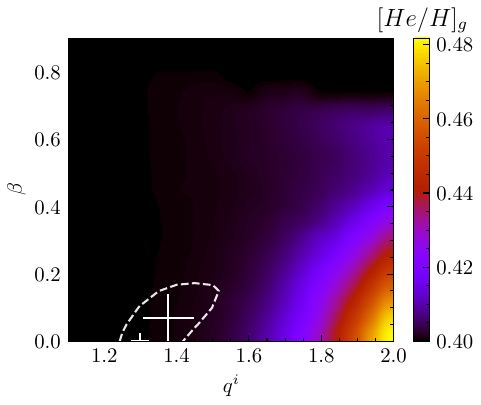}
    \caption{
    Expected surface helium-to-hydrogen ($He/H$) ratio (by mass fraction) as a function of evolutionary path for the mass donor (left panel) and mass gainer (right panel) components of Z\,Vul, modeled without the effects of thermohaline mixing. The envelope stripping does not penetrate deeply into the advanced H-burning zones where extensive helium enrichment occurs. The right panel reflects the corresponding chemical contamination on the gainer's surface resulting from the direct accretion of this mildly processed material.}
    \label{evol_ahdV}
\end{figure*}

The best-fitting evolutionary track that models the current system parameters is identified by minimizing the $\chi^2$ statistic. This rigorous $\chi^2$ test quantifies the simultaneous match between the evolutionary model's prediction and all observed absolute parameters of both components ($M, R, T_{eff}$). The $\chi^2$ function is formally defined by \cite{Nelson2001} as:

\begin{equation} 
\chi^{2}=\sum\frac{(X_{obs}-X_{th})^{2}}{\sigma_{obs}^{2}+\sigma_{th}^{2}}
\label{equation5} 
\end{equation}

This equation is central to deriving the optimal evolutionary track by comparing a chosen set of independent observed stellar parameters, $X_{obs}$, with the corresponding theoretical values, $X_{th}$, derived from a model track. In this methodology, the observational uncertainties, $\sigma_{obs}$, are represented by the standard deviations of the absolute parameters (Table \ref{tab:absolute}). Conversely, the theoretical uncertainties, $\sigma_{th}$, are derived from the grid resolution of the evolutionary models, specifically accounting for variations in the initial mass ratio ($\Delta q^i$) and mass-transfer efficiency ($\Delta \beta$). By explicitly accounting for both observational and theoretical error components, this methodology allows us to determine precise confidence limits (e.g., $1\sigma$) around the best-fit model within the $q^i - \beta$ parameter space. A noteworthy finding for Z Vul is that the resultant $\chi^2$ map exhibits a $\chi^2_{min}$ and confidence interval clustered around remarkably similar $\beta$ and $q^i$ values for both the Fast Case (AM$_1$) and the Isotropic Re-emission Case (AM$_2$) (Figure \ref{fig:evol_ahdII}). The mean parameters derived after the survival analysis across the confidence region revealed a mildly non-conservative history: for AM$_1$: $\langle q^i \rangle = 1.37 \pm 0.07$, $\langle \beta \rangle = 0.07 \pm 0.06$, and for AM$_2$: $\langle q^i \rangle = 1.41 \pm 0.1$, $\langle \beta \rangle = 0.04 \pm 0.04$. This suggests a progenitor system with mean initial parameters for AM$_1$ of $M_d^i = 5.135 \pm 0.185 M_{\odot}$, $M_g^i = 3.733 \pm 0.185 M_{\odot}$, and $P^i = 1.370 \pm 0.024$ d, and for AM$_2$; $M_d^i = 5.128 \pm 0.206 M_{\odot}$, $M_g^i = 3.656 \pm 0.109 M_{\odot}$, and $P^i = 1.363 \pm 0.026$ d. This result, which indicates a clustering of the best-fit and confidence regions around similar $\beta$ and $q^i$ values across both AM regimes, reinforces the robustness of the derived progenitor parameters and the near-conservative nature of the MT event. The derived initial parameters corresponding to the statistically robust solution space for each AM grid are listed in Table \ref{tab:survive_models}.
\par
The evolutionary states of the components are further illustrated using diagnostic diagrams. While the Hertzsprung-Russell (HR) diagram (luminosity vs. $T_{eff}$) is traditional, its utility can be limited because luminosity depends strongly on both radius and temperature ($L \propto R^2 T_{eff}^4$), potentially overemphasizing temperature variations. Given the high precision attained in the derivation of the absolute parameters for Z Vul from both spectroscopic and light-curve analyses, plotting the stellar evolution model on the Kiel Diagram ($\log g$ vs. $\log T_{eff}$) provides a superior indicator, as shown in Figure \ref{evol_ahdIII}. This diagram plots two observationally independent parameters and offers a more precise indicator of the stars' evolutionary status. The figure includes the evolutionary tracks for the best-fitting model and all surviving models within the confidence limits for both AM loss scenarios. The tracks clearly indicate that MT in Z Vul occurred during the main-sequence evolution of the original primary (Case A MT). Additionally, the Mass-Radius diagram (Figure \ref{evol_ahdIII}) demonstrates excellent agreement between the models and observed parameters for both components.
\par
A crucial aspect of our evolutionary analysis is examining the predicted surface chemical abundances, which serve as an independent test of the MT history. Figure \ref{evol_ahdIV} depicts the predicted surface carbon-to-nitrogen ($\text{C/N}$) ratio (by mass fraction) for the mass donor during its evolution. As MT proceeds, the outer layers are stripped away, exposing deeper regions where hydrogen burning via the $\text{CNO}$ cycle significantly alters the composition. This process depletes carbon and enhances nitrogen, drastically lowering the $\text{C/N}$ ratio compared to the typical cosmic abundance of $\text{C/N} \approx 3.2$. This outcome is robust and valid for both AM loss scenarios considered. Accordingly, Figure \ref{evol_ahdIV} displays the tracks corresponding to the AM$_1$ scenario, as the abundance profiles derived from both AM loss regimes are nearly identical, rendering their graphical representations indistinguishable. The nucleosynthetically altered material stripped from the donor is subsequently accreted onto the surface of the gainer. Specifically, the evolutionary models predict the expected surface yields without thermohaline mixing to be $[\text{C/N}]_d = 0.39 \pm 0.12$ and $[\text{C/N}]_g = 1.58 \pm 0.51$. This dramatic drop in the donor ratio indicates that the material has been significantly altered. The $\text{CNO}$-altered material is then accreted onto the mass gainer, contaminating its surface layers. Because the accreted material typically possesses a higher mean molecular weight than the gainer’s photosphere, an inversion in the mean molecular weight gradient is created in a thermally stable medium, triggering thermohaline mixing, a hydrodynamic instability described by \cite{Kippenhahn1980}.
\par
This process occurs when nucleosynthetically altered material, such as carbon-depleted and He-rich material from the donor, is deposited on the surface of the gainer. This process is supposed to be highly efficient in the gainer's envelope, smoothing the abundance gradient and increasing the surface $\text{C/N}$ ratio, although the final value remains below the cosmic abundance. This result is also an independent test of our evolutionary discussion because we determined the C/N ratio from spectroscopic analyses. Subsequent theoretical works, such as the comprehensive quantitative theory applied to binary MT by \cite{Ulrich1972} and the detailed time-scale estimations by \cite{Kippenhahn1980}, explored the resulting mixing rate and its implications for the stellar surface composition. However, the efficiency and timescales of thermohaline mixing remain subjects of active debate. Our model's predictions for the surface $\text{C/N}$ and $\text{He/H}$ ratios are highly dependent on the assumed efficiency of this mixing. The calculated surface yields without thermohaline mixing, where the contamination is confined to the surface, were $[\text{C/N}]_g = 1.58 \pm 0.51$ and $[\text{He/H}]_g = 0.401 \pm 0.001$. Full instantaneous mixing would drive these ratios toward solar/cosmic values. This process is supposed to be highly efficient in the gainer’s envelope, smoothing the abundance gradient and increasing the surface $\text{C/N}$ ratio, although the final value remains below the cosmic abundance. The fact that the $\text{C/N}$ value determined from our spectroscopic analyses is consistent with the predicted post-mixing value confirms the activity of thermohaline mixing, serving as an independent test of our evolutionary discussion. This result validates the hypothesis that, while active, mixing is not necessarily instantaneous or complete, preventing the ratios from returning fully to the solar standard. This approach, which uses photospheric abundances as tracers of internal processing, has been successfully employed in several previous studies of Algol \citep{Kolbas_2014, Dervisoglu_2018, Pilecki_2018}. A parallel analysis of the helium-to-hydrogen ($\text{He/H}$) ratio further corroborated these results. The donor surface $\text{He/H}$ ratio increases as deeper helium-enriched layers are exposed by the mass loss. The evolutionary models predict the expected surface yields without thermohaline mixing to be: $[\mathrm{He}/\mathrm{H}]_\odot \sim 0.4$, $[\mathrm{He}/\mathrm{H}]_d = 0.408 \pm 0.005$, $[\mathrm{He}/\mathrm{H}]_g = 0.401 \pm 0.001$. This altered material subsequently raises the gainer's surface $\text{He/H}$ ratio, even after thermohaline mixing, confirming the successful transfer and contamination by nucleosynthetically processed material. The diagrams illustrating the final states of the Z Vul components and the constraints provided by the $\text{He/H}$ ratios are presented in Figure \ref{evol_ahdV}.

\section{Concluding Remarks}
\label{sec:Concluding}
In this study, we presented a comprehensive photometric, spectroscopic, and evolutionary analysis of the hot Algol-type binary system Z~Vul. By leveraging high-resolution, high signal-to-noise spectra from the {\sc hermes} spectrograph alongside high precision space photometry from TESS, we systematically derived the absolute, atmospheric, and chemical properties of the system's components to trace its mass-transfer history.
\par
Our methodological framework utilized an iterative combination of light-curve modeling and spectral disentangling. To accurately account for the flux depressions caused by circumstellar material during the eclipses, we approximated localized obscuration using cool spot parameters. Using an \texttt{MCMC} optimization approach, we extracted highly accurate fundamental parameters, yielding a mass of $M_1 = 6.26 \pm 0.08\,M_{\odot}$ for the primary (gainer) and $M_2 = 2.44 \pm 0.03\,M_{\odot}$ for the secondary (donor). The disentangled spectra were subsequently renormalized using wavelength-dependent light contribution functions, allowing us to perform a rigorous NLTE atmospheric and abundance analysis with the \textsc{tlusty} and \textsc{synspec} codes.
\par
A critical aspect of our study is the determination of the photospheric CNO and He abundances, which serve as sensitive tracers of the thermonuclear and MT processes. Spectroscopic analysis revealed an abundance ratio of $\text{C/N} = 1.70^{+0.70}_{-0.50}$ for the primary and $\text{C/N} = 2.14^{+0.55}_{-0.44}$ for the secondary. While the primary exhibits a mild nitrogen enrichment consistent with a gainer star superficially polluted by CNO-processed material, the secondary lacks the severe C/N inversion traditionally expected for a deeply stripped donor. 
\par
To interpret these chemical signatures, we computed an extensive grid of binary evolution models using the \textsc{stars} code, testing both conservative and non-conservative (Fast and Isotropic Re-emission) AM loss regimes. Our $\chi^2$ optimization indicates that Z~Vul underwent Case A MT and evolved almost conservatively ($\beta \approx 0.04 - 0.07$). The best-fitting progenitor system consisted of a $\sim 5.1\,M_{\odot}$ primary and a $\sim 3.7\,M_{\odot}$ secondary with an initial orbital period of $P^i \approx 1.37$\,d.
\par
\begin{figure}
	\includegraphics[width=\columnwidth]{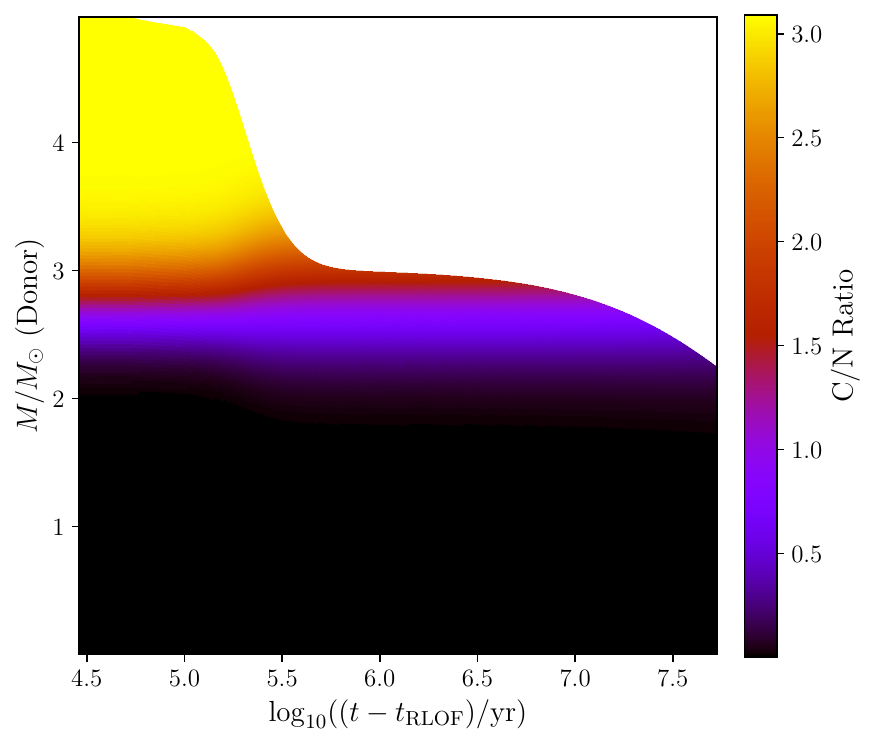}
    \includegraphics[width=\columnwidth]{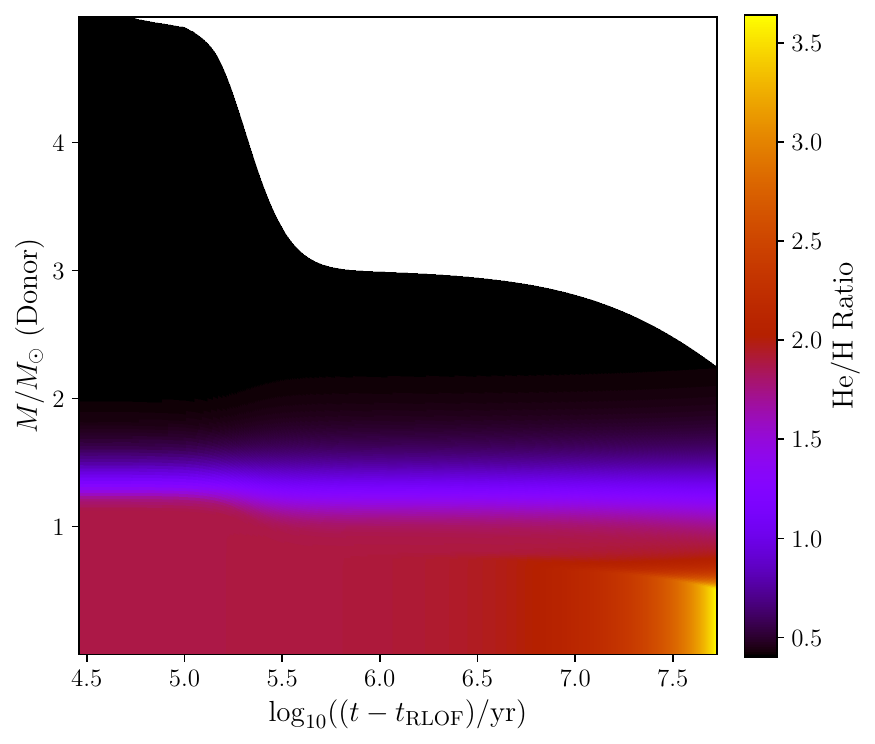}
    \caption{
    Evolution of the internal chemical profiles of the initially $5.1\,M_{\odot}$ donor star during the MT phase, displaying local $C/N$ (left panel) and $He/H$ (right panel) abundance topographies as a function of mass coordinate ($M/M_\odot$) and logarithmic time ($\log t$) since the onset of Roche Lobe Overflow (RLOF). The current observed state of the Z\,Vul secondary ($M_2 \approx 2.48\,M_{\odot}$ at $\log t \approx 7$) is indicated. The profiles demonstrate that while envelope stripping has exposed intermediate layers characterized by an altered, non-equilibrium $C/N$ ratio from incomplete CNO-processing, it has not penetrated the fully processed stellar core, leaving the surface helium abundance unaltered at its baseline value ($0.408$; i.e., mass fractions $He/H = 0.28/0.70$).}
    \label{fig:donorprofile}
\end{figure}

Crucially, our evolutionary models clarify the apparent discrepancy between the theoretically expected deep-envelope C/N ratio ($[\text{C/N}]_d = 0.39 \pm 0.12$) and our observed spectroscopic value for the donor ($\sim 2.14$). During the rapid MT phase, the initial $5.1\,M_{\odot}$ donor transferred approximately half of its mass to the companion within $10^6$ years, followed by a stage of slow mass transfer (SMT) of an additional $\sim 0.2\,M_{\odot}$ over 9 million years. As illustrated by the C/N and He/H mass profile plots of the donor star during MT (see Figure \ref{fig:donorprofile}), the mass-loss process has stripped the secondary down only to the boundary where the C/N ratio begins to deviate from the solar value. It has not reached the deeply processed stellar core. This is definitively corroborated by the He/H profile, which shows the expected surface He/H ratio of the donor to be $0.408$, exactly matching the $0.28/0.70$ mass fraction ratio of unprocessed solar surface material. Because the helium abundance remains completely unaltered, we are clearly observing an envelope region, not the core.
\par
Instead, the exposed envelope corresponds to an outer region governed by an incomplete CNO cycle. In these layers, the $^{12}\text{C} + p \rightarrow ^{14}\text{N}$ reaction, which initiates at relatively low temperatures, easily attains a C/N equilibrium ratio without activating the full high-temperature CNO cycle. This intermediate region is significantly affected by semi-convection and diffusion, making the surface C/N ratio highly sensitive to minor structural and evolutionary changes. For instance, if the donor had been stripped of just $0.2\,M_{\odot}$ less mass, the expected surface C/N would perfectly match our spectroscopic value of 2. Furthermore, because the accreted material is primarily of solar composition with only a mildly altered C/N ratio, thermohaline mixing on the gainer rapidly smooths the abundance gradient, allowing the primary's surface to naturally settle to the observed C/N ratio of $\sim 1.5 - 2$.
\par
In conclusion, our precise chemical tagging, combined with detailed binary evolution tracks, confirms that Z~Vul is a nearly conservative Case A mass-transfer system. The surface abundances of the secondary donor represent the delicately stripped outer layers of an intermediate nucleosynthetic region rather than the fully CNO-processed core. This underscores the necessity of coupling high-resolution spectroscopic abundances with interior stellar profiles to accurately reconstruct the evolutionary history of interacting binaries.

\section*{Acknowledgements}
We extend our sincere gratitude to Dr. Holger Lehmann for his valuable cooperation during the early stages of this project. We also thank Dilem Göktaş for her essential role in the initial organization and setup of the workstations used for this study. A single spectroscopic dataset obtained within the scope of the project numbered 10ARTT150-493-0 by \cite{Ibanoglu2012}, conducted using the RTT150 Telescope and the Coudé Echelle Spectrometer (CES) system at the TUG (TÜBİTAK National Observatory, Antalya) site under the Türkiye National Observatories, have been utilized, and we express our gratitude for the invaluable support provided by the Türkiye National Observatories and their personnel. This work has been supported in part by the Scientific and Technological Research Council (TÜBİTAK) by the grant number 122F111. We also acknowledge the use of ground-based NIR photometric measurements in the $JHK$ passbands observations from \cite{Lazaro2009}, which were utilized in the multi-wavelength modeling and analysis presented in this work. This paper includes data collected with the TESS mission, obtained from the MAST data archive at the Space Telescope Science Institute (STScI). Funding for the TESS mission is provided by the NASA Explorer Program. STScI is operated by the Association of Universities for Research in Astronomy, Inc., under NASA contract NAS 5–26555. We thank the referee, Dr Sergio Simón-Díaz, for a careful and constructive report whose suggestions improved the presentation of this work.

\section*{Data Availability}

The high-resolution spectroscopic data used in this article will be shared upon reasonable request to the corresponding author. The TESS time-series flux data for the target stars are publicly available from the NASA MAST archive (\hyperlink{blue}{https://mast.stsci.edu/portal/Mashup/Clients/Mast/Portal.html}). This research has also made use of the SIMBAD database, operated at CDS, Strasbourg, France.



\bibliographystyle{mnras}
\bibliography{sample} 

@ARTICLE{Martinez_2025,
       author = {{Mart{\'\i}nez-Sebasti{\'a}n}, C. and {Sim{\'o}n-D{\'\i}az}, S. and {Jin}, H. and {Keszthelyi}, Z. and {Holgado}, G. and {Langer}, N. and {Puls}, J.},
        title = "{The IACOB project: XIII. Helium enrichment in O-type stars as a tracer of past binary interaction}",
      journal = {\aap},
         year = 2025,
        month = jan,
       volume = {693},
          eid = {L10},
        pages = {L10},
          doi = {10.1051/0004-6361/202452622},
archivePrefix = {arXiv},
       eprint = {2412.14107},
 primaryClass = {astro-ph.SR},
       adsurl = {https://ui.adsabs.harvard.edu/abs/2025A&A...693L..10M}
}

@ARTICLE{Martinez_2026,
       author = {{Mart{\'\i}nez-Sebasti{\'a}n}, C. and {Holgado}, G. and {Sim{\'o}n-D{\'\i}az}, S. and {Martins}, F. and {Puls}, J.},
        title = "{The IACOB project: XVII. Nitrogen abundances in Galactic O-type stars: further hints for separating binary-interaction products from effectively single stars}",
      journal = {arXiv e-prints},
         year = 2026,
        month = apr,
          eid = {arXiv:2604.26606},
        pages = {arXiv:2604.26606},
          doi = {10.48550/arXiv.2604.26606},
archivePrefix = {arXiv},
       eprint = {2604.26606},
 primaryClass = {astro-ph.SR},
       adsurl = {https://ui.adsabs.harvard.edu/abs/2026arXiv260426606M}
}

@ARTICLE{Simon_2026,
       author = {{Sim{\'o}n-D{\'\i}az}, S. and {Holgado}, G. and {Mart{\'\i}nez-Sebasti{\'a}n}, C. and {Carretero-Castrillo}, M. and {Jin}, H. and {Urbaneja}, M.~A. and {Gamen}, R. and {Puls}, J. and {de Burgos}, A. and {Garcia}, M. and {Herrero}, A. and {Keszthelyi}, Z. and {Langer}, N. and {Najarro}, F. and {Paredes}, J.~M. and {Rib{\'o}}, M.},
        title = "{The IACOB project: XVI. Surface helium abundances in Galactic O-type stars: indications for identifying binary interaction products}",
      journal = {arXiv e-prints},
         year = 2026,
        month = jan,
          eid = {arXiv:2601.20698},
        pages = {arXiv:2601.20698},
          doi = {10.48550/arXiv.2601.20698},
archivePrefix = {arXiv},
       eprint = {2601.20698},
 primaryClass = {astro-ph.SR},
       adsurl = {https://ui.adsabs.harvard.edu/abs/2026arXiv260120698S}
}

@INPROCEEDINGS{Musaev_2000,
       author = {{Musaev}, Faig and {Bikmaev}, Ilfan},
        title = "{High-Resolution Coud{\'e}-Echelle Spectrometer for the 1. 5-m Kazan University Telescope at the Turkish National Observatory}",
    booktitle = {The Carbon Star Phenomenon},
         year = 2000,
       editor = {{Wing}, Robert F.},
       series = {IAU Symposium},
       volume = {177},
        month = jan,
        pages = {569},
       adsurl = {https://ui.adsabs.harvard.edu/abs/2000IAUS..177..569M}
}

@ARTICLE{Kunovac_2020,
       author = {{Kunovac Hod{\v{z}}i{\'c}}, Vedad and {Triaud}, Amaury H.~M.~J. and {Martin}, David V. and {Fabrycky}, Daniel C. and {Cegla}, Heather M. and {Collier Cameron}, Andrew and {Gill}, Samuel and {Hellier}, Coel and {Kostov}, Veselin B. and {Maxted}, Pierre F.~L. and {Orosz}, Jerome A. and {Pepe}, Francesco and {Pollacco}, Don and {Queloz}, Didier and {S{\'e}gransan}, Damien and {Udry}, St{\'e}phane and {Welsh}, William F.},
        title = "{The EBLM project - VII. Spin-orbit alignment for the circumbinary planet host EBLM J0608-59 A/TOI-1338 A}",
      journal = {\mnras},
         year = 2020,
        month = sep,
       volume = {497},
       number = {2},
        pages = {1627-1633},
          doi = {10.1093/mnras/staa2071},
archivePrefix = {arXiv},
       eprint = {2007.05514},
 primaryClass = {astro-ph.EP},
       adsurl = {https://ui.adsabs.harvard.edu/abs/2020MNRAS.497.1627K}
}

@ARTICLE{Cegla_2016,
       author = {{Cegla}, H.~M. and {Lovis}, C. and {Bourrier}, V. and {Beeck}, B. and {Watson}, C.~A. and {Pepe}, F.},
        title = "{The Rossiter-McLaughlin effect reloaded: Probing the 3D spin-orbit geometry, differential stellar rotation, and the spatially-resolved stellar spectrum of star-planet systems}",
      journal = {\aap},
         year = 2016,
        month = apr,
       volume = {588},
          eid = {A127},
        pages = {A127},
          doi = {10.1051/0004-6361/201527794},
archivePrefix = {arXiv},
       eprint = {1602.00322},
 primaryClass = {astro-ph.EP},
       adsurl = {https://ui.adsabs.harvard.edu/abs/2016A&A...588A.127C}
}

@ARTICLE{Xu_2025,
       author = {{Xu}, Xiao-Tian and {Langer}, Norbert and {Klencki}, Jakub and {Wang}, Chen and {Li}, Xiang-Dong},
        title = "{Stable mass transfer in massive binaries leading to merging black holes}",
      journal = {arXiv e-prints},
         year = 2025,
        month = dec,
          eid = {arXiv:2512.20054},
        pages = {arXiv:2512.20054},
          doi = {10.48550/arXiv.2512.20054},
archivePrefix = {arXiv},
       eprint = {2512.20054},
 primaryClass = {astro-ph.SR},
       adsurl = {https://ui.adsabs.harvard.edu/abs/2025arXiv251220054X}
}

@ARTICLE{Klencki_2026,
       author = {{Klencki}, Jakub and {Podsiadlowski}, Philipp and {Langer}, Norbert and {Olejak}, Aleksandra and {Justham}, Stephen and {Vigna-G{\'o}mez}, Alejandro and {de Mink}, Selma E.},
        title = "{Comfort zones of stars: A limit on orbital tightening via stable mass transfer shapes the properties of binary black hole mergers}",
      journal = {\aap},
         year = 2026,
        month = feb,
       volume = {706},
          eid = {A296},
        pages = {A296},
          doi = {10.1051/0004-6361/202555500},
archivePrefix = {arXiv},
       eprint = {2505.08860},
 primaryClass = {astro-ph.HE},
       adsurl = {https://ui.adsabs.harvard.edu/abs/2026A&A...706A.296K}
}

@ARTICLE{Olejak_2025,
       author = {{Olejak}, Aleksandra and {Klencki}, Jakub and {Vigna-Gomez}, Alejandro and {de Mink}, Selma E. and {van Son}, Lieke and {Cehula}, Jakub and {Stegmann}, Jakob and {Ryu}, Taeho and {Hendriks}, David D.},
        title = "{Non-conservative Mass Transfer as a Formation Channel for Gaia Black Hole System}",
      journal = {arXiv e-prints},
         year = 2025,
        month = nov,
          eid = {arXiv:2511.10728},
        pages = {arXiv:2511.10728},
          doi = {10.48550/arXiv.2511.10728},
archivePrefix = {arXiv},
       eprint = {2511.10728},
 primaryClass = {astro-ph.HE},
       adsurl = {https://ui.adsabs.harvard.edu/abs/2025arXiv251110728O}
}

@ARTICLE{Sen_2026,
       author = {{Sen}, Koushik and {Renzo}, Mathieu and {Jin}, Harim and {Langer}, Norbert and {Schootemeijer}, Abel and {Villase{\~n}or}, Jaime I. and {Mahy}, Laurent and {Grichener}, Aldana and {Shah}, Neev and {Wang}, Chen and {Xu}, Xiao-Tian},
        title = "{Interacting Binaries on the Main Sequence as In Situ Tracers of Mass-transfer Efficiency and Stability}",
      journal = {\apj},
         year = 2026,
        month = mar,
       volume = {1000},
       number = {1},
          eid = {2},
        pages = {2},
          doi = {10.3847/1538-4357/ae40fc},
archivePrefix = {arXiv},
       eprint = {2511.15347},
 primaryClass = {astro-ph.SR},
       adsurl = {https://ui.adsabs.harvard.edu/abs/2026ApJ..1000....2S}
}

@ARTICLE{Sen_2022,
       author = {{Sen}, K. and {Langer}, N. and {Marchant}, P. and {Menon}, A. and {de Mink}, S.~E. and {Schootemeijer}, A. and {Sch{\"u}rmann}, C. and {Mahy}, L. and {Hastings}, B. and {Nathaniel}, K. and {Sana}, H. and {Wang}, C. and {Xu}, X.~T.},
        title = "{Detailed models of interacting short-period massive binary stars}",
      journal = {\aap},
         year = 2022,
        month = mar,
       volume = {659},
          eid = {A98},
        pages = {A98},
          doi = {10.1051/0004-6361/202142574},
archivePrefix = {arXiv},
       eprint = {2111.03329},
 primaryClass = {astro-ph.SR},
       adsurl = {https://ui.adsabs.harvard.edu/abs/2022A&A...659A..98S}
}

@ARTICLE{Yucel_2026,
       author = {{Y{\"u}cel}, G{\"o}khan and {Bak{\i}{\textcommabelow s}}, Volkan and {Nitschelm}, Christian and {{\c{S}}ahin}, Timur and {G{\"u}ney}, Ferhat},
        title = "{Spectral Disentangling Reveals Deep CNO-cycle Exposure in ET Cru}",
      journal = {\aj},
         year = 2026,
        month = jul,
       volume = {172},
       number = {1},
          eid = {15},
        pages = {15},
          doi = {10.3847/1538-3881/ae6cfb},
archivePrefix = {arXiv},
       eprint = {2605.14018},
 primaryClass = {astro-ph.SR},
       adsurl = {https://ui.adsabs.harvard.edu/abs/2026AJ....172...15Y}
}

@ARTICLE{Giuricin_1981,
       author = {{Giuricin}, G. and {Mardirossian}, F.},
        title = "{Some aspects of mass loss and mass transfer in Algol variables.}",
      journal = {\apjs},
         year = 1981,
        month = may,
       volume = {46},
        pages = {1-26},
          doi = {10.1086/190732},
       adsurl = {https://ui.adsabs.harvard.edu/abs/1981ApJS...46....1G}
}

@ARTICLE{Crawford_1955,
       author = {{Crawford}, J.~A.},
        title = "{On the Subgiant Components of Eclipsing Binary Systems.}",
      journal = {\apj},
         year = 1955,
        month = jan,
       volume = {121},
        pages = {71},
       adsurl = {https://ui.adsabs.harvard.edu/abs/1955ApJ...121...71C}
}

@ARTICLE{Struve_1948,
       author = {{Struve}, Otto},
        title = "{The masses and mass-ratios of close binary systems}",
      journal = {Annales d'Astrophysique},
         year = 1948,
        month = jan,
       volume = {11},
        pages = {117},
       adsurl = {https://ui.adsabs.harvard.edu/abs/1948AnAp...11..117S}
}

@ARTICLE{Balachandran_1986,
       author = {{Balachandran}, S. and {Lambert}, D.~L. and {Tomkin}, J. and {Parthasarathy}, M.},
        title = "{The chemical compositions of algol systems - III. Beta Lyrae-nucleosynthesis revealed.}",
      journal = {\mnras},
         year = 1986,
        month = apr,
       volume = {219},
        pages = {479-494},
          doi = {10.1093/mnras/219.3.479},
       adsurl = {https://ui.adsabs.harvard.edu/abs/1986MNRAS.219..479B}
}

@ARTICLE{Tomkin_1989,
       author = {{Tomkin}, Jocelyn},
        title = "{High signal-to-noise ratio {\textemdash} The spectroscopic key to Algol systems}",
      journal = {\ssr},
         year = 1989,
        month = jun,
       volume = {50},
       number = {1-2},
        pages = {245-255},
          doi = {10.1007/BF00215934},
       adsurl = {https://ui.adsabs.harvard.edu/abs/1989SSRv...50..245T}
}

@ARTICLE{Tomkin_1993,
       author = {{Tomkin}, J. and {Lambert}, D.~L. and {Lemke}, M.},
        title = "{The chemical compositions of Algol systems - V. Confirmation of carbon deficiencies in the primaries of eight systems.}",
      journal = {\mnras},
         year = 1993,
        month = dec,
       volume = {265},
        pages = {581-587},
          doi = {10.1093/mnras/265.3.581},
       adsurl = {https://ui.adsabs.harvard.edu/abs/1993MNRAS.265..581T}
}

@ARTICLE{Parthasarathy_1983,
       author = {{Parthasarathy}, M. and {Lambert}, D.~L. and {Tomkin}, J.},
        title = "{The chemical composition of Algol systems- II. The carbon and nitrogen abundances of the secondaries of U CEP and U Sge.}",
      journal = {\mnras},
         year = 1983,
        month = jun,
       volume = {203},
        pages = {1063-1078},
          doi = {10.1093/mnras/203.4.1063},
       adsurl = {https://ui.adsabs.harvard.edu/abs/1983MNRAS.203.1063P}
}

@ARTICLE{Plavec_1982,
       author = {{Plavec}, M.~J. and {Weiland}, J.~L. and {Koch}, R.~H.},
        title = "{Energy distribution in the strongly interacting binary system SX Cas.}",
      journal = {\apj},
         year = 1982,
        month = may,
       volume = {256},
        pages = {206-221},
          doi = {10.1086/159897},
       adsurl = {https://ui.adsabs.harvard.edu/abs/1982ApJ...256..206P}
}

@ARTICLE{Plavec_1983,
       author = {{Plavec}, M.~J.},
        title = "{Far-ultraviolet emission lines in U Cephei : evidence for a hot, turbulent circumstellar envelope.}",
      journal = {\apj},
         year = 1983,
        month = dec,
       volume = {275},
        pages = {251-270},
          doi = {10.1086/161530},
       adsurl = {https://ui.adsabs.harvard.edu/abs/1983ApJ...275..251P}
}

@ARTICLE{Dobias_1985,
       author = {{Dobias}, J.~J. and {Plavec}, M.~J.},
        title = "{IUE and optical spectral scans of U Sagittae : an analysis and comparison with U Cephei.}",
      journal = {\pasp},
         year = 1985,
        month = feb,
       volume = {97},
        pages = {138-150},
          doi = {10.1086/131509},
       adsurl = {https://ui.adsabs.harvard.edu/abs/1985PASP...97..138D}
}

@ARTICLE{Cugier_1989,
       author = {{Cugier}, H.},
        title = "{Carbon abundance in the primaries of six Algol-type stars.}",
      journal = {\aap},
         year = 1989,
        month = apr,
       volume = {214},
        pages = {168-178},
       adsurl = {https://ui.adsabs.harvard.edu/abs/1989A&A...214..168C}
}

@ARTICLE{Sarna_1996,
       author = {{Sarna}, M.~J. and {De Greve}, J.-P.},
        title = "{Chemical Evolution of Algols}",
      journal = {\qjras},
         year = 1996,
        month = mar,
       volume = {37},
        pages = {11},
       adsurl = {https://ui.adsabs.harvard.edu/abs/1996QJRAS..37...11S}
}

@ARTICLE{Ibanoglu2012,
       author = {{Ibano{\v{g}}lu}, C. and {Dervi{\c{s}}o{\v{g}}lu}, A. and {{\'C}ak{\i}rl{\i}}, {\"O}. and {Sipahi}, E. and {Y{\"u}ce}, K.},
        title = "{Carbon deficiencies in the primaries of some classical Algols}",
      journal = {\mnras},
         year = 2012,
        month = jan,
       volume = {419},
       number = {2},
        pages = {1472-1479},
          doi = {10.1111/j.1365-2966.2011.19812.x},
archivePrefix = {arXiv},
       eprint = {1109.1939},
 primaryClass = {astro-ph.SR},
       adsurl = {https://ui.adsabs.harvard.edu/abs/2012MNRAS.419.1472I}
}

@ARTICLE{Dervisoglu2010,
       author = {{Dervi{\c{s}}o{\v{g}}lu}, A. and {Tout}, Christopher A. and {Ibano{\v{g}}lu}, C.},
        title = "{Spin angular momentum evolution of the long-period Algols}",
      journal = {\mnras},
         year = 2010,
        month = aug,
       volume = {406},
       number = {2},
        pages = {1071-1083},
          doi = {10.1111/j.1365-2966.2010.16732.x},
archivePrefix = {arXiv},
       eprint = {1003.4392},
 primaryClass = {astro-ph.SR},
       adsurl = {https://ui.adsabs.harvard.edu/abs/2010MNRAS.406.1071D}
}

@INPROCEEDINGS{Deschamps2014,
       author = {{Deschamps}, R. and {Siess}, L. and {Davis}, P.~J. and {Jorissen}, A.},
        title = "{Algol Evolution with Spin-down Mechanisms and Systemic Mass Loss}",
    booktitle = {Tenth Pacific Rim Conference on Stellar Astrophysics},
         year = 2014,
       editor = {{Lee}, H. -W. and {Kang}, Y.~W. and {Leung}, K. -C.},
       series = {Astronomical Society of the Pacific Conference Series},
       volume = {482},
        month = aug,
        pages = {127},
       adsurl = {https://ui.adsabs.harvard.edu/abs/2014ASPC..482..127D}
}

@ARTICLE{Deschamps2015,
       author = {{Deschamps}, R. and {Braun}, K. and {Jorissen}, A. and {Siess}, L. and {Baes}, M. and {Camps}, P.},
        title = "{Non-conservative evolution in Algols: where is the matter?}",
      journal = {\aap},
         year = 2015,
        month = may,
       volume = {577},
          eid = {A55},
        pages = {A55},
          doi = {10.1051/0004-6361/201424772},
archivePrefix = {arXiv},
       eprint = {1502.04957},
 primaryClass = {astro-ph.SR},
       adsurl = {https://ui.adsabs.harvard.edu/abs/2015A&A...577A..55D}
}

@ARTICLE{Hensberge_2000,
       author = {{Hensberge}, H. and {Pavlovski}, K. and {Verschueren}, W.},
        title = "{The eclipsing binary V578 Mon in the Rosette nebula: age and distance to NGC 2244 using Fourier disentangled component spectra}",
      journal = {\aap},
         year = 2000,
        month = jun,
       volume = {358},
        pages = {553-571},
       adsurl = {https://ui.adsabs.harvard.edu/abs/2000A&A...358..553H}
}

@ARTICLE{Pavlovski_2005,
       author = {{Pavlovski}, K. and {Hensberge}, H.},
        title = "{Abundances from disentangled component spectra: theeclipsing binary V578 Mon}",
      journal = {\aap},
         year = 2005,
        month = aug,
       volume = {439},
       number = {1},
        pages = {309-315},
          doi = {10.1051/0004-6361:20052804},
archivePrefix = {arXiv},
       eprint = {astro-ph/0504433},
 primaryClass = {astro-ph},
       adsurl = {https://ui.adsabs.harvard.edu/abs/2005A&A...439..309P}
}

@ARTICLE{Tomkin_1978,
       author = {{Tomkin}, J. and {Lambert}, D.~L.},
        title = "{Detection of the secondary of Algol.}",
      journal = {\apjl},
         year = 1978,
        month = jun,
       volume = {222},
        pages = {L119-L122},
          doi = {10.1086/182705},
       adsurl = {https://ui.adsabs.harvard.edu/abs/1978ApJ...222L.119T}
}

@ARTICLE{Kolbas_2014,
       author = {{Kolbas}, V. and {Dervi{\c{s}}o{\u{g}}lu}, A. and {Pavlovski}, K. and {Southworth}, J.},
        title = "{Tracing CNO exposed layers in the Algol-type binary system u Her}",
      journal = {\mnras},
         year = 2014,
        month = nov,
       volume = {444},
       number = {4},
        pages = {3118-3129},
          doi = {10.1093/mnras/stu1652},
archivePrefix = {arXiv},
       eprint = {1408.2681},
 primaryClass = {astro-ph.SR},
       adsurl = {https://ui.adsabs.harvard.edu/abs/2014MNRAS.444.3118K}
}

@ARTICLE{Kolbas_2015,
       author = {{Kolbas}, V. and {Pavlovski}, K. and {Southworth}, J. and {Lee}, C.-U. and {Lee}, D.-J. and {Lee}, J.~W. and {Kim}, S.-L. and {Kim}, H.-I. and {Smalley}, B. and {Tkachenko}, A.},
        title = "{Spectroscopically resolving the Algol triple system}",
      journal = {\mnras},
         year = 2015,
        month = aug,
       volume = {451},
       number = {4},
        pages = {4150-4161},
          doi = {10.1093/mnras/stv1261},
archivePrefix = {arXiv},
       eprint = {1506.01254},
 primaryClass = {astro-ph.SR},
       adsurl = {https://ui.adsabs.harvard.edu/abs/2015MNRAS.451.4150K}
}

@ARTICLE{Pavlovski_2022,
       author = {{Pavlovski}, K. and {Hummel}, C.~A. and {Tkachenko}, A. and {Dervi{\c{s}}o{\u{g}}lu}, A. and {Kayhan}, C. and {Zavala}, R.~T. and {Hutter}, D.~J. and {Tycner}, C. and {{\c{S}}ahin}, T. and {Audenaert}, J. and {Baeyens}, R. and {Bodensteiner}, J. and {Bowman}, D.~M. and {Gebruers}, S. and {Jannsen}, N.~E. and {Mombarg}, J.~S.~G.},
        title = "{Dynamical parallax, physical parameters, and evolutionary status of the components of the bright eclipsing binary {\ensuremath{\alpha}} Draconis}",
      journal = {\aap},
         year = 2022,
        month = feb,
       volume = {658},
          eid = {A92},
        pages = {A92},
          doi = {10.1051/0004-6361/202142292},
archivePrefix = {arXiv},
       eprint = {2111.03887},
 primaryClass = {astro-ph.SR},
       adsurl = {https://ui.adsabs.harvard.edu/abs/2022A&A...658A..92P}
}

@ARTICLE{Pavlovski_Southworth_2009,
       author = {{Pavlovski}, K. and {Southworth}, J.},
        title = "{Chemical evolution of high-mass stars in close binaries - I. The eclipsing binary V453Cygni}",
      journal = {\mnras},
         year = 2009,
        month = apr,
       volume = {394},
       number = {3},
        pages = {1519-1528},
          doi = {10.1111/j.1365-2966.2009.14418.x},
archivePrefix = {arXiv},
       eprint = {0812.3769},
 primaryClass = {astro-ph},
       adsurl = {https://ui.adsabs.harvard.edu/abs/2009MNRAS.394.1519P}
}

@ARTICLE{Pavlovski_2009,
       author = {{Pavlovski}, K. and {Tamajo}, E. and {Koubsk{\'y}}, P. and {Southworth}, J. and {Yang}, S. and {Kolbas}, V.},
        title = "{Chemical evolution of high-mass stars in close binaries - II. The evolved component of the eclipsing binary V380Cygni}",
      journal = {\mnras},
         year = 2009,
        month = dec,
       volume = {400},
       number = {2},
        pages = {791-804},
          doi = {10.1111/j.1365-2966.2009.15479.x},
archivePrefix = {arXiv},
       eprint = {0908.0351},
 primaryClass = {astro-ph.SR},
       adsurl = {https://ui.adsabs.harvard.edu/abs/2009MNRAS.400..791P}
}

@ARTICLE{Tkachenko_2010,
       author = {{Tkachenko}, A. and {Lehmann}, H. and {Mkrtichian}, D.},
        title = "{Spectroscopic Modeling of the Algol-type Star TW Draconis}",
      journal = {\aj},
         year = 2010,
        month = apr,
       volume = {139},
       number = {4},
        pages = {1327-1337},
          doi = {10.1088/0004-6256/139/4/1327},
       adsurl = {https://ui.adsabs.harvard.edu/abs/2010AJ....139.1327T}
}

@ARTICLE{Tkachenko_2009,
       author = {{Tkachenko}, A. and {Lehmann}, H. and {Mkrtichian}, D.~E.},
        title = "{Spectroscopic modeling of oscillating Algol-type stars. I. RZ Cassiopeia}",
      journal = {\aap},
         year = 2009,
        month = sep,
       volume = {504},
       number = {3},
        pages = {991-1001},
          doi = {10.1051/0004-6361/200911949},
       adsurl = {https://ui.adsabs.harvard.edu/abs/2009A&A...504..991T}
}

@ARTICLE{Glazunova_2011,
       author = {{Glazunova}, L.~V. and {Mkrtichian}, D.~E. and {Rostopchin}, S.~I.},
        title = "{TX UMa: new orbit, spin rotation and chemical composition of components}",
      journal = {\mnras},
         year = 2011,
        month = aug,
       volume = {415},
       number = {3},
        pages = {2238-2244},
          doi = {10.1111/j.1365-2966.2011.18854.x},
       adsurl = {https://ui.adsabs.harvard.edu/abs/2011MNRAS.415.2238G}
}

@ARTICLE{Martins_2017,
       author = {{Martins}, F. and {Mahy}, L. and {Herv{\'e}}, A.},
        title = "{Properties of six short-period massive binaries: A study of the effects of binarity on surface chemical abundances}",
      journal = {\aap},
         year = 2017,
        month = nov,
       volume = {607},
          eid = {A82},
        pages = {A82},
          doi = {10.1051/0004-6361/201731593},
archivePrefix = {arXiv},
       eprint = {1709.00937},
 primaryClass = {astro-ph.SR},
       adsurl = {https://ui.adsabs.harvard.edu/abs/2017A&A...607A..82M}
}

@ARTICLE{Plaskett1920,
       author = {{Plaskett}, J.~S.},
        title = "{The spectroscopic orbit and dimensions of ZET Vul.}",
      journal = {Publications of the Dominion Astrophysical Observatory Victoria},
         year = 1920,
        month = jan,
       volume = {1},
        pages = {251},
       adsurl = {https://ui.adsabs.harvard.edu/abs/1920PDAO....1..251P}
}

@ARTICLE{Broglia1964,
       author = {{Broglia}, P.},
        title = "{Observations photo{\'e}lectriques de la variable {\`a} {\'e}clipse Z Vulpeculae}",
      journal = {Journal des Observateurs},
         year = 1964,
        month = jan,
       volume = {47},
        pages = {99},
       adsurl = {https://ui.adsabs.harvard.edu/abs/1964JO.....47...99B}
}

@ARTICLE{Cester1977,
       author = {{Cester}, B. and {Fedel}, B. and {Giuricin}, G. and {Mardirossian}, F. and {Pucillo}, M.},
        title = "{Revised photometric elements of 12 semi-detached systems.}",
      journal = {\aap},
         year = 1977,
        month = nov,
       volume = {61},
        pages = {469-475},
       adsurl = {https://ui.adsabs.harvard.edu/abs/1977A&A....61..469C}
}

@ARTICLE{Pfeiffer1977,
       author = {{Pfeiffer}, R.~J. and {Koch}, R.~H.},
        title = "{On the linear polarization of close binaries.}",
      journal = {\pasp},
         year = 1977,
        month = apr,
       volume = {89},
        pages = {147-154},
          doi = {10.1086/130092},
       adsurl = {https://ui.adsabs.harvard.edu/abs/1977PASP...89..147P}
}

@ARTICLE{Ghoreyshi2008,
       author = {{Ghoreyshi}, S.~M.~R. and {Ghanbari}, J. and {Salehi}, F.},
        title = "{Reanalysis of two eclipsing binaries: EE Aqr and Z Vul}",
      journal = {\apss},
         year = 2008,
        month = apr,
       volume = {314},
       number = {4},
        pages = {331-340},
          doi = {10.1007/s10509-008-9774-y},
archivePrefix = {arXiv},
       eprint = {0805.0489},
 primaryClass = {astro-ph},
       adsurl = {https://ui.adsabs.harvard.edu/abs/2008Ap&SS.314..331G}
}

@ARTICLE{Simon1999,
       author = {{{\v{S}}imon}, V.},
        title = "{Variations of the orbital periods in semi-detached binary stars with radiative outer layers}",
      journal = {\aaps},
         year = 1999,
        month = jan,
       volume = {134},
        pages = {1-19},
          doi = {10.1051/aas:1999122},
       adsurl = {https://ui.adsabs.harvard.edu/abs/1999A&AS..134....1S}
}

@ARTICLE{Lazaro2009,
       author = {{Lazaro}, C. and {Arevalo}, M.~J. and {Almenara}, J.~M.},
        title = "{Absolute parameters of the Algol binary Z Vul}",
      journal = {\na},
         year = 2009,
        month = aug,
       volume = {14},
       number = {6},
        pages = {528-538},
          doi = {10.1016/j.newast.2009.01.010},
       adsurl = {https://ui.adsabs.harvard.edu/abs/2009NewA...14..528L}
}

@ARTICLE{Simon_1994,
       author = {{Simon}, K.~P. and {Sturm}, E.},
        title = "{Disentangling of composite spectra.}",
      journal = {\aap},
         year = 1994,
        month = jan,
       volume = {281},
        pages = {286-291},
       adsurl = {https://ui.adsabs.harvard.edu/abs/1994A&A...281..286S}
}

@ARTICLE{Hadrava_1995,
       author = {{Hadrava}, P.},
        title = "{Orbital elements of multiple spectroscopic stars.}",
      journal = {\aaps},
         year = 1995,
        month = dec,
       volume = {114},
        pages = {393},
       adsurl = {https://ui.adsabs.harvard.edu/abs/1995A&AS..114..393H}
}

@ARTICLE{Blanco-Cuaresma2019,
       author = {{Blanco-Cuaresma}, Sergi},
        title = "{Modern stellar spectroscopy caveats}",
      journal = {\mnras},
         year = 2019,
        month = jun,
       volume = {486},
       number = {2},
        pages = {2075-2101},
          doi = {10.1093/mnras/stz549},
archivePrefix = {arXiv},
       eprint = {1902.09558},
 primaryClass = {astro-ph.SR},
       adsurl = {https://ui.adsabs.harvard.edu/abs/2019MNRAS.486.2075B}
}

@ARTICLE{Foreman-Mackey2013,
       author = {{Foreman-Mackey}, Daniel and {Hogg}, David W. and {Lang}, Dustin and {Goodman}, Jonathan},
        title = "{emcee: The MCMC Hammer}",
      journal = {\pasp},
         year = 2013,
        month = mar,
       volume = {125},
       number = {925},
        pages = {306},
          doi = {10.1086/670067},
archivePrefix = {arXiv},
       eprint = {1202.3665},
 primaryClass = {astro-ph.IM},
       adsurl = {https://ui.adsabs.harvard.edu/abs/2013PASP..125..306F}
}

@INPROCEEDINGS{Ricker2014,
       author = {{Ricker}, George R. and {Vanderspek}, Roland Kraft and {Latham}, David W. and {Winn}, Joshua N.},
        title = "{The Transiting Exoplanet Survey Satellite Mission}",
    booktitle = {American Astronomical Society Meeting Abstracts \#224},
         year = 2014,
       series = {American Astronomical Society Meeting Abstracts},
       volume = {224},
        month = jun,
          eid = {113.02},
        pages = {113.02},
       adsurl = {https://ui.adsabs.harvard.edu/abs/2014AAS...22411302R}
}

@software{Lightkurve2018,
       author = {{Lightkurve Collaboration} and {Cardoso}, Jos{\'e} Vin{\'\i}cius de Miranda and {Hedges}, Christina and {Gully-Santiago}, Michael and {Saunders}, Nicholas and {Cody}, Ann Marie and {Barclay}, Thomas and {Hall}, Oliver and {Sagear}, Sheila and {Turtelboom}, Emma and {Zhang}, Johnny and {Tzanidakis}, Andy and {Mighell}, Ken and {Coughlin}, Jeff and {Bell}, Keaton and {Berta-Thompson}, Zach and {Williams}, Peter and {Dotson}, Jessie and {Barentsen}, Geert},
        title = "{Lightkurve: Kepler and TESS time series analysis in Python}",
 howpublished = {Astrophysics Source Code Library, record ascl:1812.013},
         year = 2018,
        month = dec,
          eid = {ascl:1812.013},
archivePrefix = {ascl},
       eprint = {1812.013},
       adsurl = {https://ui.adsabs.harvard.edu/abs/2018ascl.soft12013L}
}

@ARTICLE{Petit2014,
       author = {{Petit}, P. and {Louge}, T. and {Th{\'e}ado}, S. and {Paletou}, F. and {Manset}, N. and {Morin}, J. and {Marsden}, S.~C. and {Jeffers}, S.~V.},
        title = "{PolarBase: A Database of High-Resolution Spectropolarimetric Stellar Observations}",
      journal = {\pasp},
         year = 2014,
        month = may,
       volume = {126},
       number = {939},
        pages = {469},
          doi = {10.1086/676976},
archivePrefix = {arXiv},
       eprint = {1401.1082},
 primaryClass = {astro-ph.SR},
       adsurl = {https://ui.adsabs.harvard.edu/abs/2014PASP..126..469P}
}

@ARTICLE{Donati1997,
       author = {{Donati}, J.-F. and {Semel}, M. and {Carter}, B.~D. and {Rees}, D.~E. and {Collier Cameron}, A.},
        title = "{Spectropolarimetric observations of active stars}",
      journal = {\mnras},
         year = 1997,
        month = nov,
       volume = {291},
       number = {4},
        pages = {658-682},
          doi = {10.1093/mnras/291.4.658},
       adsurl = {https://ui.adsabs.harvard.edu/abs/1997MNRAS.291..658D}
}

@ARTICLE{Wilson1971,
       author = {{Wilson}, Robert E. and {Devinney}, Edward J.},
        title = "{Realization of Accurate Close-Binary Light Curves: Application to MR Cygni}",
      journal = {\apj},
         year = 1971,
        month = jun,
       volume = {166},
        pages = {605},
          doi = {10.1086/150986},
       adsurl = {https://ui.adsabs.harvard.edu/abs/1971ApJ...166..605W}
}

@ARTICLE{Wilson2014,
       author = {{Wilson}, R.~E. and {Van Hamme}, W.},
        title = "{Unification of Binary Star Ephemeris Solutions}",
      journal = {\apj},
         year = 2014,
        month = jan,
       volume = {780},
       number = {2},
          eid = {151},
        pages = {151},
          doi = {10.1088/0004-637X/780/2/151},
       adsurl = {https://ui.adsabs.harvard.edu/abs/2014ApJ...780..151W}
}

@ARTICLE{Güzel2020,
       author = {{G{\"u}zel}, O. and {{\"O}zdarcan}, O.},
        title = "{PyWD2015 - A new GUI for the Wilson-Devinney code}",
      journal = {Contributions of the Astronomical Observatory Skalnate Pleso},
         year = 2020,
        month = mar,
       volume = {50},
       number = {2},
        pages = {535-538},
          doi = {10.31577/caosp.2020.50.2.535},
       adsurl = {https://ui.adsabs.harvard.edu/abs/2020CoSka..50..535G}
}

@ARTICLE{van1993,
       author = {{van Hamme}, W.},
        title = "{The new Wilson reflection treatment and the nature of BF Aurigae.}",
      journal = {IAU Commission on Close Binary Stars},
         year = 1993,
        month = jan,
       volume = {21},
        pages = {53-68},
       adsurl = {https://ui.adsabs.harvard.edu/abs/1993IAUCB..21...53V}
}

@ARTICLE{Dervisoglu_2018,
       author = {{Dervi{\c{s}}o{\v{g}}lu}, A. and {Pavlovski}, K. and {Lehmann}, H. and {Southworth}, J. and {Bewsher}, D.},
        title = "{Evidence for conservative mass transfer in the classical Algol system {\ensuremath{\delta}} Librae from its surface carbon-to-nitrogen abundance ratio}",
      journal = {\mnras},
         year = 2018,
        month = dec,
       volume = {481},
       number = {4},
        pages = {5660-5674},
          doi = {10.1093/mnras/sty2684},
archivePrefix = {arXiv},
       eprint = {1810.01465},
 primaryClass = {astro-ph.SR},
       adsurl = {https://ui.adsabs.harvard.edu/abs/2018MNRAS.481.5660D}
}

@ARTICLE{Lehmann2020,
       author = {{Lehmann}, H. and {Dervi{\c{s}}o{\u{g}}lu}, A. and {Mkrtichian}, D.~E. and {Pertermann}, F. and {Tkachenko}, A. and {Tsymbal}, V.},
        title = "{Spectroscopic long-term monitoring of RZ Cas. I. Basic stellar and system parameters}",
      journal = {\aap},
         year = 2020,
        month = dec,
       volume = {644},
          eid = {A121},
        pages = {A121},
          doi = {10.1051/0004-6361/202039355},
archivePrefix = {arXiv},
       eprint = {2011.07903},
 primaryClass = {astro-ph.SR},
       adsurl = {https://ui.adsabs.harvard.edu/abs/2020A&A...644A.121L}
}

@ARTICLE{Liu2003,
       author = {{Liu}, Qing-Yao and {Yang}, Yu-Lan},
        title = "{A Possible Explanation of the O'Connell Effect in Close Binary Stars}",
      journal = {\cjaa},
         year = 2003,
        month = apr,
       volume = {3},
        pages = {142-150},
          doi = {10.1088/1009-9271/3/2/142},
       adsurl = {https://ui.adsabs.harvard.edu/abs/2003ChJAA...3..142L}
}

@ARTICLE{Deng2025,
       author = {{Deng}, Zhao-Long and {Liao}, Wen-Ping and {Zhu}, Li-Ying and {Shi}, Xiang-Dong and {Liu}, Nian-Ping and {Li}, Ping},
        title = "{V455 Car: An oscillating eclipsing Algol-type binary in triple star system}",
      journal = {\na},
         year = 2025,
        month = oct,
       volume = {119},
          eid = {102412},
        pages = {102412},
          doi = {10.1016/j.newast.2025.102412},
archivePrefix = {arXiv},
       eprint = {2506.10124},
 primaryClass = {astro-ph.SR},
       adsurl = {https://ui.adsabs.harvard.edu/abs/2025NewA..11902412D}
}

@ARTICLE{Pilecki2017,
       author = {{Pilecki}, Bogumi{\l} and {Gieren}, Wolfgang and {Smolec}, Rados{\l}aw and {Pietrzy{\'n}ski}, Grzegorz and {Thompson}, Ian B. and {Anderson}, Richard I. and {Bono}, Giuseppe and {Soszy{\'n}ski}, Igor and {Kervella}, Pierre and {Nardetto}, Nicolas and {Taormina}, M{\'o}nica and {St{\c{e}}pie{\'n}}, Kazimierz and {Wielg{\'o}rski}, Piotr},
        title = "{Mass and p-factor of the Type II Cepheid OGLE-LMC-T2CEP-098 in a Binary System}",
      journal = {\apj},
         year = 2017,
        month = jun,
       volume = {842},
       number = {2},
          eid = {110},
        pages = {110},
          doi = {10.3847/1538-4357/aa6ff7},
archivePrefix = {arXiv},
       eprint = {1704.07782},
 primaryClass = {astro-ph.SR},
       adsurl = {https://ui.adsabs.harvard.edu/abs/2017ApJ...842..110P}
}

@ARTICLE{Barbaros2023,
       author = {{Barbaros}, Emre and {Dervi{\c{s}}o{\u{g}}lu}, Ahmet},
        title = "{Tayfsal Ay{\i}rma Y{\"o}ntemlerinde MCMC Optimizasyonunun Kullan{\i}m{\i}}",
      journal = {Turkish Journal of Astronomy and Astrophysics},
         year = 2023,
        month = dec,
       volume = {4},
        pages = {323-327},
          doi = {10.55064/tjaa.1203660},
       adsurl = {https://ui.adsabs.harvard.edu/abs/2023TJAA....4S.323E}
}

@ARTICLE{Hubeny1988,
       author = {{Hubeny}, I.},
        title = "{A computer program for calculating non-LTE model stellar atmospheres}",
      journal = {Computer Physics Communications},
         year = 1988,
        month = dec,
       volume = {52},
       number = {1},
        pages = {103-132},
          doi = {10.1016/0010-4655(88)90177-4},
       adsurl = {https://ui.adsabs.harvard.edu/abs/1988CoPhC..52..103H}
}

@ARTICLE{Lanz2003,
       author = {{Lanz}, Thierry and {Hubeny}, Ivan},
        title = "{A Grid of Non-LTE Line-blanketed Model Atmospheres of O-Type Stars}",
      journal = {\apjs},
         year = 2003,
        month = jun,
       volume = {146},
       number = {2},
        pages = {417-441},
          doi = {10.1086/374373},
archivePrefix = {arXiv},
       eprint = {astro-ph/0210157},
 primaryClass = {astro-ph},
       adsurl = {https://ui.adsabs.harvard.edu/abs/2003ApJS..146..417L}
}

@ARTICLE{Lanz2007,
       author = {{Lanz}, Thierry and {Hubeny}, Ivan},
        title = "{A Grid of NLTE Line-blanketed Model Atmospheres of Early B-Type Stars}",
      journal = {\apjs},
         year = 2007,
        month = mar,
       volume = {169},
       number = {1},
        pages = {83-104},
          doi = {10.1086/511270},
archivePrefix = {arXiv},
       eprint = {astro-ph/0611891},
 primaryClass = {astro-ph},
       adsurl = {https://ui.adsabs.harvard.edu/abs/2007ApJS..169...83L}
}

@ARTICLE{Sahin2019,
       author = {{{\c{S}}ahin}, Timur and {Dervi{\textcommabelow s}o{\u{g}}lu}, Ahmet},
        title = "{High Resolution Optical Spectroscopy of a B-type Abundance Standard Candidate in Ori OB1{\textemdash}HD 35039}",
      journal = {Astronomy Letters},
         year = 2019,
        month = aug,
       volume = {45},
       number = {8},
        pages = {528-545},
          doi = {10.1134/S1063773719080073},
       adsurl = {https://ui.adsabs.harvard.edu/abs/2019AstL...45..528S}
}

@ARTICLE{Eggleton1971,
       author = {{Eggleton}, Peter P.},
        title = "{The evolution of low mass stars}",
      journal = {\mnras},
         year = 1971,
        month = jan,
       volume = {151},
        pages = {351},
          doi = {10.1093/mnras/151.3.351},
       adsurl = {https://ui.adsabs.harvard.edu/abs/1971MNRAS.151..351E}
}

@ARTICLE{Stancliffe2009,
       author = {{Stancliffe}, Richard J. and {Eldridge}, John J.},
        title = "{Modelling the binary progenitor of Supernova 1993J}",
      journal = {\mnras},
         year = 2009,
        month = jul,
       volume = {396},
       number = {3},
        pages = {1699-1708},
          doi = {10.1111/j.1365-2966.2009.14849.x},
archivePrefix = {arXiv},
       eprint = {0904.0282},
 primaryClass = {astro-ph.SR},
       adsurl = {https://ui.adsabs.harvard.edu/abs/2009MNRAS.396.1699S}
}

@ARTICLE{Paxton2015,
       author = {{Paxton}, Bill and {Marchant}, Pablo and {Schwab}, Josiah and {Bauer}, Evan B. and {Bildsten}, Lars and {Cantiello}, Matteo and {Dessart}, Luc and {Farmer}, R. and {Hu}, H. and {Langer}, N. and {Townsend}, R.~H.~D. and {Townsley}, Dean M. and {Timmes}, F.~X.},
        title = "{Modules for Experiments in Stellar Astrophysics (MESA): Binaries, Pulsations, and Explosions}",
      journal = {\apjs},
         year = 2015,
        month = sep,
       volume = {220},
       number = {1},
          eid = {15},
        pages = {15},
          doi = {10.1088/0067-0049/220/1/15},
archivePrefix = {arXiv},
       eprint = {1506.03146},
 primaryClass = {astro-ph.SR},
       adsurl = {https://ui.adsabs.harvard.edu/abs/2015ApJS..220...15P}
}

@INPROCEEDINGS{Mink2007,
       author = {{de Mink}, S.~E. and {Pols}, O.~R. and {Glebbeek}, E.},
        title = "{Critically-rotating Stars in Binaries-An Unsolved Problem}",
    booktitle = {Unsolved Problems in Stellar Physics: A Conference in Honor of Douglas Gough},
         year = 2007,
       editor = {{Stancliffe}, Richard J. and {Houdek}, Guenter and {Martin}, Rebecca G. and {Tout}, Christopher A.},
       series = {American Institute of Physics Conference Series},
       volume = {948},
        month = nov,
    publisher = {AIP},
        pages = {321-325},
          doi = {10.1063/1.2818989},
archivePrefix = {arXiv},
       eprint = {0709.2285},
 primaryClass = {astro-ph},
       adsurl = {https://ui.adsabs.harvard.edu/abs/2007AIPC..948..321D}
}

@ARTICLE{Van2016,
       author = {{Van Rensbergen}, W. and {De Greve}, J.~P.},
        title = "{Accretion disks in Algols: Progenitors and evolution}",
      journal = {\aap},
         year = 2016,
        month = aug,
       volume = {592},
          eid = {A151},
        pages = {A151},
          doi = {10.1051/0004-6361/201628798},
archivePrefix = {arXiv},
       eprint = {1604.07589},
 primaryClass = {astro-ph.SR},
       adsurl = {https://ui.adsabs.harvard.edu/abs/2016A&A...592A.151V}
}

@ARTICLE{Eggleton2000,
       author = {{Eggleton}, Peter P.},
        title = "{New labour on Algols: conservative or liberal?}",
      journal = {\nar},
         year = 2000,
        month = apr,
       volume = {44},
       number = {1-2},
        pages = {111-117},
          doi = {10.1016/S1387-6473(00)00023-3},
       adsurl = {https://ui.adsabs.harvard.edu/abs/2000NewAR..44..111E}
}

@ARTICLE{Deschamps2013,
       author = {{Deschamps}, R. and {Siess}, L. and {Davis}, P.~J. and {Jorissen}, A.},
        title = "{Critically-rotating accretors and non-conservative evolution in Algols}",
      journal = {\aap},
         year = 2013,
        month = sep,
       volume = {557},
          eid = {A40},
        pages = {A40},
          doi = {10.1051/0004-6361/201321509},
archivePrefix = {arXiv},
       eprint = {1306.1348},
 primaryClass = {astro-ph.SR},
       adsurl = {https://ui.adsabs.harvard.edu/abs/2013A&A...557A..40D}
}

@ARTICLE{Van2008,
       author = {{van Rensbergen}, W. and {De Greve}, J.~P. and {De Loore}, C. and {Mennekens}, N.},
        title = "{Spin-up and hot spots can drive mass out of a binary}",
      journal = {\aap},
         year = 2008,
        month = sep,
       volume = {487},
       number = {3},
        pages = {1129-1138},
          doi = {10.1051/0004-6361:200809943},
archivePrefix = {arXiv},
       eprint = {0804.1215},
 primaryClass = {astro-ph},
       adsurl = {https://ui.adsabs.harvard.edu/abs/2008A&A...487.1129V}
}

@ARTICLE{Soberman1997,
       author = {{Soberman}, G.~E. and {Phinney}, E.~S. and {van den Heuvel}, E.~P.~J.},
        title = "{Stability criteria for mass transfer in binary stellar evolution.}",
      journal = {\aap},
         year = 1997,
        month = nov,
       volume = {327},
        pages = {620-635},
          doi = {10.48550/arXiv.astro-ph/9703016},
archivePrefix = {arXiv},
       eprint = {astro-ph/9703016},
 primaryClass = {astro-ph},
       adsurl = {https://ui.adsabs.harvard.edu/abs/1997A&A...327..620S}
}

@ARTICLE{Hurley2002,
       author = {{Hurley}, Jarrod R. and {Tout}, Christopher A. and {Pols}, Onno R.},
        title = "{Evolution of binary stars and the effect of tides on binary populations}",
      journal = {\mnras},
         year = 2002,
        month = feb,
       volume = {329},
       number = {4},
        pages = {897-928},
          doi = {10.1046/j.1365-8711.2002.05038.x},
archivePrefix = {arXiv},
       eprint = {astro-ph/0201220},
 primaryClass = {astro-ph},
       adsurl = {https://ui.adsabs.harvard.edu/abs/2002MNRAS.329..897H}
}

@INPROCEEDINGS{Heuvel1994,
       author = {{van den Heuvel}, E.~P.~J.},
        title = "{Interacting binaries: topics in close binary evolution.}",
    booktitle = {Saas-Fee Advanced Course 22: Interacting Binaries},
         year = 1994,
       editor = {{Shore}, S.~N. and {Livio}, M. and {van den Heuvel}, Edward P.~J. and {Nussbaumer}, H. and {Orr}, Astrid},
        month = jan,
        pages = {263-474},
       adsurl = {https://ui.adsabs.harvard.edu/abs/1994inbi.conf..263V}
}

@ARTICLE{Bhattacharya1991,
       author = {{Bhattacharya}, D. and {van den Heuvel}, E.~P.~J.},
        title = "{Formation and evolution of binary and millisecond radio pulsars}",
      journal = {\physrep},
         year = 1991,
        month = jan,
       volume = {203},
       number = {1-2},
        pages = {1-124},
          doi = {10.1016/0370-1573(91)90064-S},
       adsurl = {https://ui.adsabs.harvard.edu/abs/1991PhR...203....1B}
}

@ARTICLE{Paczynski1966,
       author = {{Paczy{\'n}ski}, B.},
        title = "{Evolution of Close Binaries. I.}",
      journal = {\actaa},
         year = 1966,
        month = jan,
       volume = {16},
        pages = {231},
       adsurl = {https://ui.adsabs.harvard.edu/abs/1966AcA....16..231P}
}

@ARTICLE{Torres2010,
       author = {{Torres}, G. and {Andersen}, J. and {Gim{\'e}nez}, A.},
        title = "{Accurate masses and radii of normal stars: modern results and applications}",
      journal = {\aapr},
         year = 2010,
        month = feb,
       volume = {18},
       number = {1-2},
        pages = {67-126},
          doi = {10.1007/s00159-009-0025-1},
archivePrefix = {arXiv},
       eprint = {0908.2624},
 primaryClass = {astro-ph.SR},
       adsurl = {https://ui.adsabs.harvard.edu/abs/2010A&ARv..18...67T}
}

@ARTICLE{Nelson2001,
       author = {{Nelson}, C.~A. and {Eggleton}, P.~P.},
        title = "{A Complete Survey of Case A Binary Evolution with Comparison to Observed Algol-type Systems}",
      journal = {\apj},
         year = 2001,
        month = may,
       volume = {552},
       number = {2},
        pages = {664-678},
          doi = {10.1086/320560},
archivePrefix = {arXiv},
       eprint = {astro-ph/0009258},
 primaryClass = {astro-ph},
       adsurl = {https://ui.adsabs.harvard.edu/abs/2001ApJ...552..664N}
}

@ARTICLE{Kippenhahn1980,
       author = {{Kippenhahn}, R. and {Ruschenplatt}, G. and {Thomas}, H.-C.},
        title = "{The time scale of thermohaline mixing in stars}",
      journal = {\aap},
         year = 1980,
        month = nov,
       volume = {91},
       number = {1-2},
        pages = {175-180},
       adsurl = {https://ui.adsabs.harvard.edu/abs/1980A&A....91..175K}
}

@ARTICLE{Ulrich1972,
       author = {{Ulrich}, Roger K.},
        title = "{Thermohaline Convection in Stellar Interiors.}",
      journal = {\apj},
         year = 1972,
        month = feb,
       volume = {172},
        pages = {165},
          doi = {10.1086/151336},
       adsurl = {https://ui.adsabs.harvard.edu/abs/1972ApJ...172..165U}
}

@ARTICLE{csahin2018a,
       author = {{\c{S}}ahin, T.},
        title = "{High Resolution Optical Spectroscopy of an Intriguing High-Latitude B-Type Star HD119608}",
      journal = {Astrophysical Bulletin},
         year = 2018,
        month = jan,
       volume = {73},
       number = {1},
        pages = {35-51},
          doi = {10.1134/S1990341318010030},
       adsurl = {https://ui.adsabs.harvard.edu/abs/2018AstBu..73...35S}
}

@ARTICLE{csahin2018b,
       author = {{\c{S}}ahin, T.},
        title = "{High Resolution Optical Spectroscopy of Hot Post-AGB Star Candidates LS IV-04 1 and LB3116}",
      journal = {Astrophysical Bulletin},
         year = 2018,
        month = apr,
       volume = {73},
       number = {2},
        pages = {211-224},
          doi = {10.1134/S1990341318020074},
       adsurl = {https://ui.adsabs.harvard.edu/abs/2018AstBu..73..211S}
}

@article{hubeny1995non,
  title={Non-LTE line-blanketed model atmospheres of hot stars. 1: Hybrid complete linearization/accelerated lambda iteration method},
  author={Hubeny, Ivan and Lanz, T},
  journal={The Astrophysical Journal, Part 1 (ISSN 0004-637X), vol. 439, no. 2, p. 875-904},
  volume={439},
  pages={875--904},
  year={1995}
}

@article{nieva2012present,
  title={Present-day cosmic abundances-A comprehensive study of nearby early B-type stars and implications for stellar and Galactic evolution and interstellar dust models},
  author={Nieva, M-F and Przybilla, Norbert},
  journal={Astronomy \& Astrophysics},
  volume={539},
  pages={A143},
  year={2012},
  publisher={EDP Sciences}
}

@article{lanz2007grid,
  title={A grid of NLTE line-blanketed model atmospheres of early B-type stars},
  author={Lanz, Thierry and Hubeny, Ivan},
  journal={The Astrophysical Journal Supplement Series},
  volume={169},
  number={1},
  pages={83},
  year={2007},
  publisher={IOP Publishing}
}

@inproceedings{Ilijic2004,
  author    = {Iliji{\'c}, S. and Hensberge, H. and Pavlovski, K. and Freyhammer, L. S.},
  title     = {FDBinary: A New Tool for Spectral Disentangling},
  booktitle = {Spectroscopically and Spatially Resolving the Components of Close Binary Stars},
  year      = {2004},
  editor    = {Hilditch, R. W. and Hensberge, H. and Pavlovski, K.},
  series    = {ASP Conference Series},
  volume    = {318},
  publisher = {Astronomical Society of the Pacific},
  address   = {San Francisco},
  pages     = {111},
  adsid     = {2004ASPC..318..111I}
}

@article{asplund2009chemical,
  title={The chemical composition of the Sun},
  author={Asplund, Martin and Grevesse, Nicolas and Sauval, A Jacques and Scott, Pat},
  journal={Annual review of astronomy and astrophysics},
  volume={47},
  number={2009},
  pages={481--522},
  year={2009},
  publisher={Annual Reviews}
}

@ARTICLE{sahin2025vizier,
       author = {{G{\"u}ney}, F. and {{\c{S}}ahin}, T. and {Dervisoglu}, A.},
        title = "{VizieR Online Data Catalog: Benchmarking B-type line list (Guney+, 2026)}",
      journal = {VizieR Online Data Catalog (other)},
         year = 2026,
        month = jan,
       volume = {0840},
          eid = {J/other/PHYS/101},
        pages = {J/other/PHYS/101},
       adsurl = {https://ui.adsabs.harvard.edu/abs/2026yCatp084010101G}
}

@article{alexeeva2020neon,
  title={Neon Abundances of B Stars in the Solar Neighborhood},
  author={Alexeeva, Sofya and Chen, Tianxiang and Ryabchikova, Tatyana and Shi, Weibin and Sadakane, Kozo and Nishimura, Masayoshi and Zhao, Gang},
  journal={The Astrophysical Journal},
  volume={896},
  number={1},
  pages={59},
  year={2020},
  publisher={IOP Publishing}
}

@article{vrancken1996non,
  title={Non-LTE line formation for SII and SIII. I. Model atoms and first results.},
  author={Vrancken, M and Butler, K and Becker, SR},
  journal={Astronomy and Astrophysics, v. 311, p. 661-668},
  volume={311},
  pages={661--668},
  year={1996}
}

@article{mashonkina2020non,
  title={Non-local thermodynamic equilibrium line formation for Si i--ii--iii in A--B stars and the origin of Si ii emission lines in $\iota$ Her},
  author={Mashonkina, Lyudmila},
  journal={Monthly Notices of the Royal Astronomical Society},
  volume={493},
  number={4},
  pages={6095--6108},
  year={2020},
  publisher={Oxford University Press}
}

@article{auer1973analyses,
  title={Analyses of Light Ion Spectra in Stellar Atmospheres. IV. H II in the B Stars},
  author={Auer, Lawrence H and Mihalas, Dimitri},
  journal={Astrophysical Journal Supplement, vol. 25, p. 433 (1973)},
  volume={25},
  pages={433},
  year={1973}
}

@article{vidal1973hydrogen,
  title={Hydrogen Stark-broadening tables},
  author={Vidal, CR and Cooper, J and Smith, EW},
  journal={Astrophysical Journal Supplement, vol. 25, p. 37 (1973)},
  volume={25},
  pages={37},
  year={1973}
}

@article{barnard1974broadening,
  title={The Broadening of He I lines including ion dynamic corrections, with application to $\lambda$4471{\AA}},
  author={Barnard, AJ and Cooper, J and Smith, EW},
  journal={Journal of Quantitative Spectroscopy and Radiative Transfer},
  volume={14},
  number={10},
  pages={1025--1077},
  year={1974},
  publisher={Elsevier}
}

@article{dimitrijevic1984stark,
  title={Stark broadening of neutral helium lines},
  author={Dimitrijevic, MS and Sahal-Br{\'e}chot, S},
  journal={Journal of Quantitative Spectroscopy and Radiative Transfer},
  volume={31},
  number={4},
  pages={301--313},
  year={1984},
  publisher={Elsevier}
}

@article{shamey1969stark,
  title={Stark Broadening of Important Helium i Lines and Their Forbidden Components.},
  author={Shamey, Louis Joseph},
  journal={Ph. D. Thesis},
  year={1969}
}

@article{hubeny1994nlte,
  title={NLTE model stellar atmospheres with line blanketing near the series limits},
  author={Hubeny, I and Hummer, DG and Lanz, T},
  journal={Astronomy and Astrophysics (ISSN 0004-6361), vol. 282, no. 1, p. 151-167},
  volume={282},
  pages={151--167},
  year={1994}
}

@ARTICLE{Guney2026,
       author = {{G{\"u}ney}, Ferhat and {{\c{S}}ahin}, Timur and {Dervi{\textcommabelow s}o{\u{g}}lu}, Ahmet},
        title = "{Benchmarking a new atomic line list for high-resolution spectroscopy of B-type stars: chemical characterization of single and binary stellar systems}",
      journal = {\physscr},
         year = 2026,
        month = jan,
       volume = {101},
       number = {4},
          eid = {045004},
        pages = {045004},
          doi = {10.1088/1402-4896/ae364e},
       adsurl = {https://ui.adsabs.harvard.edu/abs/2026PhyS..101d5004G}
}

@ARTICLE{Peters_1998,
       author = {{Peters}, Geraldine J. and {Polidan}, Ronald S.},
        title = "{ORFEUS-SPAS II Observations of Algol-type Interacting Binaries}",
      journal = {\apjl},
         year = 1998,
        month = jun,
       volume = {500},
       number = {1},
        pages = {L17-L20},
          doi = {10.1086/311398},
       adsurl = {https://ui.adsabs.harvard.edu/abs/1998ApJ...500L..17P}
}

@ARTICLE{Popper_1957a,
       author = {{Popper}, Daniel M.},
        title = "{Rediscussion of Eclipsing Binaries. III. Z Vulpeculae.}",
      journal = {\apj},
         year = 1957,
        month = jul,
       volume = {126},
        pages = {53},
          doi = {10.1086/146371},
       adsurl = {https://ui.adsabs.harvard.edu/abs/1957ApJ...126...53P}
}

@ARTICLE{Popper_1957b,
       author = {{Popper}, Daniel M.},
        title = "{Photoelectric Observations of Eclipsing Binaries.}",
      journal = {\apjs},
         year = 1957,
        month = oct,
       volume = {3},
        pages = {107},
          doi = {10.1086/190034},
       adsurl = {https://ui.adsabs.harvard.edu/abs/1957ApJS....3..107P}
}

@ARTICLE{Hilditch_1975,
       author = {{Hilditch}, R.~W. and {Hill}, Graham},
        title = "{Str{\"o}mgren four-colour observations of Northern Hemisphere binary systems}",
      journal = {\memras},
         year = 1975,
        month = jan,
       volume = {79},
        pages = {101},
       adsurl = {https://ui.adsabs.harvard.edu/abs/1975MmRAS..79..101H}
}

@ARTICLE{Wood_1971,
       author = {{Wood}, D.~B.},
        title = "{An analytic model of eclipsing binary star systems.}",
      journal = {\aj},
         year = 1971,
        month = oct,
       volume = {76},
        pages = {701-710},
          doi = {10.1086/111187},
       adsurl = {https://ui.adsabs.harvard.edu/abs/1971AJ.....76..701W}
}

@ARTICLE{Wood_1973,
       author = {{Wood}, David B.},
        title = "{A Computer Program for Modeling Nonspherical Eclipsing Binary Star Systems}",
      journal = {\pasp},
         year = 1973,
        month = apr,
       volume = {85},
       number = {504},
        pages = {253},
          doi = {10.1086/129447},
       adsurl = {https://ui.adsabs.harvard.edu/abs/1973PASP...85..253W}
}

@ARTICLE{Lazaro_2002,
       author = {{L{\'a}zaro}, C. and {Ar{\'e}valo}, M.~J. and {Mart{\'\i}nez-Pais}, I.~G. and {Dom{\'\i}nguez}, R.~M.},
        title = "{BVRJK Photometry and a Spectroscopic Study of the Algol Short-Period Binary VV Ursae Majoris}",
      journal = {\aj},
         year = 2002,
        month = may,
       volume = {123},
       number = {5},
        pages = {2733-2743},
          doi = {10.1086/339835},
       adsurl = {https://ui.adsabs.harvard.edu/abs/2002AJ....123.2733L}
}

@ARTICLE{Shakhovskoi_1965,
       author = {{Shakhovskoi}, N.~M.},
        title = "{Polarization in Variable Stars. II. Eclipsing Binaries}",
      journal = {\sovast},
         year = 1965,
        month = jun,
       volume = {8},
        pages = {833},
       adsurl = {https://ui.adsabs.harvard.edu/abs/1965SvA.....8..833S}
}

@ARTICLE{Woodsworth_1977,
       author = {{Woodsworth}, A.~W. and {Hughes}, V.~A.},
        title = "{Observations of radio stars at 10.6 GHz.}",
      journal = {\aap},
         year = 1977,
        month = jun,
       volume = {58},
        pages = {105-111},
       adsurl = {https://ui.adsabs.harvard.edu/abs/1977A&A....58..105W}
}

@ARTICLE{Friedemann_1996,
       author = {{Friedemann}, C. and {Guertler}, J. and {Loewe}, M.},
        title = "{Eclipsing binaries as IRAS sources.}",
      journal = {\aaps},
         year = 1996,
        month = jun,
       volume = {117},
        pages = {205-225},
       adsurl = {https://ui.adsabs.harvard.edu/abs/1996A&AS..117..205F}
}

@ARTICLE{Raskin_2011,
       author = {{Raskin}, G. and {van Winckel}, H. and {Hensberge}, H. and {Jorissen}, A. and {Lehmann}, H. and {Waelkens}, C. and {Avila}, G. and {de Cuyper}, J.-P. and {Degroote}, P. and {Dubosson}, R. and {Dumortier}, L. and {Fr{\'e}mat}, Y. and {Laux}, U. and {Michaud}, B. and {Morren}, J. and {Perez Padilla}, J. and {Pessemier}, W. and {Prins}, S. and {Smolders}, K. and {van Eck}, S. and {Winkler}, J.},
        title = "{HERMES: a high-resolution fibre-fed spectrograph for the Mercator telescope}",
      journal = {\aap},
         year = 2011,
        month = feb,
       volume = {526},
          eid = {A69},
        pages = {A69},
          doi = {10.1051/0004-6361/201015435},
archivePrefix = {arXiv},
       eprint = {1011.0258},
 primaryClass = {astro-ph.IM},
       adsurl = {https://ui.adsabs.harvard.edu/abs/2011A&A...526A..69R}
}

@ARTICLE{Mason_2001,
       author = {{Mason}, Brian D. and {Wycoff}, Gary L. and {Hartkopf}, William I. and {Douglass}, Geoffrey G. and {Worley}, Charles E.},
        title = "{The 2001 US Naval Observatory Double Star CD-ROM. I. The Washington Double Star Catalog}",
      journal = {\aj},
         year = 2001,
        month = dec,
       volume = {122},
       number = {6},
        pages = {3466-3471},
          doi = {10.1086/323920},
       adsurl = {https://ui.adsabs.harvard.edu/abs/2001AJ....122.3466M}
}

@ARTICLE{Pilecki_2018,
       author = {{Pilecki}, Bogumi{\l} and {Dervi{\c{s}}o{\u{g}}lu}, Ahmet and {Gieren}, Wolfgang and {Smolec}, Rados{\l}aw and {Soszy{\'n}ski}, Igor and {Pietrzy{\'n}ski}, Grzegorz and {Thompson}, Ian B. and {Taormina}, M{\'o}nica},
        title = "{The Dynamical Mass and Evolutionary Status of the Type II Cepheid in the Eclipsing Binary System OGLE-LMC-T2CEP-211 with a Double-ring Disk}",
      journal = {\apj},
         year = 2018,
        month = nov,
       volume = {868},
       number = {1},
          eid = {30},
        pages = {30},
          doi = {10.3847/1538-4357/aae68f},
archivePrefix = {arXiv},
       eprint = {1810.06524},
 primaryClass = {astro-ph.SR},
       adsurl = {https://ui.adsabs.harvard.edu/abs/2018ApJ...868...30P}
}

@ARTICLE{Tonry_1979,
       author = {{Tonry}, J. and {Davis}, M.},
        title = "{A survey of galaxy redshifts. I. Data reduction techniques.}",
      journal = {\aj},
         year = 1979,
        month = oct,
       volume = {84},
        pages = {1511-1525},
          doi = {10.1086/112569},
       adsurl = {https://ui.adsabs.harvard.edu/abs/1979AJ.....84.1511T}
}

@ARTICLE{Simkin_1974,
       author = {{Simkin}, S.~M.},
        title = "{Measurements of Velocity Dispersions and Doppler Shifts from Digitized Optical Spectra}",
      journal = {\aap},
         year = 1974,
        month = mar,
       volume = {31},
        pages = {129},
       adsurl = {https://ui.adsabs.harvard.edu/abs/1974A&A....31..129S}
}

@ARTICLE{Tkachenko_2015,
       author = {{Tkachenko}, A.},
        title = "{Grid search in stellar parameters: a software for spectrum analysis of single stars and binary systems}",
      journal = {\aap},
         year = 2015,
        month = sep,
       volume = {581},
          eid = {A129},
        pages = {A129},
          doi = {10.1051/0004-6361/201526513},
archivePrefix = {arXiv},
       eprint = {1507.02864},
 primaryClass = {astro-ph.SR},
       adsurl = {https://ui.adsabs.harvard.edu/abs/2015A&A...581A.129T}
}

@ARTICLE{Tamajo_2011,
       author = {{Tamajo}, E. and {Pavlovski}, K. and {Southworth}, J.},
        title = "{Constrained fitting of disentangled binary star spectra: application to V615 Persei in the open cluster h Persei}",
      journal = {\aap},
         year = 2011,
        month = feb,
       volume = {526},
          eid = {A76},
        pages = {A76},
          doi = {10.1051/0004-6361/201015913},
archivePrefix = {arXiv},
       eprint = {1012.2244},
 primaryClass = {astro-ph.SR},
       adsurl = {https://ui.adsabs.harvard.edu/abs/2011A&A...526A..76T}
}

@ARTICLE{Pavlovski_2023,
       author = {{Pavlovski}, K. and {Southworth}, J. and {Tkachenko}, A. and {Van Reeth}, T. and {Tamajo}, E.},
        title = "{High-mass eclipsing binaries: A testbed for models of interior structure and evolution. Accurate fundamental properties and surface chemical composition for V1034 Sco, GL Car, V573 Car, and V346 Cen}",
      journal = {\aap},
         year = 2023,
        month = mar,
       volume = {671},
          eid = {A139},
        pages = {A139},
          doi = {10.1051/0004-6361/202244980},
archivePrefix = {arXiv},
       eprint = {2301.04215},
 primaryClass = {astro-ph.SR},
       adsurl = {https://ui.adsabs.harvard.edu/abs/2023A&A...671A.139P}
}

@ARTICLE{Pavlovski_2018,
       author = {{Pavlovski}, K. and {Southworth}, J. and {Tamajo}, E.},
        title = "{Physical properties and CNO abundances for high-mass stars in four main-sequence detached eclipsing binaries: V478 Cyg, AH Cep, V453 Cyg, and V578 Mon}",
      journal = {\mnras},
         year = 2018,
        month = dec,
       volume = {481},
       number = {3},
        pages = {3129-3147},
          doi = {10.1093/mnras/sty2516},
archivePrefix = {arXiv},
       eprint = {1809.04061},
 primaryClass = {astro-ph.SR},
       adsurl = {https://ui.adsabs.harvard.edu/abs/2018MNRAS.481.3129P}
}

@ARTICLE{Torres_2025,
       author = {{Torres}, Guillermo and {Tkachenko}, Andrew and {Pavlovski}, Kre{\v{s}}imir and {Gossage}, Seth and {Schaefer}, Gail H. and {Melis}, Carl and {Ireland}, Michael and {Monnier}, John D. and {Anugu}, Narsireddy and {Kraus}, Stefan and {Lanthermann}, Cyprien and {Gordon}, Kathryn and {Klement}, Robert and {Murphy}, Simon J. and {Roettenbacher}, Rachael M.},
        title = "{Orbital and Physical Properties of the Pleiades Binary 27 Tau (Atlas)}",
      journal = {\apj},
         year = 2025,
        month = sep,
       volume = {990},
       number = {2},
          eid = {107},
        pages = {107},
          doi = {10.3847/1538-4357/adf224},
archivePrefix = {arXiv},
       eprint = {2507.15933},
 primaryClass = {astro-ph.SR},
       adsurl = {https://ui.adsabs.harvard.edu/abs/2025ApJ...990..107T}
}

@ARTICLE{Torres_2015,
       author = {{Torres}, Guillermo and {Claret}, Antonio and {Pavlovski}, Kre{\v{s}}imir and {Dotter}, Aaron},
        title = "{Capella ({\ensuremath{\alpha}} Aurigae) Revisited: New Binary Orbit, Physical Properties, and Evolutionary State}",
      journal = {\apj},
         year = 2015,
        month = jul,
       volume = {807},
       number = {1},
          eid = {26},
        pages = {26},
          doi = {10.1088/0004-637X/807/1/26},
archivePrefix = {arXiv},
       eprint = {1505.07461},
 primaryClass = {astro-ph.SR},
       adsurl = {https://ui.adsabs.harvard.edu/abs/2015ApJ...807...26T}
}

@ARTICLE{Lucy_1967,
       author = {{Lucy}, L.~B.},
        title = "{Gravity-Darkening for Stars with Convective Envelopes}",
      journal = {\zap},
         year = 1967,
        month = jan,
       volume = {65},
        pages = {89},
       adsurl = {https://ui.adsabs.harvard.edu/abs/1967ZA.....65...89L}
}

@ARTICLE{Rucinski_1969,
       author = {{Ruci{\'n}ski}, S.~M.},
        title = "{The Proximity Effects in Close Binary Systems. II. The Bolometric Reflection Effect for Stars with Deep Convective Envelopes}",
      journal = {\actaa},
         year = 1969,
        month = jan,
       volume = {19},
        pages = {245},
       adsurl = {https://ui.adsabs.harvard.edu/abs/1969AcA....19..245R}
}

@INPROCEEDINGS{Jenkins_2016,
       author = {{Jenkins}, Jon M. and {Twicken}, Joseph D. and {McCauliff}, Sean and {Campbell}, Jennifer and {Sanderfer}, Dwight and {Lung}, David and {Mansouri-Samani}, Masoud and {Girouard}, Forrest and {Tenenbaum}, Peter and {Klaus}, Todd and {Smith}, Jeffrey C. and {Caldwell}, Douglas A. and {Chacon}, A.~D. and {Henze}, Christopher and {Heiges}, Cory and {Latham}, David W. and {Morgan}, Edward and {Swade}, Daryl and {Rinehart}, Stephen and {Vanderspek}, Roland},
        title = "{The TESS science processing operations center}",
    booktitle = {Software and Cyberinfrastructure for Astronomy IV},
         year = 2016,
       editor = {{Chiozzi}, Gianluca and {Guzman}, Juan C.},
       series = {Society of Photo-Optical Instrumentation Engineers (SPIE) Conference Series},
       volume = {9913},
        month = aug,
          eid = {99133E},
        pages = {99133E},
          doi = {10.1117/12.2233418},
       adsurl = {https://ui.adsabs.harvard.edu/abs/2016SPIE.9913E..3EJ}
}

@INPROCEEDINGS{Pavlovski_2010,
       author = {{Pavlovski}, K. and {Hensberge}, H.},
        title = "{Reconstruction and Analysis of Component Spectra of Binary and Multiple Stars}",
    booktitle = {Binaries - Key to Comprehension of the Universe},
         year = 2010,
       editor = {{Pr{\v{s}}a}, A. and {Zejda}, M.},
       series = {Astronomical Society of the Pacific Conference Series},
       volume = {435},
        month = dec,
        pages = {207},
          doi = {10.48550/arXiv.0909.3246},
archivePrefix = {arXiv},
       eprint = {0909.3246},
 primaryClass = {astro-ph.SR},
       adsurl = {https://ui.adsabs.harvard.edu/abs/2010ASPC..435..207P}
}


 
\appendix
\onecolumn

\section{Radial velocities}
\begin{center}
\captionof{table}{Spectral observation logs and the radial velocity measurements for Z Vul.}
\label{tab:speclog}
\begin{tabular}{ccrrcc}
\hline
\hline
        JD(Hel.)	  &Phase     &RV$_1$	         &RV$_2$            &$SNR$  &$Instrument$\\
        (2400000+)    &          &(km/s)             &(km/s)            &                            \\
        \hline
	    55345.50299   &0.23340	 &$-106.5 \pm 4.8$   &$ 200.3 \pm 9.7$	&100    &TRGöz	 \\
        56916.94212   &0.35623   &$ -75.9 \pm 3.1$	 &$ 130.1 \pm 7.5$	&1590*  &ESPaDOnS\\
        59052.58462	  &0.30645	 &$-109.4 \pm 2.0$	 &$ 197.4 \pm 6.6$	&232    &{\sc hermes}	 \\
	    59052.60037	  &0.31286	 &$-106.9 \pm 2.4$	 &$ 194.4 \pm 7.5$  &236    &{\sc hermes}	 \\
	    59052.61593	  &0.31920	 &$-104.8 \pm 2.3$	 &$ 190.1 \pm 7.2$  &231    &{\sc hermes}	 \\
	    59053.59484	  &0.71796	 &$  68.6 \pm 2.5$	 &$-240.9 \pm 7.9$  &233    &{\sc hermes}  \\
	    59053.61335	  &0.72550	 &$  68.7 \pm 2.6$   &$-244.9 \pm 9.0$  &225    &{\sc hermes}	 \\
	    59053.62893	  &0.73184	 &$  69.1 \pm 2.6$	 &$-246.4 \pm 9.3$	&227    &{\sc hermes}	 \\
	    59053.64453	  &0.73820	 &$  67.7 \pm 2.6$	 &$-244.2 \pm 9.3$  &227    &{\sc hermes}	 \\
	    59054.64184	  &0.14445	 &$ -93.3 \pm 2.1$	 &$ 152.7 \pm 6.7$  &227    &{\sc hermes}	 \\
	    59054.65729	  &0.15074	 &$ -94.4 \pm 2.2$	 &$ 157.2 \pm 7.1$  &224    &{\sc hermes}	 \\
	    59055.68391	  &0.56893	 &$  21.4 \pm 2.9$	 &$-125.9 \pm 9.5$  &197    &{\sc hermes}	 \\
	    59056.55081	  &0.92206	 &$  38.3 \pm 2.3$	 &$-111.1 \pm 7.7$  &219    &{\sc hermes}	 \\
	    59057.53297	  &0.32214	 &$-100.9 \pm 2.6$   &$ 204.8 \pm 9.8$	&192    &{\sc hermes}	 \\
	    59058.52958	  &0.72811	 &$  67.7 \pm 2.3$	 &$-249.8 \pm 7.9$	&224    &{\sc hermes}	 \\
        59105.48505	  &0.85531	 &$  56.1 \pm 2.5$	 &$-191.3 \pm 6.5$	&196    &{\sc hermes}	 \\
	    59120.38898   &0.92639	 &$  41.8 \pm 2.1$	 &$-107.4 \pm 6.8$  &242    &{\sc hermes}	 \\
	    59335.62280	  &0.60138	 &$  34.8 \pm 5.7$	 &$-147.2 \pm 21.1$ &199    &{\sc hermes}	 \\
	    59339.68070   &0.25436	 &$-109.3 \pm 2.5$   &$ 207.6 \pm 7.6$  &196    &{\sc hermes}	 \\
	    59340.71803	  &0.67691	 &$  63.4 \pm 3.9$	 &$-217.3 \pm 12.4$	&144    &{\sc hermes}	 \\
	    59390.60983   &0.00022	 &$ -24.5 \pm 3.9$	 &$ -26.6 \pm 6.1$  &231    &{\sc hermes}	 \\
	    59391.54538	  &0.38131	 &$ -78.7 \pm 2.3$	 &$ 132.2 \pm 8.5$	&223    &{\sc hermes}	 \\
	    59395.71128   &0.07828	 &$ -70.1 \pm 2.2$	 &$  66.3 \pm 7.9$  &182    &{\sc hermes}	 \\
	   	59438.47073   &0.49624	 &$ -19.4 \pm 2.8$	 &$ -11.8 \pm 10.8$	&131    &{\sc hermes}	 \\
	    59709.68796   &0.97594	 &$  42.9 \pm 5.2$	 &$ -62.8 \pm 7.9$  &209    &{\sc hermes}	 \\
	    59715.70501	  &0.42697	 &$ -58.7 \pm 2.4$	 &$  56.4 \pm 9.9$	&231    &{\sc hermes}	 \\
	    59716.59145   &0.78806	 &$  69.1 \pm 2.5$	 &$-239.3 \pm 8.6$  &230    &{\sc hermes}	 \\
	    59744.63506   &0.21156	 &$-107.9 \pm 2.3$	 &$ 201.5 \pm 6.7$	&231    &{\sc hermes}	 \\
        59745.70953   &0.64925	 &$  55.2 \pm 2.4$   &$-199.7 \pm 7.6$	&239    &{\sc hermes}	 \\
        59762.43932   &0.46409	 &$ -40.6 \pm 1.9$	 &$  20.0 \pm 7.1$	&162    &{\sc hermes}	 \\
        60032.67295   &0.54312	 &$  6.8  \pm 2.9$	 &$ -86.1 \pm 12.5$	&218    &{\sc hermes}	 \\
\hline
\hline
*Combined
\end{tabular}
\end{center}

\clearpage
\section{Photometry}

\begin{center}
\includegraphics[width=0.4\columnwidth]{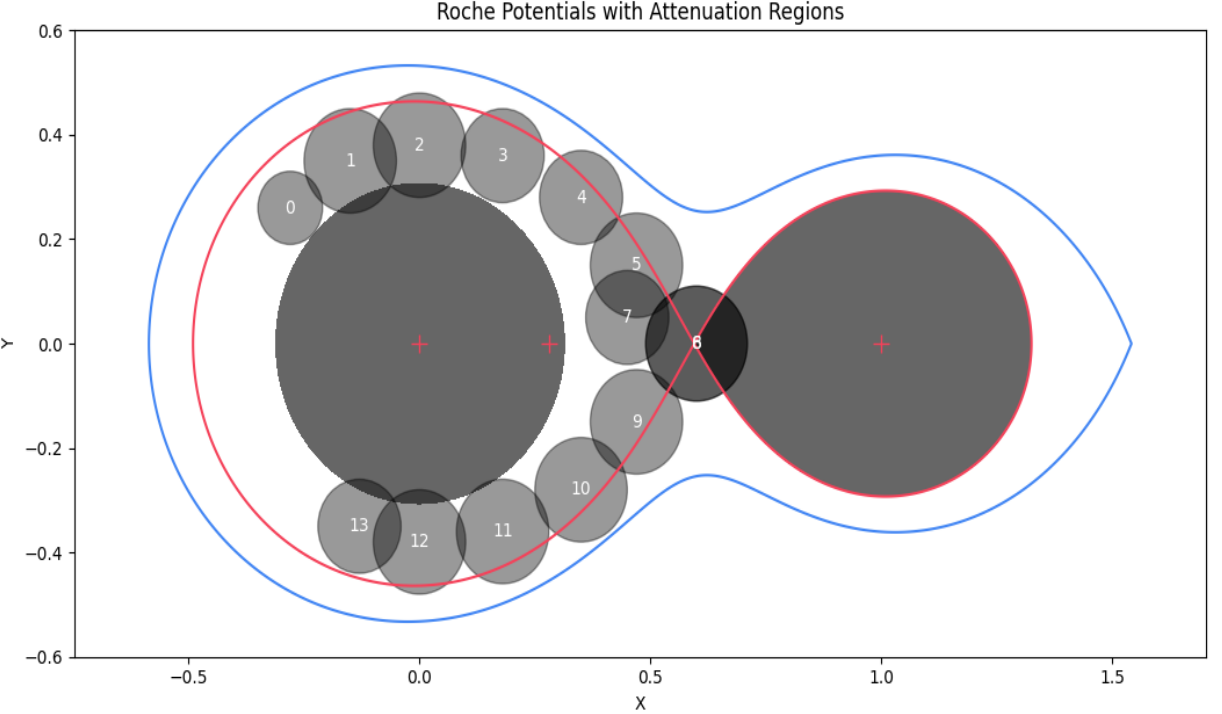}
\captionof{figure}{A schematic drawing illustrating the distribution of the added attenuation regions on the orbital plane.}
\label{fig:attenuation}
\end{center}
 
\begin{center}
\begin{tabular}{cc}
\includegraphics[width=0.45\textwidth]{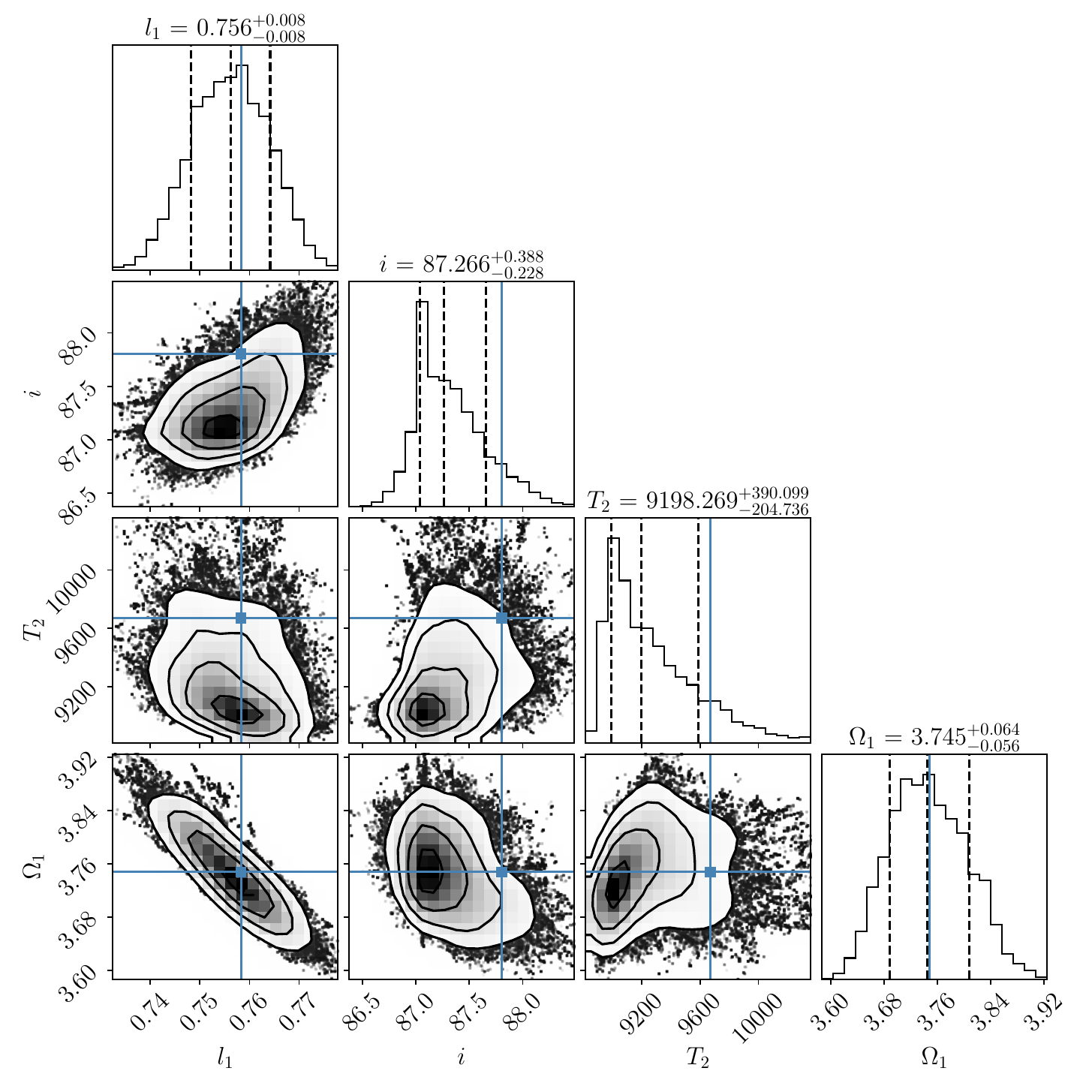} &
\includegraphics[width=0.45\textwidth]{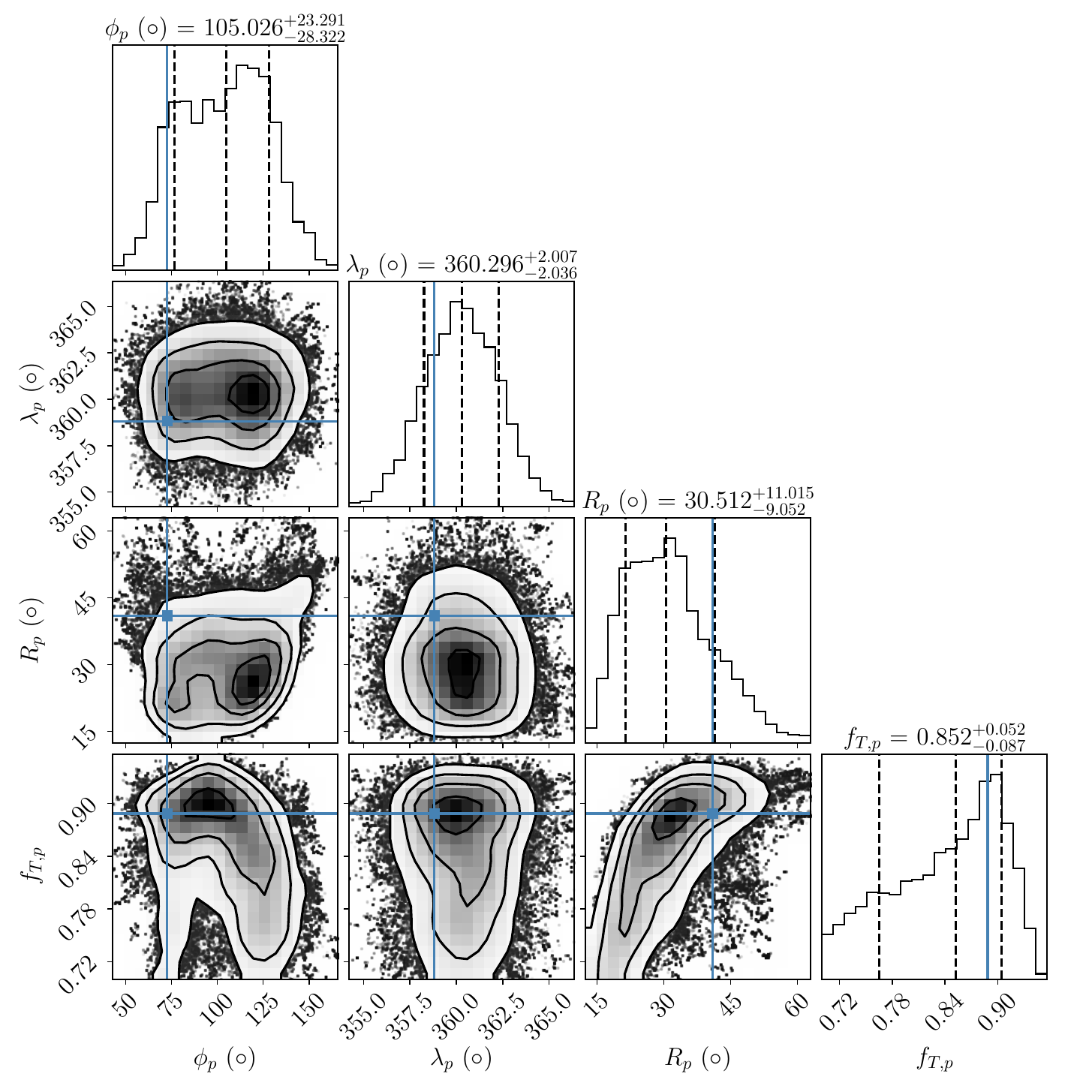} \\[0.5em]
\includegraphics[width=0.45\textwidth]{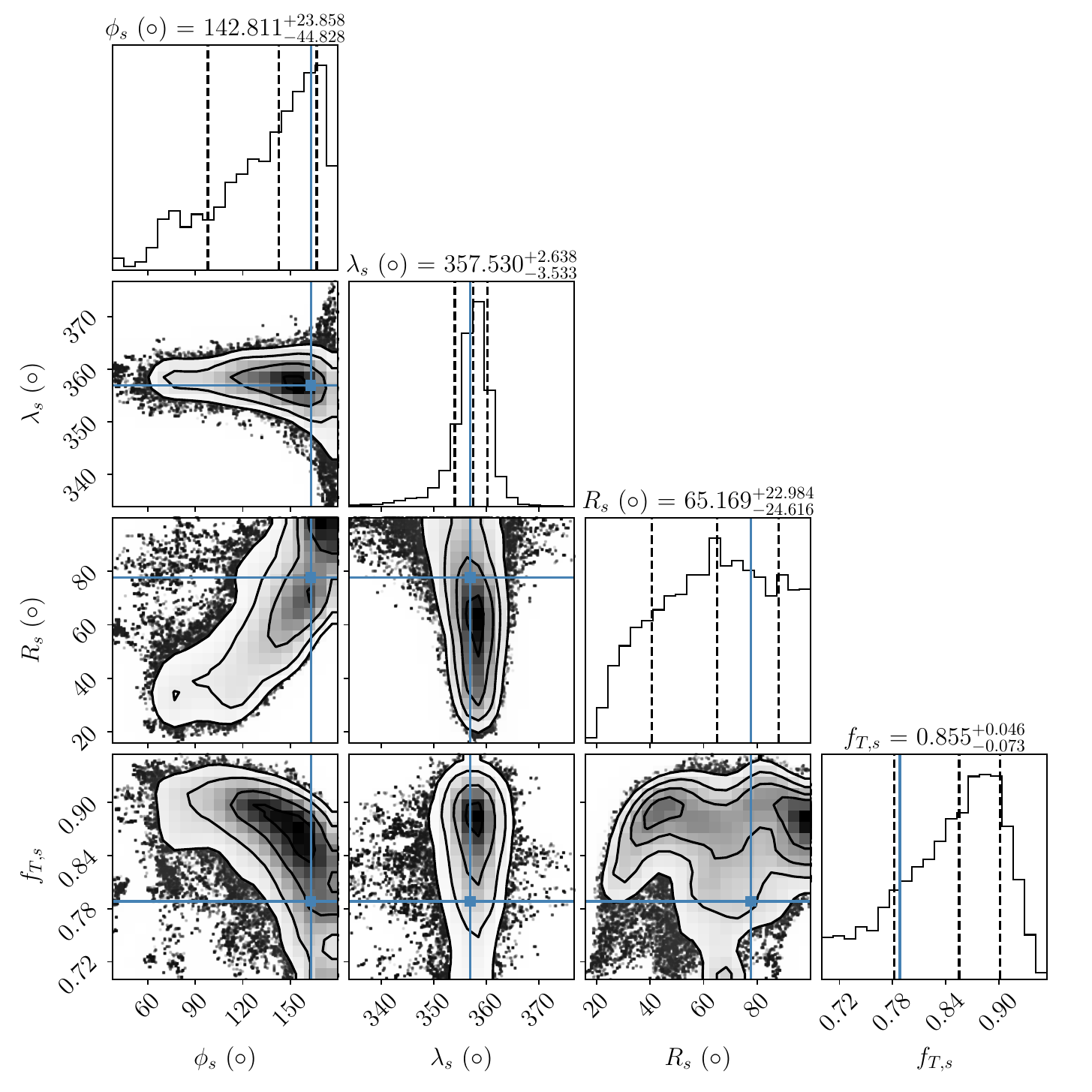} &
\includegraphics[width=0.45\textwidth]{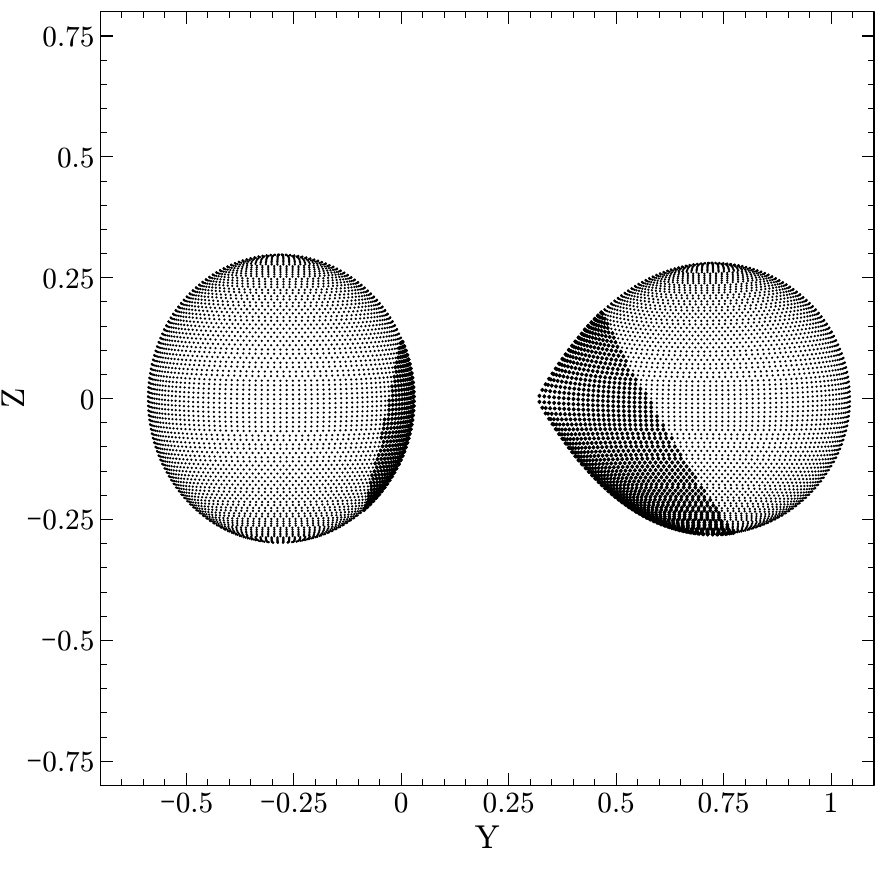} \\
\end{tabular}
\captionof{figure}{\texttt{LC$_{\texttt{MCMC}}$} results for the Z Vul system. Upper left: Posterior probability distributions and correlations of the global light curve parameters. Upper right and lower left: \texttt{MCMC} corner plots for the primary and secondary spot parameters. Lower right: 3D geometric representation of the binary components at 0.25 phase, illustrating the specific locations of the cold spots used to simulate the flux deficit from the accretion structure.}
\label{fig:LC_mcmc}
\end{center}
 
\section{Atmospheric Parameters}
\begin{center}
\includegraphics[height=0.45\textheight,keepaspectratio]{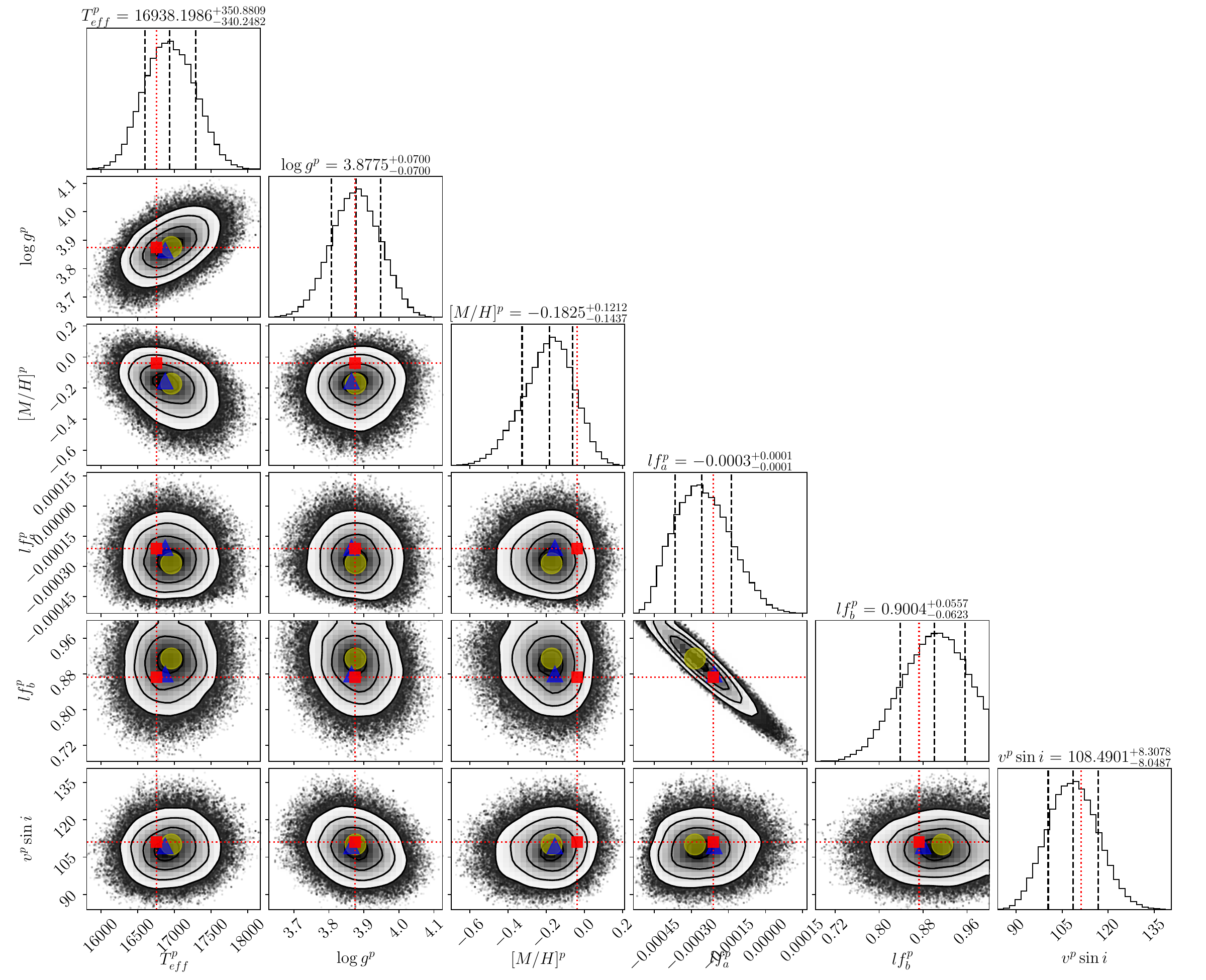}
\vfill
\includegraphics[height=0.45\textheight,keepaspectratio]{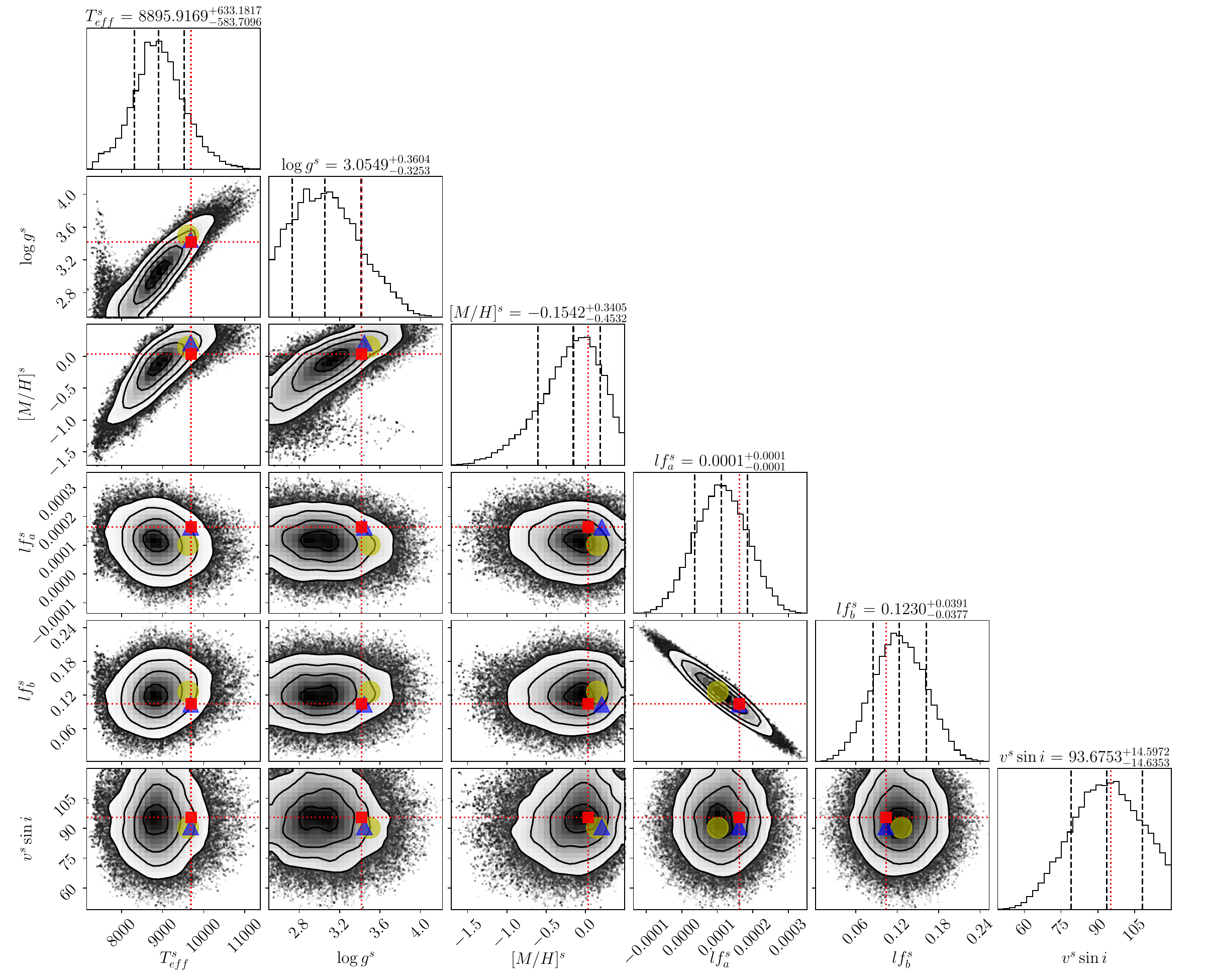}
\captionof{figure}{Atmospheric parameters ($T_{\rm eff}$, $\log g$, $[M/H]$, $v\sin i$) and fractional light contributions ($l_f$) derived for the primary and secondary components of Z\,Vul across the sequential \texttt{iSpec$_{\texttt{MCMC}}$} optimization stages. Yellow circles denote the results from Run 2, blue triangles represent Run 3, and red squares indicate the final maximum likelihood solutions from the Run 4 \texttt{MCMC} chains. The baseline unconstrained parameter space exploration (Run 1) and the individual component posterior distributions are displayed as the underlying grid contours.}
\label{fig:Atmospheric_corner}
\end{center}
 
\section{Abundance}
\begin{center}
\includegraphics[width=0.95\textwidth]{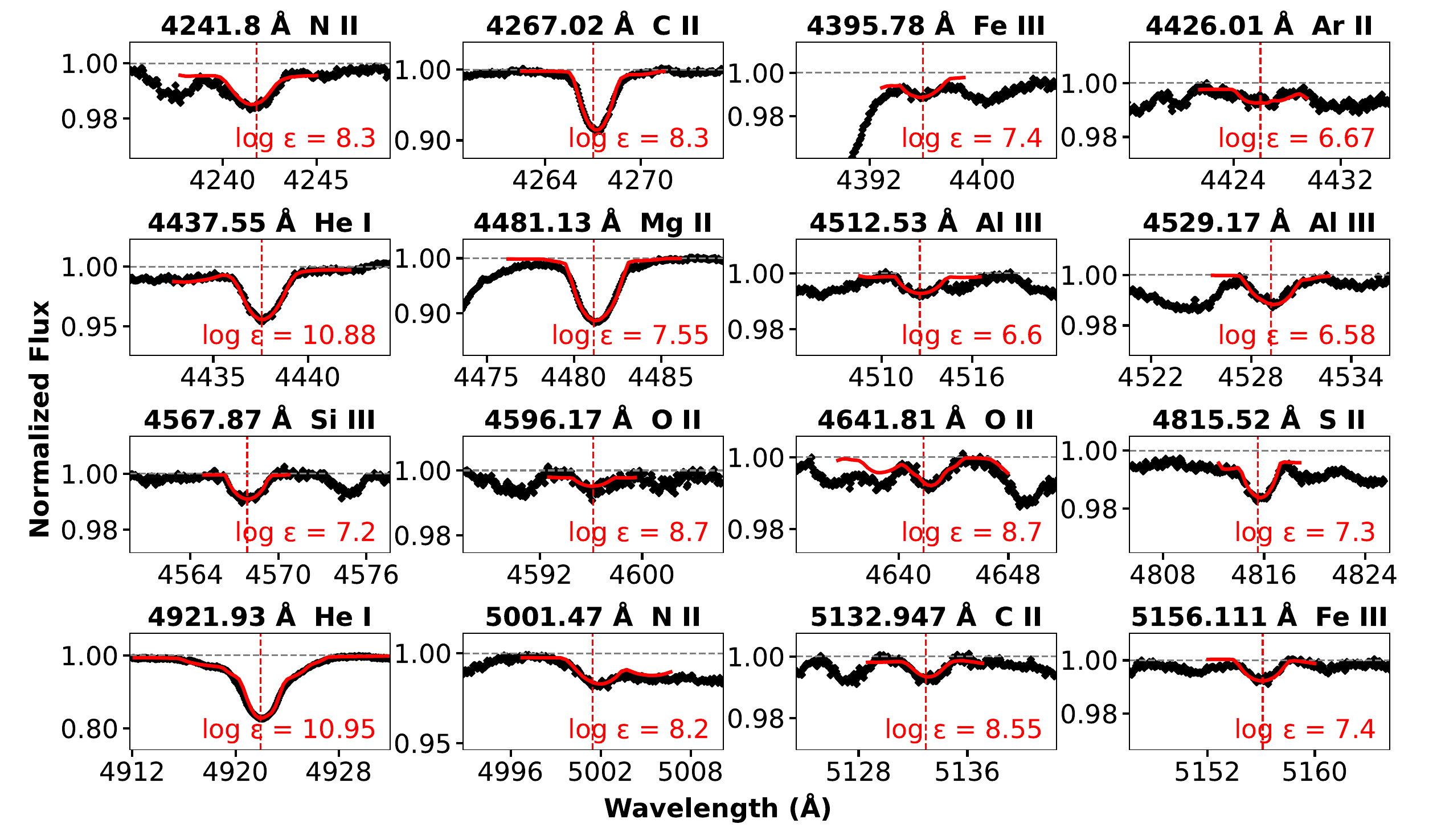}
\captionof{figure}{Comparison of observed and synthetic line profiles for the Z\,Vul primary. The observed {\sc hermes} spectrum (black dots) is shown with the best matching NLTE synthesis (red line). The continuum is indicated by the horizontal dashed line. The $log \epsilon$ label in each panel gives the final derived abundance in dex for that element.}
\label{fig:abundanceI}
\end{center}
 
\begin{center}
\includegraphics[width=0.95\textwidth]{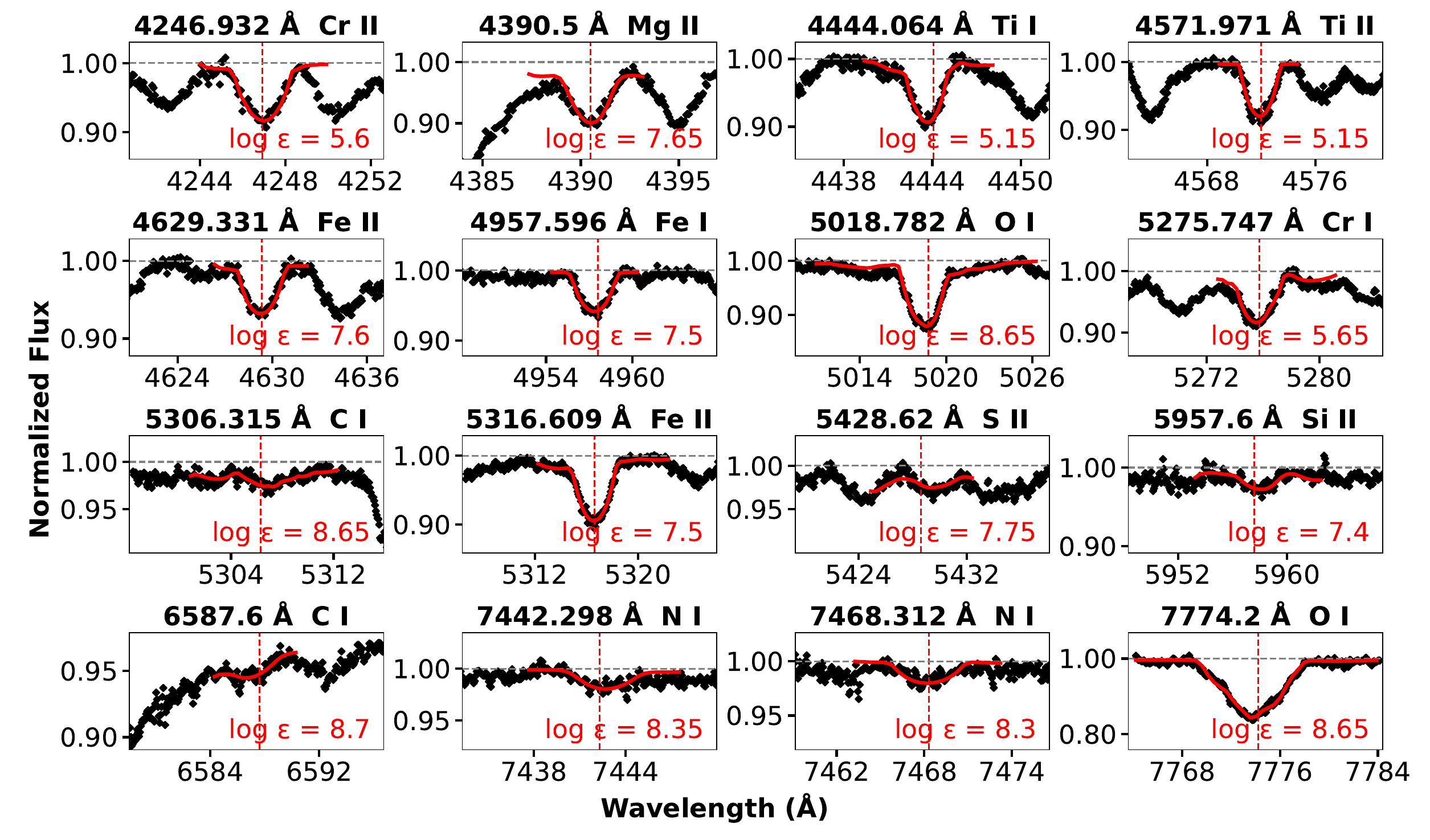}
\captionof{figure}{Comparison of observed and synthetic line profiles for the Z\,Vul secondary. The observed {\sc hermes} spectrum (black dots) is shown with the best-matching NLTE synthesis (red line). The continuum is indicated by a horizontal dashed line. The $log \epsilon$ label in each panel gives the final derived abundance in dex for that element.}
\label{fig:abundanceII}
\end{center}
 
\section{Evolutionary Models}

\begin{center}
\captionof{table}{Derived initial progenitor parameters for the statistically robust solution space of Z\,Vul, listing the properties of the evolutionary models that survived the $\chi^2$ confidence filtering across both AM loss regimes (AM$_1$ and AM$_2$). The tabulated values comprise the initial component masses ($M_{\rm d}^{\rm i}$, $M_{\rm g}^{\rm i}$), initial orbital periods ($P^{\rm i}$), initial mass ratios ($q^{\rm i}$), and mass-transfer efficiencies ($\beta$) corresponding to the $1\sigma$ confidence limits derived via survival analysis within the $q^{\rm i}$--$\beta$ parameter space.}
\label{tab:survive_models}
\resizebox{\linewidth}{!}{%
\begin{tabular}{ccccccccccccccc}
\toprule
       $q^i$        & $\beta$  & $M_d^i$   & $M_g^i$   & $M_{tot}$  & $P^i$     & $q^f$     & $M_d^f$  & $M_g^f$   & $P^f$     & $\chi^2_R$ & [C/N]$_d$ & [C/N]$_g$ & [He/H]$_d$ & [He/H]$_g$\\
\midrule
   \multicolumn{15}{c}{\textbf{AM$_1$}} \\
    \midrule
	      1.25	      &0.00	     &4.817	     &3.853	     &8.670	      &1.339      &0.410      &2.522	 &6.148	     &2.297      &1.792	      &0.790	  &2.479	  &0.402	   &0.400\\
	    1.25	    &0.05	   &4.885	   &3.908	   &8.793	    &1.328	    &0.413	    &2.537	   &6.138	   &2.278	   &2.280       &0.740	    &2.505	    &0.402	     &0.400\\
	      1.30	      &0.00	     &4.900	     &3.770	     &8.670	      &1.358      &0.398      &2.470	 &6.200	     &2.383      &1.000	      &0.588	  &2.143	  &0.403	   &0.400\\
	    1.30	    &0.05	   &4.972	   &3.825	   &8.797	    &1.346	    &0.402	    &2.489	   &6.184	   &2.355	   &1.277       &0.532	    &2.172	    &0.403	     &0.400\\
	      1.30	      &0.10	     &5.048	     &3.883	     &8.932	      &1.333      &0.408      &2.517	 &6.162	     &2.314      &1.934	      &0.479	  &2.209	  &0.404	   &0.400\\
	      1.35	      &0.00	     &4.981	     &3.689	     &8.670	      &1.379      &0.388      &2.425	 &6.245	     &2.465      &1.040	      &0.422	  &1.729	  &0.404	   &0.400\\
	      1.35	      &0.05	     &5.056	     &3.745	     &8.801	      &1.365      &0.394      &2.451	 &6.220	     &2.418      &1.046	      &0.383	  &1.779	  &0.405	   &0.400\\
	      1.35	      &0.10	     &5.136	     &3.805	     &8.941	      &1.350      &0.400      &2.478	 &6.197	     &2.375      &1.405	      &0.361      &1.860	  &0.406	   &0.400\\
	      1.35	      &0.15	     &5.221	     &3.868	     &9.089	      &1.335      &0.406      &2.506	 &6.176	     &2.336      &2.063	      &0.342	  &1.969	  &0.407	   &0.400\\
	      1.40	      &0.05	     &5.136	     &3.669	     &8.805	      &1.387      &0.386      &2.415	 &6.255	     &2.482      &1.255	      &0.291	  &1.410	  &0.409	   &0.401\\
	    1.40	    &0.10	   &5.220	   &3.729	   &8.949	    &1.370	    &0.392	    &2.442	   &6.229	   &2.434	   &1.302       &0.275	    &1.515	    &0.409	     &0.401\\
	    1.40	    &0.00	   &5.058	   &3.612	   &8.670	    &1.403	    &0.380	    &2.387	   &6.283	   &2.537	   &1.648       &0.307	    &1.342	    &0.408	     &0.401\\
	      1.40	      &0.15	     &5.309	     &3.792	     &9.102	      &1.352      &0.399      &2.474	 &6.202	     &2.383      &1.740	      &0.266	  &1.625	  &0.411	   &0.401\\
	    1.40	    &0.20	   &5.405	   &3.860	   &9.265	    &1.333	    &0.406	    &2.506	   &6.179	   &2.339	   &2.589       &0.254	    &1.774	    &0.411	     &0.401\\
	    1.45	    &0.10	   &5.301	   &3.656	   &8.957	    &1.392	    &0.385	    &2.410	   &6.258	   &2.489	   &1.514       &0.214	    &1.191	    &0.414	     &0.401\\
	    1.45	    &0.15	   &5.394	   &3.720	   &9.115	    &1.372	    &0.392	    &2.442	   &6.229	   &2.434      &1.678       &0.205	    &1.308	    &0.415	     &0.401\\
	    1.45	    &0.05	   &5.214	   &3.596	   &8.809	    &1.411	    &0.378	    &2.378	   &6.290	   &2.552	   &1.872       &0.220	    &1.094	    &0.413	     &0.401\\
	    1.45	    &0.20	   &5.494	   &3.789	   &9.283	    &1.351	    &0.400	    &2.480	   &6.200	   &2.374	   &2.478       &0.203	    &1.402	    &0.418	     &0.401\\
	    1.50	    &0.15	   &5.476	   &3.651	   &9.127	    &1.393	    &0.386	    &2.413	   &6.254	   &2.482	   &1.891       &0.160	    &1.017	    &0.421	     &0.401\\
	    1.50	    &0.10	   &5.379	   &3.586	   &8.965	    &1.416	    &0.378	    &2.378	   &6.287	   &2.547	   &2.066       &0.166	    &0.917	    &0.420	     &0.402\\
	    1.55	    &0.15	   &5.555	   &3.584	   &9.139	    &1.416	    &0.380	    &2.384	   &6.279	   &2.534	   &2.355       &0.124	    &0.780	    &0.428	     &0.402\\
	    1.55	    &0.20	   &5.663	   &3.654	   &9.317	    &1.390	    &0.388	    &2.424	   &6.244	   &2.463	   &2.526       &0.121	    &0.892	    &0.429	     &0.402\\
\midrule
\multicolumn{15}{c}{\textbf{Possible progenitor parameters of Z Vul for AM$_1$ regime}}\\
\midrule
        $1.370$     &$0.070$    &$5.135$    &$3.733$    &$8.868$    &$1.370$	   &$0.393$	   &$2.447$	   &$6.224$	   &$2.427$	   &$1.00$ &$0.366$	   &$1.658$	   &$0.408$	    &$0.401$\\
        $\pm0.070$ &$\pm0.056$ &$\pm0.185$ &$\pm0.185$ &$\pm0.165$ &$\pm0.024$ &$\pm0.010$ &$\pm0.044$ &$\pm0.042$ &$\pm0.076$ &  &$\pm0.172$ &$\pm0.446$ &$\pm0.005$  &$\pm0.001$\\
\midrule
\multicolumn{15}{c}{\textbf{AM$_2$}} \\
\midrule
	    1.25       &0.00	   &4.817	   &3.853	   &8.670	   &1.339	   &0.410	   &2.522	   &6.148	   &2.297	   &1.792  &0.790	   &2.479	   &0.402	   &0.400\\
	    1.30       &0.00	   &4.900	   &3.770	   &8.670	   &1.358	   &0.398	   &2.470	   &6.200	   &2.383	   &1.000  &0.588	   &2.143	   &0.403	   &0.400\\
	    1.35       &0.00	   &4.981	   &3.689	   &8.670	   &1.379	   &0.388	   &2.425	   &6.245	   &2.465	   &1.040  &0.422	   &1.729	   &0.404	   &0.400\\
	    1.35       &0.05	   &5.056	   &3.745	   &8.801	   &1.326	   &0.406	   &2.505	   &6.169	   &2.322	   &1.791  &0.456	   &1.992	   &0.404	   &0.400\\
	    1.40       &0.00	   &5.058	   &3.612	   &8.670	   &1.403	   &0.380	   &2.387	   &6.283	   &2.537	   &1.648  &0.307	   &1.342	   &0.408	   &0.401\\
	    1.40       &0.05	   &5.136	   &3.669	   &8.805	   &1.348	   &0.397	   &2.464	   &6.208	   &2.393	   &1.296  &0.344	   &1.635	   &0.406	   &0.401\\
	    1.45       &0.05	   &5.214	   &3.596	   &8.809	   &1.373	   &0.389	   &2.426	   &6.243	   &2.462	   &1.303  &0.263	   &1.284	   &0.410	   &0.401\\
	    1.45       &0.10	   &5.301	   &3.656	   &8.957	   &1.316	   &0.406	   &2.506	   &6.171	   &2.318	   &2.498  &0.300	   &1.570	   &0.408	   &0.401\\
	    1.50       &0.05	   &5.288	   &3.525	   &8.813	   &1.400	   &0.380	   &2.389	   &6.279	   &2.534	   &1.791  &0.202	   &0.983	   &0.415	   &0.401\\
	    1.50       &0.10	   &5.379	   &3.586	   &8.965	   &1.342	   &0.398	   &2.470	   &6.204	   &2.382	   &1.962  &0.232	   &1.232	   &0.413	   &0.401\\
	    1.55       &0.05	   &5.359	   &3.457	   &8.816	   &1.428	   &0.373	   &2.355	   &6.311	   &2.604	   &2.667  &0.157	   &0.750	   &0.421	   &0.402\\
	    1.55       &0.10	   &5.454	   &3.519	   &8.972	   &1.369	   &0.390	   &2.433	   &6.237	   &2.450	   &1.841  &0.179	   &0.945	   &0.418	   &0.401\\
	    1.60       &0.10	   &5.526	   &3.454	   &8.980	   &1.398	   &0.383	   &2.403	   &6.265	   &2.511	   &2.150  &0.144	   &0.728	   &0.426	   &0.403\\
\midrule
\multicolumn{15}{c}{\textbf{Possible progenitor parameters of Z Vul for AM$_2$ regime}}\\
\midrule
        $1.405$    &$0.040$    &$5.128$    &$3.656$    &$8.785$    &$1.363$	   &$0.394$	   &$2.449$	   &$6.222$	   &$2.423$	   &$1.00$ &$0.378$	   &$1.576$	   &$0.408$	   &$0.401$\\
        $\pm0.096$ &$\pm0.039$ &$\pm0.206$ &$\pm0.109$ &$\pm0.117$ &$\pm0.026$ &$\pm0.010$ &$\pm0.045$ &$\pm0.044$ &$\pm0.082$ &  &$\pm0.192$ &$\pm0.510$ &$\pm0.005$ &$\pm0.001$\\
\bottomrule
\end{tabular}}
\end{center}


\bsp	
\label{lastpage}
\end{document}